\documentclass{amsart}
\def\cal{\mathcal}

\theoremstyle{definition}

\theoremstyle{remark}

\numberwithin{equation}{section} \font\teneurm=eurm10
\font\seveneurm=eurm7 \font\fiveeurm=eurm5
\newfam\eurmfam
\textfont\eurmfam=\teneurm \scriptfont\eurmfam=\seveneurm
\scriptscriptfont\eurmfam=\fiveeurm

 \font\teneusm=eusm10 \font\seveneusm=eusm7 \font\fiveeusm=eusm5
\newfam\eusmfam
\textfont\eusmfam=\teneusm \scriptfont\eusmfam=\seveneusm
\scriptscriptfont\eusmfam=\fiveeusm
\def\eusm#1{{\fam\eusmfam\relax#1}}
\font\tencmmib=cmmib10 \skewchar\tencmmib='177
\font\sevencmmib=cmmib7 \skewchar\sevencmmib='177
\font\fivecmmib=cmmib5 \skewchar\fivecmmib='177
\newfam\cmmibfam
\textfont\cmmibfam=\tencmmib \scriptfont\cmmibfam=\sevencmmib
\scriptscriptfont\cmmibfam=\fivecmmib
\def\cmmib#1{{\fam\cmmibfam\relax#1}}
\def\EUN{\eusm N}
\def\R{{\Bbb{R}}}\def\Z{{\Bbb{Z}}}
\def\EUB{\eusm B}
\begin{document}

\title{Gauge Theory And Wild Ramification}

\author{Edward Witten}
\address{School of Natural Sciences, Institute for Advanced Study, Princeton NJ 08540}
\email{witten@ias.edu}
\thanks{Supported in part by NF Grant PhD-0503584.}

%    General info
%\subjclass[2000]{81Q99, 53D45}

\date{February, 2007}

\def\Bbb{\mathbb}
%\keywords{geometric Langlands, electric-magnetic duality,
%isomonodromy, irregular singularities}

\begin{abstract}
The gauge theory approach to the geometric Langlands program is
extended to the case of wild ramification.  The new ingredients that
are required, relative to the tamely ramified case, are differential
operators with irregular singularities, Stokes phenomena,
isomonodromic deformation, and, from a physical point of view, new
surface operators associated with higher order singularities.
\end{abstract}

\maketitle

\input epsf
\section{Introduction}
\label{intro}
\def\hat{\widehat}
\def\bar{\overline}
\def\C{{\Bbb{C}}}

\def\MH{{{\eusm M}_H}}
\def\hat{\widehat}
\def\tilde{\widetilde}
\def\neg{\negthinspace}
 The geometric Langlands program describes an analog in
geometry of the Langlands program of number theory and is
intimately connected with many topics in mathematical physics.  It
has been extensively studied via two-dimensional conformal field
theory \cite{BD}, \cite{FG}, \cite{F}, and more recently via
four-dimensional gauge theory  with electric-magnetic duality
\cite{KW}.  That last paper contains more detailed references.

The gauge theory in question is a topologically twisted version of
$\EUN=4$ super Yang-Mills theory. As was first argued  in
\cite{BJV}, \cite{HMS}, electric-magnetic duality in this situation
reduces in two dimensions to mirror symmetry of Hitchin fibrations
\cite{H}, \cite{H2}. The particular case of mirror symmetry that is
relevant here was first studied mathematically in \cite{HT}.

The simplest form of the geometric Langlands program deals with a
flat connection on a Riemann surface $C$.  However, the analogy
with number theory motivates the extension to incorporate
``ramification,'' that is to consider a flat connection on $C$
with singularities of a prescribed nature.  The singularities may
be simple poles, corresponding to\footnote{Sometimes the term
``tame ramification'' is used more narrowly to refer to the case
that the residues of the poles are nilpotent.  For our purposes,
it simply means that the singularities are simple poles, or more
precisely that this is so if one suitably extends the holomorphic
structure of the bundle over the singular points.} ``tame
ramification,'' or poles of higher order, in which case one speaks
of ``wild ramification.''

The gauge theory approach to tame ramification has been described in
detail recently \cite{GW}.  The main novelty, from a physics point
of view, is the need to enrich $\EUN=4$ super Yang-Mills theory with
``surface operators,'' which are characterized by prescribed
singularities on codimension two surfaces in spacetime. The
appropriate singularities appear in solutions of Hitchin's equations
with tame singularities; these solutions were first described in
\cite{S}. After supplementing the parameters that appear in the
classical description of these singularities with certain quantum
parameters (theta-like angles), it was possible in \cite{GW} to
describe an action of electric-magnetic duality on a certain family
of (half-BPS) supersymmetric  surface operators. This led to a
natural gauge theory description of the tame case of the geometric
Langlands correspondence.  For an elegant supergravity analysis of
the relevant family of surface operators, see \cite{Gom}.

The purpose of the present paper is to extend the gauge theory
approach to the case of wild ramification.  This depends on
overcoming two major obstacles and interpreting the results in
quantum field theory.  As it turns out, at the classical level the
two obstacles have already been dealt with in the literature.

The first obstacle is that at first sight the higher order
singularities relevant to wild ramification  look incompatible
with Hitchin's equations. We can write Hitchin's equations very
schematically in the form $d\Phi+\Phi^2=0$, where $\Phi$ combines
the connection and Higgs field (which we usually denote as $A$ and
$\phi$, respectively). For now, it is not necessary to describe
Hitchin's equations more precisely.  Tame ramification means that
at a point on a Riemann surface that is labeled as $z=0$ in terms
of some local parameter $z$, we have a singularity with
$|\Phi|\sim 1/|z|$ (possibly up to logarithms). With  this
behavior of $\Phi$, both $d\Phi$ and $\Phi^2$ are of order
$1/|z|^2$ for small $z$, so it is natural, as in \cite{S}, to look
for solutions of Hitchin's equations of this form.

But for wild ramification, we want $|\Phi|\sim 1/|z|^n$ with
$n>1$, and then the equation $d\Phi+\Phi^2=0$ looks unnatural, as
it seems that $\Phi^2$ will be more singular than $d\Phi$.  As
shown in \cite{BB}, following earlier work \cite{Sa}, the
resolution of this point is simply that the relevant singular
behavior of Hitchin's equations can be modeled by abelian
solutions, in which $d\Phi$ and $\Phi^2$ both vanish. There is no
problem in finding abelian solutions of Hitchin's equations with
poles of arbitrary order. It is perhaps surprising that abelian
solutions (possibly twisted by an element of the Weyl group) are
sufficient for modeling the local singularity, but this follows
from classical facts about irregular singularities.

The second problem that must be overcome is particularly vexing at
first sight, although again, the resolution involves facts that
are known and are summarized or developed in \cite{B}. In the
unramified case, the geometric Langlands correspondence begins
with a flat connection on a Riemann surface $C$. Such a flat
connection has a topological interpretation, independent of the
complex structure on $C$, since it determines a homomorphism to
$G_\C$ of the fundamental group of $C$. Likewise, in the tamely
ramified case, one deals with flat connections on a punctured
Riemann surface $C'=C\backslash \{p_1,p_2,\dots,p_s\}$ (that is,
$C$ with the $s$ points $p_1,p_2,\dots,p_s$ omitted) whose
singularities are simple poles at the punctures.  Again, such a
connection has a topological interpretation, in terms of a
homomorphism to $G_\C$ of the fundamental group of $C'$.

This is all in accord with the fact that the gauge theory approach
to the geometric Langlands correspondence begins with a twisted
topological field theory in four dimensions.  The underlying
topological invariance means that the ingredients that appear
after reduction to two dimensions must have a topological
interpretation.

In the wildly ramified case, however, the starting point is a flat
connection whose singularities are poles of order greater than 1.
Such flat connections depend on parameters that in general cannot
be given a topological interpretation.  For example, consider in
the holomorphic setting\footnote{Exactly how to relate a solution
of Hitchin's equations to a connection in this holomorphic sense
is explained at the beginning of section \ref{backgr}.} a
connection with a pole at $z=0$:
\begin{equation}\label{dogus}{\mathcal A}=dz\left(\frac{T_n}
{z^n}+\frac{T_{n-1}}{z^{n-1}}+\dots +\frac{T_1}{
z}+\dots\right).\end{equation}
 Here $T_1,T_2,\dots,T_n$ are elements of
$\mathfrak g_\C$, the Lie algebra of $G_\C$.

A flat connection on $C'$  which is allowed to have singularities
of this nature depends on more parameters, namely $T_2,\dots,T_n$,
than a flat connection with only simple poles. But whatever the
order of the poles, the only obvious topological invariant of  a
flat connection is the monodromy, or in other words the
representation of the fundamental group of $C'$, that it
determines.   So the extra information associated with wild
ramification does not appear to have any topological meaning.

Related to this, $T_1$, being the residue of the holomorphic
differential $\mathcal A$, is independent of the choice of local
parameter $z$. But the remaining elements $T_i$, $i\geq 2$, do
depend on the choice of  local parameter.  So they scarcely can be
meaningful parameters in a topological field theory.

The resolution of this conundrum involves the theory of Stokes
phenomena in ordinary differential equations with irregular
singularity. Some of the information contained in a flat
connection with irregular singularity does have a topological
meaning, in terms of a generalized monodromy that includes the
Stokes matrices.  The remaining information that characterizes an
irregular singularity can be varied by a natural process of
isomonodromic deformation \cite{JMU}, preserving the generalized
monodromy.  A natural quantum field theory argument shows that the
relevant information in our problem is invariant under
isomonodromy.

The generalized monodromy parametrizes a variety  that, just like
the more familiar moduli spaces of representations of the
fundamental group, has a natural complex symplectic structure.
Moreover, and crucially for our application, this structure is
invariant under isomonodromy \cite{B}, a fact which turns out to
have a natural interpretation in $\EUN=4$ super Yang-Mills theory.
Additionally, the relevant variety can be interpreted \cite{BB} as
a moduli space of solutions of Hitchin's equations with a suitable
singularity. This leads to a Hitchin fibration and mirror
symmetry, just as in the unramified or tamely ramified case.  This
instance of mirror symmetry is related to four-dimensional gauge
theory, and leads to the geometric Langlands duality, just as in
those cases. The duality commutes with isomonodromic deformation.

Section 2 of this paper is devoted mainly to an introduction to
Stokes phenomena; none of this material is new. However, section
\ref{strategy} contains a further explanation of the strategy of
this paper.

In section 3, we introduce surface operators associated with wild
ramification.  In section 4, we offer a supersymmetric perspective
on isomonodromy.  Sections 5 and 6 describe the application to
geometric Langlands. In section 5, we consider the case that the
coefficient $T_n$ of the leading singularity is regular semi-simple;
in section 6, this assumption is relaxed.  Finally, some examples of
Stokes phenomena are described in an appendix.

I would like to thank O. Biquard, P. Boalch, E. Frenkel, and D.
Gaitsgory for explanations of their work and P. Deligne, S. Gukov,
N. Hitchin, and I. M. Singer for helpful discussions.

\section{Review Of Stokes Phenomena}
\label{backgr}

\def\Tr{{\rm Tr}}
\def\CA{{\mathcal A}}
\def\CF{{\mathcal F}}
We consider $\EUN=4$ super Yang-Mills theory with gauge group $G$
on a four-manifold $M$ that is a product of Riemann surfaces,
$M=\Sigma\times C$; $C$ is the Riemann surface on which we will
study the geometric Langlands program. The theory works for any
compact $G$, but for brevity we frequently take $G$ to be simple
and connected.

The fields that will be important in the discussion are a
connection $A$ on a $G$-bundle $E\to C$, and a field $\phi$ that
is a one-form on $C$ with values in ${\rm ad}(E)$.  Physically,
$\phi$ arises by the twisting procedure applied to some of the
scalar fields of $\EUN=4$ super Yang-Mills theory.  Asking for a
pair $(A,\phi)$ to preserve supersymmetry gives Hitchin's
equations:
\begin{align}\label{hitchin}
F-\phi\wedge\phi &= 0 \nonumber \\
D\phi = D\star \phi &= 0. \\ \nonumber \end{align} Here $\star$ is
the Hodge star operator.

Hitchin's equations imply among other things that the complex-valued
connection $\CA=A+i\phi$ is flat, or in other words that its
curvature $\CF=d\CA+\CA\wedge \CA$ vanishes.  The gauge-covariant
exterior derivative $d_\CA$ can be decomposed as
$d_\CA=\partial_\CA+\bar\partial_\CA$, where $\partial_\CA$ and
$\bar\partial_\CA$ are of types $(1,0)$ and $(0,1)$, respectively.
The operator $\bar\partial_\CA$ determines a complex structure on
the bundle\footnote{In our notation, we generally do not distinguish
$E$ from its complexification.} $E$, at least away from possible
singularities.

The present section will be devoted to describing some aspects of
the behavior in the presence of singularities.  We consider a
solution of Hitchin's equations with a singularity at a point $p$.
We choose a  local coordinate $z$ so that $p$ is the point $z=0$.
The bundle $E$ can be extended over $p$ as a holomorphic bundle,
though not as a flat bundle.  (The holomorphic extension is not
quite unique, something that played an important role in \cite{GW}
and will be incorporated below.)  Like any holomorphic bundle, $E$
is trivial locally.  Once a trivialization is picked in a
neighborhood of $p$, the operator $\bar\partial_\CA$ reduces in
that neighborhood to the standard operator $\bar\partial=d\bar z
\,\partial/\partial \bar z$.  Flatness of the connection $\CA$ is
now equivalent to the statement that $\CA_z$, defined by
$\partial_\CA=dz\left(\partial_z+\CA_z\right)$, is holomorphic
away from $p$.  $\CA_z$ may be singular at $p$, since the bundle
$E$ is only flat away from $p$.  We are interested in the case
that the singularity of $\CA_z$ is a pole:
\begin{equation}\label{tongo}\CA_z=\frac{T_n}{z^n}+\frac{T_{n-1}}{z^{n-1}}+\dots+\frac
{T_1}{z}+\dots \end{equation} for some positive integer $n$. The
last ellipses refer to regular terms.

Away from the point $z=0$, a covariantly constant section $\Psi$
of the flat bundle $E$ must be annihilated by
$\bar\partial_A=\bar\partial$, and thus is holomorphic in the
usual sense. It also must obey $\partial_A\Psi=0$ or
\begin{equation}\label{diffop}
\left(\frac{\partial}{\partial
z}+\CA_z\right)\Psi=0.\end{equation} The differential equation
(\ref{diffop}) for the holomorphic object $\Psi$ is said to have a
regular singularity at $z=0$ if $n=1$, and an irregular
singularity if $n>1$.

In the rest of this section, we will give a brief synopsis of a
few facts about linear differential equations with such an
irregular singularity.  We first explain a few basic facts about
Stokes phenomena; much more can be found in classical references
such as \cite{Wa}, \cite{BJL}. Then we briefly describe the notion
of isomonodromic deformation \cite{JMU} and its symplectic nature
\cite{B}.  (Papers \cite{JMU} and \cite{B} also contain
introductions to the Stokes phenomena.  See also \cite{M},
\cite{MMS} for string theory papers with some applications of
Stokes phenomena.) The aim is only to explain the minimum that is
needed for the rest of this paper.

As in many treatments of irregular singularities, we will make in
much of this paper the simplifying assumption  that $T_n$ is
regular and semisimple.  If $G_\C$ is $SL(N,\C)$ or $GL(N,\C)$,
this means that $T_n$ can be diagonalized and has distinct
eigenvalues. For any simple $G_\C$, it means that $T_n$ can be
conjugated to a Cartan subalgebra, and that the subgroup of $G_\C$
that commutes with $T_n$ is precisely $\Bbb{T}_\C$. Assuming that
$T_n$ is regular and semisimple will enable us to describe a
little more simply the main points of the gauge theory approach to
wild ramification.  In section \ref{ugo}, we sketch what is
involved in relaxing the assumption about $T_n$.

In our very schematic introduction to Stokes phenomena, to avoid
an inessential extra layer of abstraction, we will assume that
$G_\C$ is $GL(N,\C)$ or $SL(N,\C)$.  The general case is similar
with triangular matrices replaced by elements of Borel subgroups.
See section 2 of \cite{Btwo} for a brief explanation.

\subsection{Preliminaries}\label{prelim}

$C^*$ will denote a small disc in the complex $z$-plane with the
point $z=0$ omitted.  We consider a differential equation with an
irregular singularity at $z=0$.
 The first question to consider is up to what type of
equivalences such singularities should be classified.

{}From a topological point of view, if we allow arbitrary gauge
transformations, the only invariant of a flat connection on $C^*$ is
the holonomy or monodromy around the origin.  From a holomorphic
point of view, the analogous statement is that, if we allow
holomorphic gauge transformations of the holomorphic differential
equation (\ref{diffop}) that may have essential singularities at
$z=0$, then the monodromy is the only invariant.

In fact, if the monodromy is trivial, then integrating along a path
gives a $G_\C$-valued function
$g(z)=P\exp\left(-\int_{z_0}^z\CA(z')\right)$.  Here, the point $z$,
the base-point $z_0$, and the path of integration are taken to lie
in $C^*$; $g(z)$ is independent of the path because the monodromy
vanishes. A gauge transformation by $g^{-1}$ sets $\CA=0$. A similar
procedure shows that any two holomorphic connections with the same
monodromy are gauge-equivalent if we allow gauge transformations of
this type.

However, in the case of an irregular singularity, $g(z)$ has an
essential singularity at $z=0$.  In studying irregular
singularities, we do not want to allow gauge transformations with
an essential singularity, since as we have just seen this will not
lead to an interesting theory. Instead, we allow only gauge
transformations that are meromorphic\footnote{Since we will
ultimately study wild ramification via gauge theory and
topological field theory, we also need to know the gauge theory
equivalent of this restriction. One version is described in
section 4 of \cite{B}.   Another version, which involves Hitchin's
equations and restriction to unitary gauge transformations, is
described in \cite{BB} and reviewed in section \ref{hitchwild}
below.} at $z=0$.

Stokes phenomena arise because there is a crucial difference
between gauge transformations that are meromorphic in a
neighborhood of $z=0$ and gauge transformations that can only be
defined in a formal Laurent series near $z=0$. We will give a
simple example to show why this must be so.

Given our assumption that the coefficient $T_n$ of the leading
singularity is regular and semisimple, it is possible order by
order in powers of $z$ to make $\CA$ diagonal. In explaining why,
we take $G_\C=SL(2,\C)$ to keep the notation simple. Since $T_n$
is regular and semisimple, we have
\begin{equation}\label{mork}\CA_z=\frac{1}{z^n}\begin{pmatrix}w& 0\\ 0 & -w\\
\end{pmatrix}+\dots \end{equation}
where $w\not=0$, and the ellipses refer to terms less singular
than $z^{-n}$. Consider a gauge transformation generated by
\begin{equation}\label{ork}\epsilon=\begin{pmatrix}0& \sum_{r=1}^\infty f_rz^r\\
\sum_{r=1}^\infty g_rz^r & 0 \\ \end{pmatrix},\end{equation} with
formal power series $\sum_{r=1}^\infty f_rz^r$ and
$\sum_{r=1}^\infty g_rz^r$.  Under a gauge transformation, to
first order $\CA$ transforms by
$\delta\CA=-\partial_\CA\epsilon=-\partial\epsilon/\partial
z-[\CA_z,\epsilon].$  For $n>1$ (or for $n=1$ if $2w\notin
\Bbb{Z}$) the coefficients $f_r$ and $g_r$ can be determined
inductively to set the off-diagonal part of $\CA$ to zero.  The
key point is that, since $w\not=0$, for any $a,b$ (representing
off-diagonal terms in $\CA$ that we wish to eliminate) one can
find $f,g$ (two of the coefficients in
eqn. (\ref{ork})) such that
\begin{equation}\label{funky}\left[\begin{pmatrix}w& 0\\0 & -w \\ \end{pmatrix},\begin{pmatrix}0& f\\
g & 0 \\ \end{pmatrix}\right]=\begin{pmatrix}0&a\\
b & 0 \\ \end{pmatrix}.\end{equation}

Now, let us see why the formal power series that diagonalizes $\CA$
cannot possibly converge, in general, in any neighborhood of $z=0$.
To illustrate the point, we will consider a special case with
$\CA_z=T_2/z^2+T_1/z$ exactly.  Moreover, we diagonalize $T_2$ as in
(\ref{mork}), and write out $T_1$ as an explicit $2\times 2$ matrix:
\begin{equation}\label{zork}\CA_z=\frac{1}{z^2}\begin{pmatrix}w& 0\\ 0 & -w\\
\end{pmatrix}+\frac{1}{z}\begin{pmatrix}v& b\\ c & -v\\
\end{pmatrix}.\end{equation}
We write $T_{1,D}={\rm diag}(v,-v)$ for the diagonal part of
$T_1$.  The formal diagonalization procedure replaces $\CA_z$ by
\begin{equation}\label{repl}\CA'_z=
\begin{pmatrix}\frac{w}{z^2}+\frac{v}{z}+\dots & 0\\ 0 & -\frac{w}{z^2}-\frac{v}{z}-\dots\\
\end{pmatrix},\end{equation} where the point is that modulo
regular terms, $\CA'_z$ coincides with the diagonal part of
$\CA_z$.  The regular terms in $\CA'_z$ do not coincide with the
analogous diagonal terms in $\CA_z$.

Because $\CA'_z$ is diagonal, the monodromy of the modified
connection $\CA'$ is trivial to compute and is $\exp(-2\pi i
T_{1,D})$.  However, this does not coincide with the monodromy of
the original connection $\CA$.  In fact, the conjugacy class of
the monodromy of $\CA$ is easily determined.  Because $\CA$ is
holomorphic throughout the whole punctured $z$-plane, its
monodromy around $z=0$ can be evaluated on a large circle at
infinity. To do so, we observe that $\CA=T_2/z^2+T_1/z$ can be
replaced by $\CA''=T_1/z$, since $T_2/z^2$ vanishes too rapidly
near infinity to contribute to the monodromy.  The monodromy of
$\CA''$ is just $\exp(-2\pi i T_1)$, and this gives the conjugacy
class of the monodromy of $\CA$.

Generically, $\exp(-2\pi i T_1)$ and $\exp(-2\pi i T_{1,D})$ are
not conjugate, so the holonomies of $\CA$ and $\CA'$ are
different.  What has gone wrong with the reduction from $\CA$ to
$\CA'$ is that the formal power series used to diagonalize $\CA$
has zero radius of convergence. (This can be verified explicitly
in some examples treated in the appendix.)

To classify irregular singularities, we want to consider not
formal power series, but only gauge transformations that are
meromorphic in a punctured neighborhood of $z=0$.  By such a gauge
transformation, we cannot make $\CA$ diagonal.  But there is no
problem with the diagonalization procedure up to any desired
finite order.  In particular, given that $T_n$ was assumed to be
regular semisimple, we can assume $\CA_z$ to take the form
\begin{equation}\label{jurfy}\CA_z=\frac{T_n}{z^n}+\frac{T_{n-1}}{z^{n-1}}+\dots+\frac
{T_1}{z}+{\mathcal B},
\end{equation}
where $T_n,\dots,T_1$ are diagonal and  ${\mathcal B}$ is regular
at $z=0$. It is convenient to write
\begin{equation}\label{convwr}\frac{T_n}{z^n}+\frac{T_{n-1}}{z^{n-1}}+\dots+\frac
{T_1}{z}={\rm diag}(R_1(z),R_2(z),\dots,R_N(z)),\end{equation}
with explicitly
\begin{equation}\label{onvwr} R_j(z)=
\frac{q_{jn}}{z^n}+\frac{q_{j,n-1}}{z^{n-1}}+\dots
+\frac{q_{j1}}{z},~~j=1,\dots,N.\end{equation} Further, we let
\begin{equation}\label{bonvr}Q_j(z)=\frac{q_{jn}}{(n-1)z^{n-1}}+\frac{q_{j,n-1}}{(n-2)z^{n-2}}+\dots
+{q_{j1}}(-\ln z), \end{equation} so that $dQ_j/dz=-R_j$.  To
define $Q_j$, it is necessary to pick a branch of $\ln z$, but the
choice will not be important.

Even after making them diagonal, the matrices
$T_n,T_{n-1},\dots,T_1$ are not quite uniquely determined.  A
meromorphic gauge transformation by \begin{equation}g={\rm
diag}(z^{s_1},z^{s_2},\dots,z^{s_N})\end{equation} with integer
exponents would change the eigenvalues of $T_1$ by the integers
$s_1,s_2,\dots, s_N$. Once we pick a particular $T_1$, we can limit
ourselves to gauge transformations that are holomorphic and
invertible at $z=0$. Making a choice of $T_1$ is equivalent to
picking a particular holomorphic extension over the singular point
at $z=0$ of the original flat bundle $E$ on the punctured disc
$C^*$.  After picking such an extension, we are still free to modify
$\CA$ by permuting the eigenvalues of $T_n$, that is, by a Weyl
transformation, and by holomorphic gauge transformations that are
diagonal up to order $z^n$.  In particular, we can make a gauge
transformation by a constant diagonal matrix, and this will be
important later in counting parameters.

\subsection{Stokes Rays}
\label{stokeslines}

Now we want to study covariantly constant sections $\Psi$ of the
flat bundle $E$, or equivalently holomorphic sections that obey
the differential equation $(\partial_z+\CA_z)\Psi=0$. If $\mathcal
B=0$ in (\ref{jurfy}), then a basis of such sections is given by
\begin{equation}\label{basis}\Psi_j=h_j
\,\exp(Q_j(z)),~~j=1,\dots,N,\end{equation} where the column
vector $h_j$ has a 1 in the $j^{th}$ position, with other entries
zero.  In general, with $\mathcal B$ regular at $z=0$ but not
necessarily zero, there is for each $j$ a unique formal power
series $ H_j(z)=h_j+{\mathcal O}(z)$ such that a basis of
solutions (in a formal power series) is given by
\begin{equation}\label{asis}\Psi_j(z)=H_j(z)\exp(Q_j(z)). \end{equation} The
existence of the formal power series $H_j(z)$ is more or less
equivalent to the statement that a gauge transformation
diagonalizing $\CA$ can be found as a formal power series.  All
entries of $H_j(z)$ are non-zero in general, but $H_j(0)=h_j$.

If the series $H_j$ have  nonzero radii of convergence, the
monodromy of the flat connection $\CA$ can be computed from the
covariantly constant sections (\ref{asis}) and equals $\exp(-2\pi
i T_1)$.  For this reason, $T_1$ is known as the exponent of
formal monodromy of the connection $\CA$.  As we have seen, in
general the actual monodromy does not coincide with $\exp(-2\pi i
T_1)$, so the formal power series $H_j$ are not convergent.

\begin{figure}[tb]
{\epsfxsize=3in\epsfbox{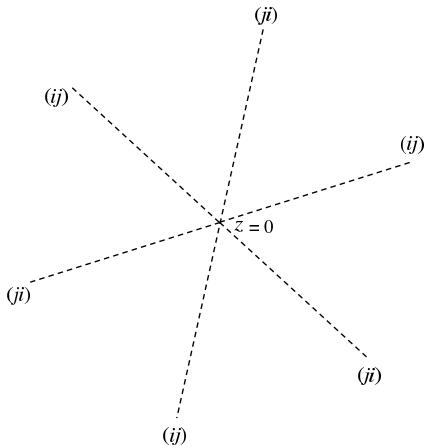}}
\begin{center}
\end{center}
 \caption{For each pair $i,j$, there are $n-1$ Stokes rays of type $(ij)$ and an equal number of
 type $(ji)$. They alternate and are equally spaced, as shown here for $n=4$. To avoid clutter, only the
 Stokes rays associated with one pair $i,j$ are shown. }
 \label{tbone}
\end{figure}

Before proceeding, we need to discuss the asymptotic behavior near
$z=0$ of the functions $\exp(Q_j)$.  This is determined by the
real part of the leading terms $q_j/(n-1)z^{n-1}$ in the $Q_j$.
Whether the real part of this expression is positive or negative
depends on the direction that one approaches the point $z=0$ in
the complex plane.  We say that a ``Stokes ray'' of type $(ij)$ is
a ray in the complex plane along which $(q_i-q_j)/z^{n-1}$ takes
values on the negative imaginary axis. The sign of $q_{ij}={\rm
Re}\left((q_i-q_j)/z^{n-1}\right)$ changes from positive to
negative as $z$ crosses the Stokes ray in the counterclockwise
direction. So ``before'' crossing the Stokes ray, one  has $|\exp
(Q_i(z))|>>|\exp(Q_j(z))|$ for $z\to 0$,  while ``after'' crossing
it (in the counterclockwise direction) this inequality is
reversed. There are a total of $n-1$ Stokes rays of type $(ij)$,
for each ordered pair $i,j$, as sketched in fig. \ref{tbone}.

By an angular sector in the disc $C^*$, we mean a sector defined by
$\theta_a\leq {\rm Arg}\,z\leq \theta_b$, for some $\theta_a,$
$\theta_b$.  A basic result about differential equations with
irregular singularities  is that in any sufficiently small angular
sector $S$ in the complex $z$-plane, after possibly replacing $C^*$
by a smaller disc around the origin, there are holomorphic sections
$\hat H_{j,S}$, asymptotic to $H_j$ as $z\to 0$ in the sector $S$,
such that
\begin{equation}\label{hh}\Psi_{j,S}=\hat{H}_{j,S}\exp(Q_j(z))
\end{equation}
give a basis of covariantly constant sections of the bundle $E$.
(For some explicit examples of construction of the $\hat H_{j,S}$
in simple cases, and verification of their asymptotic behavior,
see the appendix.) For the case that $T_n$ is regular and
semisimple, this is part of Theorem 12.3 of \cite{Wa}, which also
asserts that the $\Psi_{j,S}$ exist with the claimed asymptotic
behavior as long as
\begin{equation}\label{gg} \theta_b-\theta_a\leq {\pi\over
n-1}.\end{equation} The importance of the value in (\ref{gg}) will
become clear. The generalization in which $T_n$ is not assumed to
be regular and semisimple is Theorem 19.1 of \cite{Wa}.  (In the
generalization, one needs to suitably modify the definition of
$Q_j$ and $H_j$ to reflect the asymptotic behavior of solutions of
the differential equation.)

Since $H_{j,S}=h_j+{\mathcal O}(z)$, the fact that $\hat H_{j,S}$
is asymptotic to $H_{j,S}$ for small $z$ implies that
\begin{equation}\label{aseq}\Psi_{j,S}\sim h_j\exp(Q_j(z)),~~z\to
0.\end{equation} Now let us determine to what extent the
$\Psi_{j,S}$ are uniquely determined by their asymptotic behavior
(plus the differential equation that they obey).  If $q_{jk}>0$,
then $|\exp(Q_j(z))|>>|\exp(Q_k(z))|$ for small $z$.  This being so,
we can add to $\Psi_{j,S}$ a multiple of $\Psi_{k,S}$ without
changing its asymptotic behavior  for $z\to 0$ in the sector $S$. We
cannot do the opposite; adding to $\Psi_{k,S}$ a multiple of
$\Psi_{j,S}$ would change its asymptotic behavior for $z\to 0$. If
the sector $S$ contains no Stokes rays, we can order the eigenvalues
of $T_n$ so that throughout $S$, $q_{jk}>0$ if $j>k$. In that case,
the indeterminacy  is precisely that we can add to each $\Psi_{j,S}$
a linear combination of the $\Psi_{k,S}$ with $k<j$. Equivalently,
the row vector
\begin{equation}\label{colvec}\begin{pmatrix}\Psi_{1,S}& \Psi_{2,S}&
\cdots & \Psi_{N,S}\\ \end{pmatrix}\end{equation} can be
multiplied on the right by an upper triangular matrix
\begin{equation}\label{olvec}M=\begin{pmatrix} 1 & * & * & \dots& * \\
                                             0 & 1 & *& \dots &
                                             *\\ 0&0&1&\dots&*\\
                                             &&\ddots &&\\
                                              0&0&0&\dots&1\\
                                              \end{pmatrix}\end{equation}

An $N\times N$ matrix $Y$ whose columns are a basis of solutions
of the differential equation $(\partial_z+\CA_z)\Psi=0$ is called
a fundamental matrix solution.  For example, we can take $Y$ to
have columns $\Psi_{1,S},\Psi_{2,S},\dots,\Psi_{N,S}$.  Write $H$
for the matrix of formal power series whose columns are
$H_1,H_2,\dots,H_N$, so in particular $H=1$ at $z=0$. And write
$Q$ for the matrix $Q={\rm diag}(Q_1,Q_2,\dots,Q_N)$. Then the
asymptotic behavior for $z\to 0$ in the sector $S$ of the
fundamental matrix solution $Y$ is
\begin{equation}Y\sim H \exp(Q).\end{equation}
The result of the last paragraph can be restated to say that a
fundamental matrix solution with this asymptotic behavior is
unique up to $Y\to YM$, where $M$ is a constant matrix, and, as in
(\ref{olvec}), $M-1$ is strictly upper triangular.  (In the
general theory, for an arbitrary simple Lie group, $M$ takes
values in the unipotent radical of a suitable Borel subgroup, as
explained in \cite{Btwo}, section 2.)

Now suppose instead that the sector $S$ contains a Stokes ray of
type $(ij)$ or $(ji)$.  Then $\Psi_{i,S}$ and $\Psi_{j,S}$
exchange dominance in crossing the Stokes ray.  So we cannot add a
multiple of one to the other without spoiling the asymptotic
behavior on one side or the other of the Stokes ray. Thus, if $S$
contains a Stokes ray, the indeterminacy of the solutions
$\Psi_{i,S}$ is reduced.

For an important application of this, pick a sector $S$ whose
boundary rays are not Stokes rays and whose angular width is
precisely $\pi/(n-1)$, the maximum value in eqn. (\ref{gg}).  This
is the same as the spacing between adjacent Stokes rays of type
$(ij)$ and $(ji)$. So for each unordered pair $i,j$, the sector
$S$ contains precisely one Stokes ray of one of these two type,
and we cannot change either $\Psi_{i,S}$ or $\Psi_{j,S}$ by a
multiple of the other.  Hence, in a sector $S$ of this special
type, the solutions $\Psi_{i,S}$ of the differential equation are
uniquely determined by their required asymptotic behavior.

\def\T{{\Bbb{T}}}
It is important to clarify exactly what this uniqueness means.
Once we pick a holomorphic extension of the bundle $E$ over the
singular point, and further make a gauge transformation to put the
connection in the form (\ref{jurfy}), the $\Psi_{i,S}$ are
uniquely determined. The condition (\ref{aseq}) that determines
$\Psi_{i,S}$ is preserved by a gauge transformation that is 1 at
$z=0$. But in general, a gauge transformation that preserves the
form (\ref{jurfy}) need not be 1 at $z=0$; rather, at $z=0$, it
can be an arbitrary (invertible) diagonal matrix -- that is, an
element of the complex maximal torus $\T_\C$ of $G_\C$. The choice
of the $\Psi_{i,S}$ is not invariant under the action of $\T_\C$,
and we will have to allow for this in classifying irregular
singularities.

\subsection{Enlarging The Sector}
\label{sector}

Let $\Psi_{i,S}$ and $\Psi_{j,S}$ be as above and suppose that the
sector $S$ contains a Stokes ray of type $(ij)$.  And let
$\tilde\Psi_{i,S}=\Psi_{i,S}+\lambda \Psi_{j,S}$ for some constant
$\lambda$. Consider the asymptotic behavior of $\tilde\Psi_{i,S}$
along a ray $\ell$ that approaches $z=0$ in the sector $S$.

Let us suppose that $q_{ij}>0$ if  $\ell$ is ``before'' the Stokes
ray (in a counterclockwise sense).  Then in that region
$\exp(Q_j)$ is subdominant relative to $\exp(Q_i)$, so
$\tilde\Psi_{i,S}$ has the same asymptotic behavior as
$\Psi_{i,S}$:
\begin{equation}\label{gurto}\tilde\Psi_{i,S}\to
H_i\,\exp(Q_i),~z\to 0.\end{equation} But if $\ell$ is ``after''
the Stokes ray, the term $\lambda \Psi_{j,S}$ dominates
$\tilde\Psi_{i,S}$ for $z\to 0$, and the asymptotic behavior is
\begin{equation}\label{ogurto}\tilde\Psi_{i,S}\to \lambda
H_j\,\exp(Q_j),~z\to 0.\end{equation} This demonstrates an
important phenomenon: the asymptotic behavior of a solution of the
differential equation for $z\to 0$ can change as one crosses a
Stokes ray.

This statement has an equally important converse: the asymptotic
behavior of such a solution can change {\it only} in crossing a
Stokes ray.  To see this, we consider a sector $S$ with sections
$\Psi_{i,S}$ that obey the differential equation and the
asymptotic condition (\ref{aseq}), and we suppose that one of the
boundary lines $\ell_0$ of sector $S$ is not a Stokes ray.  We
want to show that under this condition, the $\Psi_{i,S}$ can be
analytically continued beyond $\ell_0$, with the asymptotic
condition remaining valid. We order the eigenvalues of $T_n$ so
that along $\ell_0$,
\begin{equation}\label{hevoc}q_{ij}>0~{\rm for}~i>j.\end{equation}
  Let $\tilde S$ be a sector containing
$\ell_0$ in its interior and sufficiently small to contain no
Stokes ray.  The latter condition ensures that eqn. (\ref{hevoc})
holds throughout $\tilde S$.  Also, it means that $\tilde S$ is
sufficiently small that we can invoke Theorem 12.3 of \cite{Wa}
and find solutions $\Psi_{i,\tilde S}$ of the differential
equation in sector $\tilde S$ obeying the asymptotic condition
(\ref{aseq}) in that sector. The intersection $S\cap \tilde S$ is
non-empty, and in this sector, the $\Psi_{i,S}$ are related to
$\Psi_{i,\tilde S}$ by a triangular matrix, as in eqn.
(\ref{olvec}):
\begin{equation}\label{tired}\begin{pmatrix}\Psi_{1,S}& \Psi_{2,S}& \cdots
& \Psi_{N,S}\\ \end{pmatrix} =\begin{pmatrix}\Psi_{1,\tilde S}&
\Psi_{2,\tilde S}&  \cdots &
\Psi_{N,\tilde S}\\ \end{pmatrix}\begin{pmatrix} 1 & * & * & \dots& * \\
                                             0 & 1 & *& \dots &
                                             *\\ 0&0&1&\dots&*\\
                                             &&\ddots &&\\
                                              0&0&0&\dots&1\\
                                              \end{pmatrix}
\end{equation} Since the
$\Psi_{i,\tilde S}$ are holomorphic in the sector $\tilde S$, this
gives an analytic continuation of the $\Psi_{i,S}$ throughout
$\tilde S$.  Since the condition (\ref{hevoc}) holds throughout
$\tilde S$, the fact that the $\Psi_{i,\tilde S}$ obey the
asymptotic condition (\ref{aseq}) throughout $\tilde S$ plus the
fact that the $\Psi_{i,S}$ are related to them by an upper
triangular matrix means that $\Psi_{i,S}$ obey the asymptotic
condition throughout $\tilde S$.

This process can be continued until a Stokes ray is reached.  Even
in crossing a Stokes ray, the above argument for analytic
continuation still works; but the asymptotic condition
(\ref{aseq}) fails, since if $\tilde S$ contains a Stokes ray, we
cannot assume (\ref{hevoc}) throughout $\tilde S$.

\subsection{Stokes Matrices}
\label{stokesmat}

Pick a sector $S^0_1$ of angular width $\pi/(n-1)$ whose boundary
rays are not Stokes rays. By rotating it through an angle that is
an integer multiple of $\pi/(n-1)$, we get $2n-1$ additional
sectors $S^0_2,S^0_3,\dots,S^0_{2n-2}$.  Each of these has width
$\pi/(n-1)$ and boundary rays that are not Stokes rays.

In each of the sectors $S_\alpha^0$, $\alpha=1,\dots,2n-2$, there
are solutions of the differential equation
$\Psi_{j,\alpha},~j=1,\dots,N$ that are uniquely determined by
requiring that they obey the asymptotic condition (\ref{aseq}) in
the sector $S^0_\alpha$. Each of these can be continued to angular
sectors $S_\alpha$ that are slightly larger than $S^0_\alpha$,
still obeying the same asymptotic condition.  The sectors
$S_\alpha$ are wide enough to give a covering of the punctured
disc.

We can label the eigenvectors of $Q$  and the $S_\alpha$ so that
the inequalities (\ref{hevoc}) are obeyed on the intersection
$S_\alpha\cap S_{\alpha+1}$ if $\alpha$ is odd.  In that case, if
$\alpha$ is even, the opposite inequalities are obeyed on
$S_\alpha\cap S_{\alpha+1}$:
\begin{equation}\label{nevoc}q_{ij}>0~{\rm if}~i<j.\end{equation}
On each sector $S_\alpha$, we define a fundamental matrix solution
$Y_\alpha$ whose columns are the $\Psi_{j,\alpha}$:
\begin{equation}\label{bevo}Y_\alpha=\begin{pmatrix}\Psi_{1,\alpha}&
\Psi_{2,\alpha}& \cdots & \Psi_{N,\alpha}\\
\end{pmatrix}.\end{equation}
On the intersection of the two sectors $S_\alpha$ and
$S_{\alpha+1}$, the two fundamental matrix solutions $Y_\alpha$
and $Y_{\alpha+1}$, which both obey the same asymptotic condition,
are related by
\begin{equation}Y_{\alpha+1}=Y_\alpha M_\alpha
.\end{equation} Here $M_\alpha$ is a triangular matrix with 1's on
the diagonal.  It is upper triangular if $\alpha$ is odd (and the
inequalities (\ref{hevoc}) are obeyed on $S_\alpha\cap
S_{\alpha+1}$). It is lower triangular if $\alpha$ is even (and
the opposite inequalities (\ref{nevoc}) are obeyed on the
intersection).

The matrices $M_\alpha$ are known as Stokes matrices (or Stokes
multipliers). They are uniquely determined up to conjugation by a
common diagonal matrix -- which arises from the freedom to make a
diagonal gauge transformation of the connection $\CA$, preserving
the form (\ref{jurfy}).

To compute the monodromy around the singularity at $z=0$, we must
take the product of Stokes matrices $M_{2n-2}M_{2n-1}\cdots M_1$.
But this is not quite the whole story.  The  asymptotic condition
(\ref{aseq}) determines the $z\to 0$ asymptotic behavior of the
solutions $\Psi_{j,\alpha}$ in terms of $\exp(Q_j(z))$, which
itself has a monodromy, because of the logarithmic term in $Q_j$.
These logarithmic terms alone would lead to a monodromy
$\exp(-2\pi iT_1)$ (which is the monodromy of the formal solutions
(\ref{asis}) that were constructed as formal power series times
$\exp(Q_j)$). The actual monodromy $\hat M$ is the product of the
monodromy built into the condition (\ref{aseq}) times the
monodromy coming from the product of the Stokes matrices:
\begin{equation}\label{polgo}\hat M=\exp(-2\pi
iT_1)M_{2n-2}M_{2n-1}\cdots M_1.\end{equation}

We think of the Stokes matrices and the exponent $T_1$ of formal
monodromy, or equivalently the Stokes matrices and the actual
monodromy $\hat M$, as the generalized monodromy data near the
singularity at $z=0$.  To classify the generalized monodromy up to
gauge equivalence, this data must be taken modulo the action of
the diagonal matrices, that is the action of the maximal torus
$\T_\C$. Let us count the parameters in the generalized monodromy
in the neighborhood of a single irregular singularity.

In our derivation, the complexified gauge group $G_\C$ is
$SL(N,\C)$ or $GL(N,\C)$.  The complex dimension of $G_\C$, which
we denote at ${\rm dim}\,G_\C$,  is $N^2-1$ or $N^2$, and the
rank, which we call $r$, is equal to $N-1$ or $N$.  A pair
$M_\alpha$, $M_{\alpha+1}$ of successive Stokes matrices depends
on ${\rm dim}\,G_\C-r$ complex parameters, and we have $n-1$ such
pairs. To this we must add $r$ parameters for the exponent of
formal monodromy. But we must also subtract $r$ parameters for
dividing by the action of $\T_\C$.  So altogether in a local
description near an irregular singularity, the generalized
monodromy is parametrized by
\begin{equation}\label{covec} c_n=(n-1)({\rm
dim}(G_\C)-r)\end{equation} complex parameters.

Though our derivation has been for $SL(N,\C)$ or $GL(N,\C)$, the
general case is similar, as explained in \cite{Btwo}, section 2.
Groups of upper or lower triangular matrices are replaced with
suitable Borel subgroups.  Most of the discussion has a close analog
for general $G$, and in particular the number of parameters in the
generalized monodromy is still given by (\ref{covec}).

\subsection{Classification Of Irregular Singularities}
\label{classir}

The Stokes matrices plus the diagonal matrix-valued function $Q$
give a complete set of local invariants of an irregular
singularity.  (We need not mention separately the exponent of
formal monodromy as it appears in $Q$.)

To prove this last statement, suppose we are given two different
connections $\CA$ and $\tilde \CA$, that have the same Stokes
matrices and the same $Q$.  Let $S^{0}_1$ be a sector of angular
width $\pi/(n-1)$ whose boundary rays are not Stokes rays for
either connection, and as above rotate it to get additional
sectors $S^0_2,\dots,S^0_{2n-2}$, and thicken these slightly to
sectors $S_1,\dots,S_{2n-2}$ whose intersections contain no Stokes
lines. The connections $\CA$ and $\tilde \CA$ lead to two
differential equations, each of which can be analyzed as above.
Let $Y_\alpha$ and $\tilde Y_\alpha$ be the fundamental matrix
solutions of the two equations in sector $S_\alpha$ with
asymptotic behavior
\begin{align}\label{murkier} Y_\alpha & \sim H \exp(Q) \\ \nonumber \tilde Y_\alpha & \sim
\tilde H\exp(Q), \end{align} where $H$ and $\tilde H$ are formal
power series with $H(0)=\tilde H(0)=1$.  We have
\begin{equation}\label{yuyu}Y_{\alpha+1}=Y_\alpha M_\alpha,
~~\tilde Y_{\alpha+1}=\tilde Y_\alpha M_\alpha, \end{equation}
with by hypothesis the same Stokes matrices for the two
connections. This implies that $g=Y_{\alpha}\tilde
Y_{\alpha}{}^{-1}$ is independent of $\alpha$. (This remains valid
after going all the way around the circle, since the two exponents
of formal monodromy are also the same.) Moreover, the asymptotic
condition (\ref{murkier}) shows that $g(0)=1$.  $g$ is a gauge
transformation that maps one connection $\tilde\CA$ to the other
one $\CA$.

This statement also has a converse.  For given $Q$, one can find
an $\CA$ that realizes any required set of Stokes matrices.  This
is shown in \cite{BJL}, following \cite{Si}.

\subsection{A More Global View}\label{global}

\def\Y{{\cal Y}}
Now we are going to embed this local description in a global
context.  We consider a compact Riemann surface $C$ of genus $g_C$
with a flat $G_\C$ bundle $E$ with connection $\CA$.  From a
holomorphic point of view, the $(0,1)$ part of the connection
endows $E$ with a holomorphic structure, and then the $(1,0)$ part
of the connection is a holomorphic one-form, locally $dz\,\CA_z$,
valued in ${\rm ad}(E)$.  We are interested in the case that this
one-form has a pole of order $n$ near a point $p$. We want to
describe the appropriate generalized monodromy data and count the
parameters that it depends on.  (The generalization to several
irregular singularities is straightforward.)

First let us review what happens in the absence of the singularity.
We pick a basepoint $q\in C$.  We let $A_1,\dots,A_g$ and
$B_1,\dots,B_g$ be loops (``$A$-cycles'' and ``$B$-cycles'')
starting and ending at $q$ and generating in the usual way the first
homology group of $C$. Taking the monodromy of $\CA$ around the
$A$-cycles and $B$-cycles, we get elements of $G_\C$ that we denote
as $U_1,\dots,U_g$ and $V_1,\dots, V_g$.  They obey one relation
\begin{equation}\label{turko}1=U_1V_1U_1^{-1}V_1^{-1}\cdots
U_gV_gU_g^{-1}V_g^{-1}.\end{equation} In addition, they are only
defined up to conjugation by a common element of $G_\C$ (coming
from the action of gauge transformations at the basepoint $q$).
The number of complex parameters is therefore
\begin{equation}\label{pl}d_\Y=(2g-2)\,{\rm dim}(G).\end{equation}
This is the complex dimension of the moduli space\footnote{In
section \ref{hitchwild}, we will introduce Hitchin's equations and
interpret $\Y$ as a hyper-Kahler manifold that parametrizes
solutions of those equations and is related to either flat
connections or Higgs bundles. In that context, we will call it
$\MH$. For now, we view it solely as the moduli space of
representations of the fundamental group, and denote it as $\Y$.}
$\Y$ of flat $G_\C$-bundles on $C$.

\begin{figure}[tb]
{\epsfxsize=3in\epsfbox{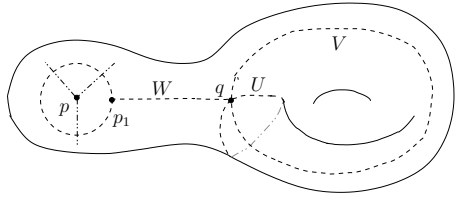}}
\begin{center}
\end{center}
 \caption{A Riemann surface $C$, here taken to be of genus $g_C=1$, with an irregular
 singularity at a point $p$.  A basepoint is taken at $q$.  Show are the Stokes rays near $p$
 and the important paths in defining the generalized monodromy data.}
 \label{tolgo}
\end{figure}

Now we incorporate an irregular singularity at a point $p\in C$.
Restricting to a small punctured disc $C^*$ containing  $p$, we
analyze the local behavior by covering
 $C^*$ with sectors $S_1,\dots,S_{2n-2}$, as in section \ref{stokesmat}.  Let $p_1$ be a
point in the sector $S_1$.   To describe a flat connection on $C$
up to gauge equivalence of the desired sort, we repeat the
analysis with a few corrections to account for the singularity. We
must include one more group element $W$ to account for parallel
transport from $q$ to $p_1$ along some chosen path (fig.
\ref{tolgo}), and then according to (\ref{covec}) we have
$c_n=(n-1)({\rm dim}(G)-r)$ parameters to account for the local
behavior near $p$.  These parameters comprise the monodromy $\hat
M$ on a small loop circling the singularity in the disc $C^*$, as
well as the Stokes matrices $M_\alpha$ that involve the asymptotic
behavior near $p$. So the total number of extra complex parameters
required to describe the situation in the presence of an irregular
singularity is
\begin{equation}\label{hobox}\hat c_n={\rm dim}(G)+c_n=n\,{\rm
dim}(G)-(n-1)r.\end{equation}

The monodromy data $U_i,V_j,$ and $W$, together with the local data
at the singularity, obey one relation, as was the case in the
absence of the singularity.  But now, instead of (\ref{turko}), this
relation is more complicated:
\begin{align}\label{juko}1=&U_1V_1U_1^{-1}V_1^{-1}\cdots
U_gV_gU_g^{-1}V_g^{-1}W\hat MW^{-1}\\
\nonumber =&U_1V_1U_1^{-1}V_1^{-1}\cdots U_gV_gU_g^{-1}V_g^{-1}W
\exp(-2\pi T_1)M_{2n-2}M_{2n-1}\cdots M_1W^{-1}.\end{align} We have
written this relation both in terms of the monodromy $\hat M$ around
the singular point, and more explicitly in terms of the formal
monodromy and the Stokes matrices. And now, the group of
equivalences that acts on this data is $G_\C\times \T_\C$, where the
first factor acts by gauge transformations at $q$ and the second by
gauge transformations at $p_1$.  An element $g\in G_\C$ acts by
$U_i\to gU_ig^{-1}$, $V_i\to gV_ig^{-1}$, and $W\to gW$. And an
element $S\in \T_C$ acts by $W\to WS^{-1}$, $M_\alpha\to SM_\alpha
S^{-1}$.

\subsection{Topological Interpretation}
\label{topint}

 As above, we write
$\Y$ for the moduli space of $G_\C$-valued flat connections on $C$,
up to gauge transformation. And we write $\Y^*$ for the space that
parametrizes the generalized monodromy data in the presence of an
irregular singularity at $p$ (or more generally in the presence of
several irregular singularities).

$\Y$ can be defined purely topologically, since it can be
interpreted as a moduli space of representations of the
fundamental group of $C$.  The topological nature of $\Y$ is
explicit in the equation (\ref{turko}), which does not depend on
the complex structure of $C$.  A flat connection up to gauge
transformation is equivalent to a set of elements $U_i,V_j\in
G_\C$ obeying (\ref{turko}), up to conjugation.  So $\Y$ can be
defined purely in topological terms.

The same is true of $\Y^*$, though this may be surprising at
first. To describe, up to isomorphism, the generalized monodromy
data of a flat connection on $C\backslash p$ with an irregular
singularity at $p$, we must specify a larger set of group
elements, namely $U_i,V_j\in G_\C$, $\exp(-2\pi iT_1)\in \T_\C$,
and the Stokes matrices $M_\alpha$; the latter take values in
groups of unipotent upper or lower triangular matrices (or, for
general $G_\C$, in the unipotent radicals of appropriate Borel
subgroups).  The number of these elements, the subgroups in which
they take values, and the equation (\ref{juko}) that they obey are
all completely independent of the complex structure on $C$.  So
$\Y^*$, like $\Y$, can be defined in purely topological terms.
Moreover, except for the exponent of formal monodromy, $\Y^*$ is
independent of the function $Q(z)$ that enters the description of
the singularity.

What may make this surprising is that the whole discussion of
Stokes matrices and generalized monodromy seems to depend on
viewing $C$ as a complex manifold and considering the function
$\exp(Q)$. However, if we change slightly the complex structure of
$C$, the position of the point $p$, or the leading singular term
$T_n/z^n$ of the connection (preserving the condition that $T_n$
is regular and semisimple), the Stokes rays will move, but they
will not change in number.\footnote{To be more precise, the number
of Stokes rays of any given type $(ij)$ will not change.  Stokes
rays of different types may cross as we vary $T_n$, but this does
not affect the analysis.} The Stokes matrices will still take
values in the same group of upper or lower triangular matrices,
and they will still appear in the same equation (\ref{juko}).

By comparing the additional variables that enter the description of
$\Y^*$, relative to those that entered in describing $\Y$, we see
that the difference in complex dimension between $\Y^*$ and $\Y$ is
\begin{equation}\label{diffdem}\hat c_n={\rm dim}(G_\C)+c_n=n\,{\rm
dim}(G_\C)-(n-1)r.\end{equation} Though we have described the case
of one irregular singularity, the generalization to the case of
several such singularities is straightforward.  Each singularity
associated with a pole of order $n$ increases the dimension by $\hat
c_n$.

Let us compare this to the total number of parameters needed to
describe an irregular singularity.  If we permit the $(1,0)$ part
of a connection $\CA$ to have a pole of order $n$ at a point $p$,
then the singular behavior takes the familiar form
$T_n/z^n+T_{n-1}/z^{n-1}+\dots +T_1/z$, and is described by $n$
elements of the Lie algebra ${\mathfrak g}_\C$.  In all it takes
$n\,{\rm dim}\,G_\C$ parameters to specify $T_1,\dots,T_n$.

Of a total of $n\,{\rm dim}\,G_\C$ parameters, the generalized
monodromy data give  a  topological interpretation to $\hat
c_n=n\,{\rm dim}\,G_C-(n-1)r$ parameters.  We seem to be left, for
each irregular singularity, with $(n-1)r$ parameters that do {\it
not} have a topological interpretation.  What are these?

In our previous analysis, we have in fact encountered certain
parameters associated with each irregular singularity that at least
appear not to have a topological interpretation.  As a preliminary
step in the analysis, we picked a local parameter $z$ near the
singularity, and put the connection in the form
\begin{equation}\label{jurfyx}\CA_z=\frac{T_n}{z^n}+\frac{T_{n-1}}{z^{n-1}}+\dots+\frac
{T_1}{z}+{\mathcal B},
\end{equation}
with $T_n,\dots,T_1\in {\mathfrak t}_\C$, the Lie algebra of
$\T_\C$, and $\mathcal B$ regular.  Here $T_1$ is independent of
the choice of local coordinate, since it is the residue of the
differential form $dz \,\CA_z$.  But $T_2,\dots,T_n$ do depend on
the choice of coordinate, so it would be hard to give them any
topological interpretation. They depend on a total of
\begin{equation}\label{gorky}\delta_n=(n-1)r\end{equation} parameters,
since ${\mathfrak t}_\C$ has dimension $r$.  These are the
parameters that characterize the irregular singularity and are
{\it not} captured by the generalized monodromy.  How to vary
these parameters while keeping fixed the generalized monodromy is
shown in the theory of isomonodromic deformation for irregular
singularities, developed by Miwa, Jimbo, and Ueno \cite{JMU}.

\subsubsection{ Action Of Braid Group\hskip .3cm}\label{brgroup}
The assertion that $\Y$ or $\Y^*$ can be defined purely
topologically must be clarified in one respect.  Let us first give
an analogy. If we vary the complex structure of $C$ slightly, $\Y$
in a natural sense does not vary. However, if we consider
arbitrary families of complex structures on $C$, then $\Y$ will in
general acquire a monodromy, involving an action of the mapping
class group of $C$.  For $C$ of genus zero and a flat connection
with regular singularities, this type of deformation is described
by Schlesinger's equation; for example, see \cite{Mal}, \cite{H3}.
Now let us consider varying the polar coefficients $T_n,\dots,T_2$
of an irregular singularity. Let ${\mathfrak t}_{\C}^{\rm reg}$ be
the space of regular  elements of $\mathfrak t_\C$.  The space
$\Y^*$ can be defined for any $T_n\in{\mathfrak t}_{\C}^{\rm
reg}$. As $T_n$ is varied, the spaces $\Y^*$ are locally constant
-- they vary as fibers of a flat bundle over ${\mathfrak
t}_{\C}^{\rm reg}$. But globally there is a monodromy, via which
the fundamental group of ${\mathfrak t}_{\C}^{\rm reg}$ acts on
$\Y^*$. The monodromy arises because the choice of a sector $S_1$
that is not bounded by Stokes rays cannot be made globally. (But
${\mathfrak t}_\C^{\rm reg}$ can be covered by small open sets, in
each of which one can make such a choice, so $\Y^*$ is naturally
invariant under a small change of $T_n$.) The fundamental group of
${\mathfrak t}_{\C}^{\rm reg}$ is called the braid group of $G$;
we will denote it as $B(G)$. Its monodromy action on $\Y^*$ was
exploited in \cite{Btwo}.

\subsection{Isomonodromic Deformation And Symplectic Structure}\label{iso}

In the theory of isomonodromic deformation \cite{JMU}, one
constructs meromorphic differential equations by which one can vary
the parameters contained in $T_2,T_3,\dots,T_n$ without changing the
generalized monodromy.  This description of isomonodromy has many
applications in two-dimensional integrable systems.  It may well be
eventually relevant to geometric Langlands, but in this paper we
will use instead  (section \ref{physper}) a gauge theory approach to
isomonodromy, more similar to that in \cite{B}.

The possibility of isomonodromic deformation makes it clear that the
complex structure of the variety $\Y^*$ that parametrizes the
generalized monodromy data must be independent of $T_2,\dots,T_n$.
This particular point is clear more directly from the explicit
description of $\Y^*$ via the equation (\ref{juko}), which does not
depend on the choice of $T_2,\dots,T_n$.

To go farther, we need to recall that the moduli space $\Y$ of
homomorphisms of the fundamental group of $C$ to a simple complex
Lie group $G_\C$ has (up to a multiplicative constant) a natural
symplectic structure, which can be defined in gauge theory
 by the formula \cite{AB}
\begin{equation}\label{delfo}\Omega= -{i\over 4\pi}\int_C
\,\Tr\,\delta \CA\wedge \delta \CA.\end{equation} (Here $-\Tr$ is an
invariant quadratic form on the Lie algebra $\mathfrak g_\C$; we
normalize it so that short coroots have length squared 2.) This is a
symplectic structure in the holomorphic sense;  $\Omega$ is a
closed, holomorphic, and nondegenerate  $(2,0)$-form with respect to
the complex structure of $\Y$.

It was shown in \cite{B}, section 5, that in the presence of an
irregular singularity, the same formula can be used to define a
complex symplectic structure.  But now we define the complex
symplectic structure not on $\Y^*$, but rather on what we might call
$\Y^*(T_1)$, the subvariety of $\Y^*$ in which $T_1$, the exponent
of formal monodromy, is kept fixed.  The idea here is that in
defining $\Y^*(T_1)$, we keep fixed all the coefficients
$T_1,T_2,\dots,T_n$ of singular terms in $\CA$.  $T_2,\dots, T_n$
are kept fixed in defining $\Y^*$, and additionally $T_1$ is kept
fixed in defining $\Y^*(T_1)$.  So, although $\CA$ has a
singularity, its variation $\delta\CA$ does not, as  a result of
which the formula (\ref{delfo}) makes sense and has its usual
properties.

It is fairly obvious that the holomorphic symplectic form $\Omega$
on $\Y$ or $\Y^*(T_1)$ does not depend on a choice of complex
structure of $C$; indeed, no such complex structure is used in the
definition (\ref{delfo}).  Also true, but much less obvious, is that
the symplectic structure of $\Y^*(T_1)$ does not depend on
$T_2,\dots,T_n$.  This is the main result of \cite{B} (see Theorems
7.1 and 7.3), where it is proved using gauge theory.  For
alternative approaches, see \cite{Wo}, \cite{Kr}, \cite{Bthree}. In
applying $\EUN=4$ super Yang-Mills theory to wild ramification, a
natural quantum field theory explanation of the fact that the
symplectic structure of $\Y^*(T_1)$ is independent of
$T_2,\dots,T_n$ will emerge (section \ref{physper}).  When made
explicit, this will lead to an argument similar to that in \cite{B}.

The complex structure and symplectic structure of $\Y^*(T_1)$ do
depend on $T_1$. The fact that one must hold $T_1$ fixed to define a
symplectic manifold and that the resulting symplectic structure
depends on $T_1$ has nothing to do with irregular singularities;
these statements also hold for $n=1$, which is the case of a regular
singularity.  The fact that the complex and symplectic structures
should naturally depend on $T_1$ will be clear in the quantum field
theory approach.

For future use, let us note that since $T_1$ is kept fixed in
defining $\Y^*(T_1)$, the dimension of $\Y^*(T_1)$ is less than that
of $\Y^*$ by $r$, the rank of $G$.  So from (\ref{diffdem}), we get
that
\begin{equation}\label{huffy}{\rm dim}\,\Y^*(T_1)={\rm
dim}\,\Y+\hat c_n-r=(2g-2){\rm dim}(G_\C)+n({\rm
dim}G_\C-r).\end{equation} For example, for $G_\C=SL(2,\C)$, we get
\begin{equation}\label{uffy}{\rm
dim}\,\Y^*(T_1)=6g-6+2n.\end{equation}

\subsection{Strategy Of This Paper}\label{strategy}

\def\CP{{\Bbb{CP}}}
Now we can explain the strategy of the present paper.  In the
process, it will hopefully become clearer why we have begun the
paper with a review of the theory of Stokes phenomena.

Let us first recall what was done in \cite{KW} in the unramified
case, or in \cite{GW} with tame ramification.  If $\Y(G,C)$ denotes
the moduli space of flat $G_\C$ bundles on $C$, with structure group
$G_\C$, then to $\Y(G,C)$ we can associate a pair of topological
field theories, namely the $B$-model defined using the natural
complex structure of $\Y(G,C)$ and the $A$-model defined using the
real symplectic structure ${\rm Re}\,\Omega$. These theories do not
depend on the complex structure of $C$, since as a complex
symplectic manifold, $\Y(G,C)$ has no such dependence. (These are
actually two points in a larger family of topological field theories
described in \cite{KW}, and parametrized by $\CP^1$, but we will not
emphasize the generalization in the present paper.)

Similarly, if we replace $G$ with the dual group $^L\neg G$, we can
define a $B$-model and an $A$-model with target $\Y({}^L\neg G,C)$.
One might wonder if there is some kind of duality between the
topological field theories associated with $G$ and with $^L\neg G$.
But even once it is asked, this question is hard to answer without
some additional structure.

However, if one interprets $\Y(G,C)$ and $\Y({}^L\neg G,C)$ as
moduli spaces of solutions of Hitchin's equations, then one has a
hyper-Kahler structure, and, using a different complex structure on
these spaces (not the natural one that we have used up to this
point) one can define the Hitchin fibration \cite{H}, \cite{H2}. As
was first described mathematically in \cite{HT}, the Hitchin
fibration in this situation can be interpreted as a special
Lagrangian fibration \cite{SYZ} that establishes a mirror symmetry
between $\Y(G,C)$ and $\Y({}^L\neg G,C)$.

This framework can be derived from four-dimensional $\EUN=4$ super
Yang-Mills theory with electric-magnetic duality,  as first
considered in \cite{BJV}, \cite{HMS}.  The idea of \cite{KW} was
that by incorporating additional ingredients of the physics, such
as the Wilson and 't Hooft operators and various special branes,
one can get a natural understanding of geometric Langlands
duality. This duality maps a flat connection on $C$, with gauge
group $^L\neg G$, to a ${\mathcal D}$-module on the moduli space
of $G$-bundles.

This approach was extended to the case of tame ramification in
\cite{GW}.  In this case, one must consider flat bundles with
ramification (monodromy around marked points).  The appropriate
moduli spaces can again be interpreted \cite{S} as moduli spaces
of solutions of Hitchin's equation.  This leads to a Hitchin
fibration and a mirror symmetry, and ultimately to an
understanding of geometric Langlands duality by the same logic as
in \cite{KW}.  The details are a little more elaborate, however,
because the dependence on the ramification parameters leads to the
existence of noncommutative monodromy symmetries that commute with
the duality.

In the present paper, we extend this to the case of wild
ramification.  Here, the basic symmetry is a mirror symmetry between
the extended monodromy manifolds $\Y^*(T_1)$ with gauge groups $G$
and $^L\neg G$.  The mirror symmetry follows as usual from the fact
\cite{BB} that $\Y^*(T_1)$ can be interpreted as a moduli space of
solutions of Hitchin's equations.  The rest of the gauge theory
machinery can then be applied, as in \cite{KW}, to argue a geometric
Langlands correspondence.

However, the fact that as a complex symplectic manifold, $\Y^*(T_1)$
is independent of the parameters $T_2,\dots,T_n$ that appear in a
flat connection with irregular singularity shows that we must be
careful in stating the geometric Langlands correspondence, if we
want it to be a natural one-to-one correspondence between two kinds
of object.  Both the left and right hand sides of the correspondence
are invariant under isomonodromic deformation, and the duality
between them also commutes with isomonodromic deformation.  Two flat
$^L\neg G_\C$ connections with irregular singularity that are
equivalent under isomonodromic deformation have equivalent duals. So
 if we want the geometric Langlands
correspondence to be a natural correspondence between two types of
object, one  approach might be to consider the starting point to be
a flat $^L\neg G_\C$ connection with irregular singularity, modulo
isomonodromic deformation.

But this would force us to identify two flat connections with
irregular singularity that have the same values of $T_2,\dots,T_n$
and differ by the action of the braid group $B(G)$. This will
probably not work nicely, since $\Y^*(T_1)$ is unlikely to have a
nice quotient by the action of $B(G)$.  Hence, it is probably
better not to try to divide by isomonodromic deformation but
simply to assert that the duality commutes with such deformation.
{}From an algebraic point of view, there is another reason to
formulate things this way. The isomonodromy equations \cite{JMU}
are algebraic, but their solutions are not algebraic (in the usual
algebraic structure   relevant to geometric Langlands). So in the
algebraic setting, isomonodromy gives an infinitesimal way of
varying $T_2,\dots,T_n$, commuting with geometric Langlands
duality, but cannot be exponentiated to an actual map between
objects with different values of $T_2,\dots,T_n$.

\section{Surface Operators With Wild Ramification}
\label{hitchwild}

\subsection{Local Model Of Abelian Singularity}\label{locmod}
As explained in section \ref{strategy}, to find a mirror symmetry
for connections with irregular singularity, we need a relation
between such connections and solutions of Hitchin's equations:
\begin{align}\label{uhitchin}
F-\phi\wedge\phi &= 0 \nonumber \\
D\phi = D\star \phi &= 0. \\ \nonumber \end{align}

Consider an irregular singularity at a point $p$ defined as $z=0$
in terms of some local parameter $z$.  Near $p$, the fields
$A,\phi$ are more singular than $1/|z|$. Hitchin's equations are
schematically $d\Phi+\Phi^2=0$, where $\Phi=(A,\phi)$, and are not
compatible with having $\Phi$ more singular than $1/|z|$ unless
the singular parts of $d\Phi$ and $\Phi^2$ both vanish. This means
that the singular part of the solution must be abelian. And
indeed, this assumption leads to a good theory \cite{Sa},
\cite{BB} of solutions of Hitchin's equations with irregular
singularity.

We write $z=r e^{i\theta}$, and we let $\mathfrak t$ denote the
Lie algebra of a maximal torus $\Bbb{T}$ of the compact Lie group
$G$, and $\mathfrak t_\C$ its complexification.  We pick elements
$\alpha\in \mathfrak t$ and $u_1,\dots,u_n\in \mathfrak t_\C$, and
consider the following explicit solution of Hitchin's equations on
a trivial $G$-bundle $E$ over the punctured complex $z$-plane:
\begin{align}\label{jugid}A&=\alpha \,d\theta\\
\nonumber
\phi&=\frac{dz}{2}\left(\frac{u_n}{z^n}+\frac{u_{n-1}}{z^{n-1}}+\dots+\frac{u_1}{z}\right)
 +\frac{d\bar z}{2}\left(\frac{\bar u_n}{\bar z^n}+\frac{\bar u_{n-1}}{\bar z^{n-1}}+\dots+\frac{\bar u_1}
 {\bar z}\right).
\end{align}
$\bar u_k$ is the complex conjugate of $u_k$, so $\phi$ is real,
that is, it is a $\mathfrak t$-valued one-form.

For the regular case, $n=1$, this reduces to the local model of a
singular solution used in \cite{S} in studying Higgs bundles with
regular singularity, and in \cite{GW} to define surface operators
in $\EUN=4$ super Yang-Mills theory.  To make this explicit, we
write
\begin{equation}\label{tebone}u_1=\beta+i\gamma,\end{equation}
with $\beta,\gamma\in \mathfrak t$.  Then eqn. (\ref{jugid})
becomes
\begin{align}\label{ugid}A&=\alpha\,d\theta\\ \nonumber \phi&=
\beta\,\frac{dr}{r}-\gamma \,d\theta, \end{align} which was the
starting point in eqn. (2.2) of \cite{GW}.

Now we return to the general case.  A solution of Hitchin's
equations can be interpreted in terms of either a Higgs bundle or
a complex-valued flat connection.  Let us work this out in the
present situation.

To get a Higgs bundle, we endow the bundle $E$ with a holomorphic
structure using the $(0,1)$ part of the connection $A$.  Then,
writing $\varphi$ for the $(1,0)$ part of $\phi$, $\varphi$ is a
holomorphic section of ${\rm ad}(E)\otimes K$ (here $K$ is the
canonical line bundle of the punctured $z$-plane) and the pair
$(E,\varphi)$ is our Higgs bundle.  Explicitly, upon conjugation
by the $G_\C$-valued function $r^{i\alpha}$, the operator
$\bar\partial_A=d\bar z(\partial_{\bar z}+A_{\bar z})$ reduces to
the standard operator $\bar\partial =d\bar z\partial_{\bar z}$.
This gives a trivialization of the holomorphic structure of $E$
near $z=0$ and an extension of $E$ across the singularity.  With
this trivialization, the Higgs field is simply
\begin{equation}\label{murty}\varphi=
\frac{dz}{2}\left(\frac{u_n}{z^n}+\frac{u_{n-1}}{z^{n-1}}+\dots+\frac{u_1}{z}\right).\end{equation}
This is unchanged from the $(1,0)$ part of $\phi$ as presented in
(\ref{jugid}), because $\phi$ is $\mathfrak t_\C$-valued and hence
unchanged by conjugation by $r^{i\alpha}$.

Alternatively, we can consider the  $G_\C$-valued connection
$\CA=A+i\phi$, which is flat by virtue of Hitchin's equations. Now
we make a unitary gauge transformation $\CA\to \CA-d\epsilon$ with
the real (that is $\mathfrak t$-valued) gauge parameter
\begin{align}\label{turkey}\epsilon=&-i\left(\frac{u_n}{(n-1)z^{n-1}}+\frac{u_{n-1}}{(n-2)z^{n-2}}
+\dots  +\frac{u_2}{ z}\right)\\ \nonumber &+i\left(\frac{\bar
u_n}{(n-1)\bar z^{n-1}}+\frac{\bar u_{n-1}}{(n-2)\bar z^{n-2}}
+\dots +\frac{\bar u_2}{ \bar z}\right)-\beta\,\ln r.
\end{align} After this gauge transformation, we get
\begin{equation}\label{urkey}\CA=
dz\left(\frac{u_n}{z^n}+\frac{u_{n-1}}{z^{n-1}}+\dots
+\frac{u_2}{z^2}\right)+(\alpha-i\gamma)\,d\theta.
\end{equation} Finally, a non-unitary ($\T_\C$-valued) gauge
transformation $d_\CA\to g d_\CA g^{-1}$ with
\begin{equation}\label{onery} g=r^{i(\alpha-i\gamma)}
\end{equation}
puts the connection in the form familiar from section
\ref{backgr}, namely $\CA=dz\,\CA_z$ with
\begin{equation}\label{nery}\CA_z=\frac{u_n}{z^n}+\frac{u_{n-1}}{z^{n-1}}+
\dots+\frac{u_2}{z^2}-i\frac{\alpha-i\gamma}{z} .\end{equation}
This is the standard form (\ref{jurfy}) of an irregular
singularity, with
\begin{align}\label{ery} T_1& =-i(\alpha-i\gamma) \\ \nonumber
T_k&=u_k, ~~k>1.\end{align}

As we know from section \ref{backgr}, the singular part of any
connection $\CA$  such that $T_n$ is regular and semisimple can be
put in this form.  So if we make this restriction on $T_n$ -- as
we will until section \ref{ugo} -- the abelian ansatz
(\ref{jugid}), which was forced upon us by the nonlinear nature of
Hitchin's equations, is sufficiently general to give a local model
for any irregular singularity.

\subsection{Hitchin Moduli Space}\label{hitchmod}
\def\EUW{\eusm W}
\def\EUM{\eusm M}
\def\EUG{\eusm G}
\def\MH{{{\EUM}_H}}
Now we can state the main result that was obtained in \cite{BB},
following earlier results in \cite{Sa}.  We will formulate this
result in terms of a hyper-Kahler quotient.  Let $C$ be a compact
Riemann surface and $E$ a smooth $G$-bundle over $C$. $G$ is a
compact Lie group with complexification $G_\C$. $A$ and $\phi$
will denote respectively a connection on $E$ and an
$\rm{ad}(E)$-valued one-form.

Let $p$ be a point in $C$ described as $z=0$ in terms of some
local parameter $z=re^{i\theta}$. We consider pairs $(A,\phi)$
with a singularity of the type described in section \ref{locmod}.
Thus, we fix $\alpha\in \mathfrak t$ and $u_1,\dots,u_n\in
\mathfrak t_\C$, and let $\EUW_p$ be the space of pairs $(A,\phi)$
that have a singularity at $p$ with the local behavior
\begin{align}\label{juggid}A&=\alpha \,d\theta+\dots\\
\nonumber
\phi&=\frac{dz}{2}\left(\frac{u_n}{z^n}+\frac{u_{n-1}}{z^{n-1}}+\dots+\frac{u_1}{z}\right)
 +\frac{d\bar z}{2}\left(\frac{\bar u_n}{\bar z^n}+\frac{\bar u_{n-1}}{\bar z^{n-1}}+\dots+\frac{\bar u_1}
 {\bar z}\right)+\dots,
\end{align}
where the ellipses refer to terms that are bounded at $z=0$.  And we
let $\EUG_p$ be the group of $G$-valued gauge transformations that
are $\Bbb{T}$-valued modulo terms of order $|z|^n$, and hence
preserve this form of $A,\phi$. For a more precise description, see
\cite{BB}.

Just as in the unramified case \cite{H}, or the tamely ramified
case \cite{S}, \cite{Kon}, the space $\EUW_p$ has a natural
hyper-Kahler structure, and $\EUG_p$ acts on $\EUW_p$ preserving
this structure.  The action of $\EUG_p$ has a hyper-Kahler moment
map $\vec \mu$, which is simply the left hand side of Hitchin's
equations. The space of solutions of Hitchin's equations, modulo
the action of $\EUG_p$, can be interpreted as the hyper-Kahler
quotient of $\EUW_p$ by $\EUG_p$. We denote this moduli space of
Hitchin's equations as $\MH$. (When we want to specify the gauge
group $G$, the Riemann surface $C$, or the parameters $\alpha$ and
$u_1,\dots,u_n$, we write more specifically $\MH(G,C)$,
$\MH(\alpha,u_1,\dots,u_n)$, etc.)

The result of \cite{BB} is to construct $\MH$ as a hyper-Kahler
manifold, which can be identified either with a suitable moduli
space of Higgs bundles, or with a moduli space of flat bundles
with irregular singularity.  Following the notation of \cite{H}
for the complex structures, in complex structure $I$, $\MH$ is a
moduli space of Higgs bundles with a singularity described locally
in eqn. (\ref{murty}), while in complex structure $J$, $\MH$ is a
moduli space of flat bundles with an irregular singularity of the
form (\ref{nery}). According to \cite{BB}, all general properties
of the moduli space of solutions of Hitchin's equations hold in
this situation, just as in the unramified or tamely ramified
cases. Hence, as we will spell out in more detail, all arguments
in \cite{KW} and \cite{GW} concerning the application to the
geometric Langlands program have close analogs.

We write $\omega_I,\omega_J,$ and $\omega_K$ for the three Kahler
forms on $\MH$.  Thus, $\omega_I$ is a Kahler form in complex
structure $I$, and similarly for $\omega_J$ or $\omega_K$. The
holomorphic symplectic form in complex structure $I$ is
$\Omega_I=\omega_J+i\omega_K$.  In the other complex structures,
the holomorphic two-forms are obtained by cyclic permutations of
$I,J,K$: $\Omega_J=\omega_K+i\omega_I$,
$\Omega_K=\omega_I+i\omega_J$. The symplectic forms $\omega_I,$
$\omega_J$, and $\omega_K$ are all defined by their standard gauge
theory formulas, described in detail in \cite{KW}, section 4.1.

\subsubsection{Nearly Abelian Structure}\label{nearab}

We will now explain an important detail (see \cite{BB}, Lemma 4.6,
for a more precise account). To define $\MH$ in the presence of an
irregular singularity, we require that the off-diagonal parts of
$A$ and $\phi$ are regular at $z=0$. In fact, Hitchin's equations
then require that the off-diagonal parts vanish near $z=0$ faster
than any power of $z$.

To see this, we start with the singular abelian model solution
(\ref{jugid}) and consider a perturbation $(\delta A,\delta\phi)$.
If we impose a gauge condition $D_z\delta A_{\bar z}+D_{\bar
z}\delta A_z+[\phi_z,\delta\phi_{\bar z}]+[\phi_{\bar
z},\delta\phi_z]=0$, then the linearization of Hitchin's equations
gives
\begin{equation}\begin{pmatrix} D_z & [\phi_{\bar z},\,\cdot\,]\\
          -[\phi_z,\,\cdot\,]& D_{\bar z}\end{pmatrix}\begin{pmatrix}\delta A_{\bar z}\\
          \delta\phi_z\end{pmatrix}=0. \end{equation}
So
\begin{equation}\begin{pmatrix} D_{\bar z} & -[\phi_{\bar z},\,\cdot\,]\\
          [\phi_z,\,\cdot\,]& D_{ z}\end{pmatrix}
          \begin{pmatrix} D_z & [\phi_{\bar z},\,\cdot\,]\\
          -[\phi_z,\,\cdot\,]& D_{\bar z}\end{pmatrix}\begin{pmatrix}\delta A_{\bar z}\\
          \delta\phi_z\end{pmatrix}=0. \end{equation}
Equivalently,
\begin{equation}\begin{pmatrix}D_{\bar z}D_z +[\phi_{\bar z},[\phi_z,\,\cdot\,]\,]& [D_{\bar z}\phi_{\bar
z},\,\cdot\,]\\ -[D_z\phi_z,\,\cdot\,] & D_zD_{\bar
z}+[\phi_z,[\phi_{\bar
z},\,\cdot\,]\,]\end{pmatrix}\begin{pmatrix}\delta A_{\bar z}\\
          \delta\phi_z\end{pmatrix}=0.\end{equation}
In the case of an irregular singularity, we have $|\phi|\sim
1/|z|^n$ for some $n>1$, so the terms $[D\phi,\,\cdot\,] $ are less
singular than $[\phi,[\phi,\,\cdot\,]\,]$.  For analyzing the
behavior near $z=0$ of the off-diagonal part of, for example,
$\delta\phi_z$, we can omit these terms and consider the equation
\begin{equation}\left(D_zD_{\bar z}+[\phi_z,[\phi_{\bar
z},\,\cdot\,]\,]\right)\delta\phi_z=0.\end{equation}

Now to explain why the off-diagonal parts of $\delta\phi$ and
$\delta A$ vanish very rapidly near $z=0$, let us consider the case
that $G=SU(2)$ and the most singular part of $\phi_z$ is
\begin{equation}\phi_z\sim \begin{pmatrix} w & 0 \\ 0 & -w
\end{pmatrix}\frac{1}{z^n},\end{equation}
with some $n>1$. The general case is similar. The leading
singularity of $\phi_{\bar z}$ is then
\begin{equation}\phi_{\bar z}\sim -\begin{pmatrix} \bar w & 0 \\ 0 & -\bar w
\end{pmatrix}\frac{1}{\bar z^n},\end{equation}
where the minus sign ensures that $\phi=\phi_z\,dz+\phi_{\bar
z}\,d\bar z$ is anti-hermitian.  Subleading terms in $\phi$ will
not be important near the singularity. Now, let us look at the
behavior of an off-diagonal matrix element of $\delta\phi_z$, say
\begin{equation}\delta\phi_z=\begin{pmatrix}0 & f \\ 0 & 0
\end{pmatrix}.\end{equation}
The behavior of $f$ near $z=0$ is governed by the equation
\begin{equation}\left(-\frac{\partial^2}{\partial z\partial \bar z}
+\frac{4|w|^2}{|\bar z z|^n}\right)f=0.\end{equation} (The
singularity of the connection $A=\alpha\,d\theta$ is too weak to be
relevant, so we have set $\alpha=0$.) The leading behavior of the
solution near $z=0$ is
\begin{equation}f\sim \exp\left(-\frac{4|w|}{(n-1)|\bar z z|^{(n-1)/2}}\right),\end{equation}
showing as claimed that $f$ vanishes near $z=0$ faster than any
power of $z$.

Since the classical analysis that we have just made is the starting
point for quantum mechanical perturbation theory, a similar result
holds quantum mechanically for the appropriate surface operators
(which will be introduced in section \ref{supersurf}): the
off-diagonal parts of the fields vanish very rapidly near the
support of a surface operator with wild ramification. Consequently,
the nonlinear effects are very small near such a surface operator,
rather than being very large, as one might have surmised.  In a
sense, this is the secret of wild ramification.

\subsubsection{Complex Structure J}\label{cpxj}
We will next discuss complex structures $I$ and $J$ in more detail.

In complex structure $J$, $\MH$ parametrizes flat $G_\C$ bundles
with a singularity of the type considered in section \ref{backgr}.
So $\MH(\alpha,u_1,\dots,u_n)$ coincides, as a complex manifold,
with the complex manifold ${\Y}^*(T_1)$, described in section
\ref{iso}, that parametrizes the generalized monodromy data. The
relationship between the parameters was given in (\ref{ery}):
$T_1=-i(\alpha-i\gamma)=-i(\alpha-i\,{\rm Im}\,u_1)$. Moreover,
the holomorphic form $\Omega_J$ of $\MH$ in complex structure $J$
coincides with the complex symplectic form $\Omega$ of
${\Y}^*(T_1)$, defined via gauge theory in eqn. (\ref{delfo}).

In the definition of ${\Y}^*(T_1)$, one considers irregular
singularities specified by a choice of $T_1,T_2,\dots,T_n$, all of
which are kept fixed.  However, the structure of ${\Y}^*(T_1)$ as
a complex symplectic manifold turns out to be independent of
$T_2,\dots,T_n$, while varying holomorphically with $T_1$.
Equivalently, then, $\MH(\alpha,u_1,\dots,u_n)$, as a complex
symplectic manifold in complex structure $J$, is independent of
$u_2,\dots,u_n$.  It is likewise independent of $\beta={\rm
Re}\,u_1$, as in the tame case \cite{GW}.  (We give alternative
explanations of this and similar statements in section
\ref{physper}.) Thus, as a complex symplectic manifold in complex
structure $J$, $\MH$ is independent of all the parameters that
specify the singularity, except the exponent of formal monodromy
$T_1=-i(\alpha-i\,{\rm Im}\,u_1)$, with which it varies
holomorphically.  $\beta$ does control the Kahler class, as in the
tame case.

As explained in section \ref{strategy}, the geometric Langlands
correspondence is derived by comparing the $B$-model of $\MH$ in
complex structure $J$ to the corresponding $A$-model defined using
the symplectic structure $\omega_K={\rm Re}\,\Omega_J$.  Away from
singularities of the moduli spaces (where recourse to the full
four-dimensional gauge theory is useful), these models can be
described as two-dimensional sigma models in which the target
space is the complex manifold ${\Y}^*(T_1)$ that parametrizes the
generalized monodromy data. No reference to Hitchin's equations is
needed.  As explained in section \ref{strategy}, what we gain from
Hitchin's equations is the knowledge that the parameter space of
the generalized monodromies has additional structure.  We describe
this next.

\subsubsection{Complex Structure $I$}\label{compi}
In complex structure $I$, $\MH$ parametrizes Higgs bundles
$(E,\varphi)$, where $E$ is a holomorphic $G_\C$-bundle over $C$
and $\varphi$ is a section  of ${\rm ad}(E)\otimes K_C$  that is
holomorphic away from the point $p$ (here $K_C$ is the canonical
bundle of $C$).  Near $p$, the singular behavior of $\varphi$ is
\begin{equation}\label{nearp}\varphi=\frac{u_n}{z^n}+\frac{u_{n-1}}{z^{n-1}}+\dots+\frac{u_1}{z}+\dots,
\end{equation}
where the last ellipses denote terms that are regular at $z=0$. As
a complex manifold in complex structure $I$, $\MH$ depends
holomorphically on $u_1,\dots, u_n$.  It is independent of
$\alpha$, which controls the cohomology class of the Kahler form
$\omega_I$.

The theory of Hitchin fibrations, originally developed \cite{H},
\cite{H2} for holomorphic Higgs fields, extends naturally to the
case of  Higgs fields with poles, as described in \cite{Be},
\cite{MD}. For Higgs bundles with simple poles, a short explanation
is given in section 3.9 of \cite{GW}.  That explanation focused
mainly on the simple example of $G=SU(2)$, and we will here briefly
extend it to the case of poles of higher order.

\def\EUBB{\cmmib B}
\def\CMR{\cmmib R}
The Hitchin fibration is defined in general by taking the
characteristic polynomial of the Higgs field $\varphi$.  For
$G=SU(2)$, this just means that we consider the object
$\Tr\,\varphi^2$, which is a quadratic differential on
$C\backslash p$ (that is, on $C$ with the point $p$ removed) with
a pole at $p$.  In view of (\ref{nearp}), the behavior of
$\Tr\,\varphi^2$ near $p$ is
\begin{equation}\label{zarp}\Tr\,\varphi^2=\frac{\Tr\,u_n^2}{z^{2n}}+
\frac{2\,\Tr\,u_nu_{n-1}}{z^{2n-1}}+\dots+\frac{2\,\Tr\,(u_nu_1+u_{n-1}u_2+\dots)}{z^{n+1}}+\dots,
\end{equation}
where the terms that are more singular than $z^{-n}$ depend only
on the polar part of $\varphi$, but the terms that are no more
singular than $z^{-n}$ depend also on the nonsingular part of
$\varphi$.

Let us write $\EUBB$ for the space of quadratic differentials on
$C\backslash p$ that take the form indicated in eqn. (\ref{zarp}).
Thus, a point in $\EUBB$ labels a quadratic differential that has
a pole of order $2n$ at $p$, such that the first $n$ coefficients
in a Laurent expansion near $p$ are as indicated in (\ref{zarp}).
The Hitchin fibration in the present situation is the map
$\MH\to\EUBB$ that maps a pair $(E,\varphi)$ to the point in
$\EUBB$ that is specified by $\Tr\,\varphi^2$.

 $\EUBB$ is an affine space isomorphic to $\Bbb{C}^{3g-3+n}$.  Indeed, two points in
 $\EUBB$ differ by a quadratic differential with a possible pole of order
 $n$ at $p$, that is, by an element of $H^0(C,K_C^2\otimes {\cal
 O}(p)^n)$.  This is a vector space of dimension $3g-3+n$, since the
 space of  quadratic differentials without pole has dimension
 $3g-3$, and allowing a pole of order $n$ increases the dimension
 by $n$.

 The usual general arguments about the Hitchin fibration apply in this situation.
 If we think of $\MH$ as a complex symplectic manifold in complex
 structure $I$, with the holomorphic symplectic form $\Omega_I$,
 then the functions on $\EUBB$ are Poisson-commuting.\footnote{The reason for
 this, as explained more fully in section 4.3 of \cite{KW},
 is that if one defines Poisson brackets using the holomorphic symplectic structure $\Omega_I$,
 then $\varphi$ commutes
 with itself (though not with $A$).  But $\Tr\,\varphi^2$ is, of course,
 a function only of $\varphi$.}
 The $3g-3+n$ independent linear functions on $\EUBB$ can thus be
 interpreted as Poisson-commuting Hamiltonians.
 There are precisely enough of these commuting Hamiltonians to
 establish the complete integrability of $\MH$.  Indeed, the
 dimension of $\MH$, according to (\ref{uffy}), is $6g-6+2n$, just
 twice the dimension of $\EUBB$.

The functions on $\EUBB$ generate, via Poisson brackets, a family
of commuting flows on the fibers of the Hitchin fibration
$\pi:\MH\to \EUBB$.  This strongly suggests that the generic
fibers will be complex tori, and this is so. Indeed, for
$G_\C=SL(N,\C)$, the fibers of the Hitchin fibration are Prym
varieties of a suitable spectral curve \cite{Be}, \cite{MD}, as in
the unramified case \cite{H}.

\def\CMF{{\cmmib F}}
The Hitchin fibration is a holomorphic map in complex structure
$I$, so the fibers are complex submanifolds in this complex
structure.  Being defined by the values of a maximal set of
commuting Hamiltonians, the fibers  are Lagrangian with respect to
the complex symplectic structure $\Omega_I$, or equivalently
(since $\Omega_I=\omega_J+i\omega_K$) with respect to the real
symplectic structures $\omega_J$ and $\omega_K$. Thus a fiber
$\CMF$ of the Hitchin fibration (endowed with a trivial Chan-Paton
line bundle, or more generally a flat one) is in the language of
\cite{KW} a brane of type $(B,A,A)$, that is, it is holomorphic in
complex structure $I$ and Lagrangian in symplectic structure
$\omega_J$ or $\omega_K$.

\subsubsection{Duality Of Hitchin Fibrations}\label{dualhit} As in
\cite{HT}, let us view the situation in complex structure $J$,
with Kahler form $\omega_J$.  The fibers of the Hitchin fibration
are Lagrangian with respect to $\omega_J$, as we have just
observed. They are actually special Lagrangian submanifolds, since
being holomorphic in complex structure $I$, they have minimal
volume.

So the Hitchin fibration is a fibration of $\MH(G,C)$ by special
Lagrangian tori. Such a fibration, according to \cite{SYZ}, is
precisely the input for mirror symmetry.  Thus the question arises
of what is the mirror of $\MH(G,C)$.  The answer to this question
turns out to be that the mirror of $\MH(G,C)$ (in complex
structure $J$) is $\MH({}^L\neg G,C)$ (in symplectic structure
$\omega_K$). This follows from the statement that the fibers of
the Hitchin fibrations for $G$ and $^L\neg G$, over corresponding
points in the base,\footnote{Once one picks a $G$-invariant metric
on the Lie algebra $\mathfrak g$, one gets a natural
identification between the bases of the Hitchin fibrations for $G$
and $^L\neg G$. Physically, a choice of $G$-invariant metric is
part of the definition of the theory since it is needed to define
the gauge theory action.} are dual tori. In the unramified case,
this was established in \cite{HT} for $G=SU(N)$ by directly
showing the duality between the fibers of the two Hitchin
fibrations.  It was subsequently proved in general \cite{DP} and
by a very direct argument in \cite{H4} for gauge group $G_2$.

{}From the point of view of four-dimensional $\EUN=4$ super
Yang-Mills theory, the SYZ duality between the Hitchin fibrations
of $G$ and $^L\neg G$ follows from electric-magnetic duality. This
was explained in section 5.5 of \cite{KW}, following earlier more
qualitative arguments \cite{BJV}, \cite{HMS}.  The arguments were
originally formulated for the unramified case, but  extend to
allow for ramification once one incorporates surface operators in
the formulation of electric-magnetic duality, as was done in
\cite{GW} in the tamely ramified case and as we will do next for
wild ramification.

\subsection{Surface Operators With Wild Ramification}\label{supersurf}

We now want to define supersymmetric surface operators in $\EUN=4$
super Yang-Mills theory that are appropriate for  wild
ramification. As in the tamely ramified case \cite{GW}, the main
ingredient is the singularity (\ref{juggid}) of a wildly ramified
solution of Hitchin's equations.

We consider  $\EUN=4$ super Yang-Mills theory, with the GL
topological twist described in \cite{KW}, on a four-manifold $M$
with Riemannian metric $g$.  The fields that are most important in
our discussion are a connection $A$ on a $G$-bundle $E\to M$ and
an ${\rm ad}(E)$-valued one-form $\phi$.  The general equations
for supersymmetry depend on a twisting parameter $t$ and are
\begin{align}\label{doilf} (F-\phi\wedge\phi +tD\phi)^+&=0\\
\nonumber (F-\phi\wedge\phi-t^{-1}D\phi)^-&= 0 \\
D\star \phi & = 0 \end{align} as in eqn. (3.29) of \cite{KW}. For
solutions that are pulled back from two dimensions, these
equations reduce to Hitchin's equations, independent of $t$.

We let $D$ be a codimension two submanifold of $M$, with an
oriented normal bundle $N$. If the metric on $M$ is near $D$ a
product $D\times N$, which will be the case in our application to
the geometric Langlands program, we proceed as follows.  We pick a
local parameter $z$ on $N$, such that $z=0$ along $D$.  We pick
parameters $\alpha\in \mathfrak t$ and $u_1,\dots,u_n\in \mathfrak
t_\C$, such that $u_n$ is regular. Then we consider $\EUN=4$ super
Yang-Mills theory on $M$ with fields that are singular along $D$,
with a singularity that in the normal plane to $D$ takes
everywhere the familiar form:
\begin{align}\label{nuggid}A&=\alpha \,d\theta+\dots\\
\nonumber
\phi&=\frac{dz}{2}\left(\frac{u_n}{z^n}+\frac{u_{n-1}}{z^{n-1}}+\dots+\frac{u_1}{z}\right)
 +\frac{d\bar z}{2}\left(\frac{\bar u_n}{\bar z^n}+\frac{\bar u_{n-1}}{\bar z^{n-1}}+\dots+\frac{\bar u_1}
 {\bar z}\right)+\dots,
\end{align} where the ellipses represent terms that are bounded for $z\to 0$.
Amplitudes for $\EUN=4$ super Yang-Mills theory on $M$, with a
surface operator on $D$, are computed by evaluating the standard
path integral, with the usual action, for fields with a
singularity of this kind.  This is analogous to the usual
definition of 't Hooft operators.

One point to verify here is that despite the singularities along
$D$, the gauge theory action is well-defined. This follows from
the fact that the singular parts of the fields obey Hitchin's
equations. The bosonic part of the action of GL-twisted $\EUN=4$
super Yang-Mills theory can be written, as in eqn. (3.33) of
\cite{KW}, as the integral of a sum of squares of the expressions
that appear on the left hand side of (\ref{doilf}). Those
expressions are all nonsingular near the singularities, because
the singularities are characterized by a solution of Hitchin's
equations. So the integral defining the action is convergent.

Though our emphasis is on the GL-twisted theory, the construction
is also natural in the underlying physical $\EUN=4$ super
Yang-Mills theory.  If we take $M=\R^4$, $D=\R^2$, then the
surface operators defined by the above construction preserve half
of the supersymmetry, since Hitchin's equations have this
property. Thus, as in \cite{GW}, these are half-BPS surface
operators, analogous to the half-BPS Wilson and 't Hooft line
operators of $\EUN=4$ super Yang-Mills theory.

\subsubsection{More on the Normal Behavior}\label{morenormal} Even if the metric on
$M$ is a product $D\times N$ near $D$, the choice of local
parameter $z$ is only natural up to a multiplicative constant.
Choosing the parameter is equivalent to trivializing $N$.
Regarding $N$ as a complex line bundle, a more canonical
formulation of the above is to say that $u_k$, for $k=2,\dots, n$,
takes values not in $\mathfrak g_\C$ but in $\mathfrak
g_\C\otimes_\C N^{k-1}$. For applications to geometric Langlands,
this underscores the fact, already emphasized in the introduction
and in section \ref{iso}, that the parameters $u_2,\dots,u_n$ (or
equivalently $T_2,\dots,T_n$) are not topological invariants. As a
result, a theory of isomonodromic deformation will play an
essential role. To keep our expressions simple, we usually suppose
that $N$ has been trivialized and just think of the $u_k$ as
taking values in $\mathfrak g_\C$.

Though adequate for application to geometric Langlands, the
assumption that the metric of $M$ looks like a product near $D$ is
somewhat unnatural.  This assumption will be relaxed in section
\ref{helfo}.

\subsection{Quantum Parameters And Action Of
Duality}\label{quantpar}
 The next step is to generalize the
definition of the surface operators to include certain quantum
parameters, and to determine the action of the electric-magnetic
duality group.  These arguments closely follow sections 2.3 and
2.4 of \cite{GW}, so we will be brief.

The form of the singularity (\ref{nuggid}) reduces the structure
group of the bundle $E$ along $D$ from $G$ to the maximal torus
$\Bbb{T}$.  For $G=SU(2)$, we have $\Bbb{T}=U(1)$, so $E$ is
equivalent along $D$ to a  $U(1)$-bundle ${\mathcal L}$, which has
a first Chern class $c_1({\mathcal L})$.  We can include in the
path integral an extra factor $\exp(2\pi i\eta c_1({\mathcal
L}))$, with $\eta\in \R/\Z$.  Since this factor is a topological
invariant, including it in the path integral preserves
supersymmetry.

In the general case with $G$ of rank $r$, we have $\Bbb{T}\cong
U(1)^r$.  Accordingly, the analog of $\eta$ now takes values in
$(\R/\Z)^r$, with one angular variable for each $U(1)$ factor.

As explained in section 2.3 of \cite{GW}, the torus in which
$\eta$ takes values can be canonically identified as $^L\neg\T$,
the maximal torus of the dual group $^L\neg G$.  We have $^L\neg
\T=\,^L\neg\mathfrak t/\Lambda_{\rm char}$, where $^L\neg\mathfrak
t$ is the Lie algebra of $^L\neg \T$, and $\Lambda_{\rm char}$ is
the character lattice of $G$.  Furthermore, $^L\neg\mathfrak t$
coincides with $\mathfrak t^\vee$, the dual of $\mathfrak t$.

Dually, although we introduced $\alpha$ as an element of
$\mathfrak t$, it is more precise to think of $\alpha$ as an
element of $\mathfrak t/\Lambda_{\rm cochar}=\T$, where
$\Lambda_{\rm cochar}$ is the cocharacter lattice of $G$.  The
reason for this is that by a $\T$-valued gauge transformation that
is singular along $D$, $\alpha$ can be shifted by an element of
$\Lambda_{\rm cochar}$. The gauge-invariant information contained
in $\alpha$ is the holonomy around $D$ of the unitary connection
$A$; this holonomy is $\exp(-2\pi\alpha)\in \T$.

The complete set of quantum parameters of our surface operator are
thus $\alpha\in \T$, $\eta\in {}^L\neg \T$, and $u_1,\dots,u_n\in
\mathfrak t_\C$, with $u_n$ constrained to be regular.  All of
these parameters are subject to the action of the Weyl group. The
Weyl group action was very important in \cite{GW}, leading
eventually to an action of the affine braid group  commuting with
the geometric Langlands duality. It will be a little less
important in the present paper, because of the restriction to
regular $u_n$. Even when we relax this restriction in section
\ref{gcase}, we will just get a similar story with fewer
variables, rather than a close analog of the role of the affine
braid group in the tamely ramified case.

\subsubsection{Action Of Duality}\label{actual} Now assuming that
this class of surface operators is mapped to itself by
electric-magnetic duality, we have to ask how the parameters
transform.  This question can be answered precisely as in
\cite{GW}.

First of all, the transformation of $u_1,\dots,u_n$ is exactly
like the transformation of $u_1=\beta+i\gamma$ in the tamely
ramified case (and is of secondary importance, as in that case).
It is determined by the transformation under duality of the
characteristic polynomial of the Higgs field, since
$u_1,\dots,u_n$ are determined (up to a Weyl transformation) by
the singular part of this characteristic polynomial, as we see for
$G=SU(2)$ in eqn. (\ref{zarp}).  So we can borrow the result of
eqn. (2.22) of \cite{GW}.  Let $x\to x^*$ be the map from
$\mathfrak t$ to $^L\neg \mathfrak t=\mathfrak t^\vee$ that comes
from the metric on $\mathfrak t$ in which a short coroot has
length squared 2. And let $\tau=\theta/2\pi + 4\pi i/g^2$ be the
gauge coupling parameter of $\EUN=4$ super Yang-Mills theory. The
basic electric-magnetic duality transformation $S$ maps a gauge
theory with gauge group $G$ and coupling parameter $\tau$ to one
with gauge group $^L\neg G$ and coupling parameter $^L\neg
\tau=-1/n_{\mathfrak g}\tau$ (here $n_{\mathfrak g}$ is the ratio
of length squared of long and short roots of $G$). In the process,
$u_1,\dots,u_n$ map to $^L\neg u_1,\dots,{}^L\neg u_n$ with
\begin{equation}\label{dinky}^L\neg u_k=|\tau|\,u_k^*,~k=1,\dots,n.\end{equation}

As in the tamely ramified case, the important transformation law
is that of $\alpha$ and $\eta$. They take values, respectively, in
$\T=\mathfrak t/\Lambda_{\rm cochar}$ and in
$^L\neg\T=^L\neg\mathfrak t/\Lambda_{\rm char}$. These groups are
exchanged under the basic electric-magnetic duality transformation
$S:\tau\to -1/n_{\mathfrak g}\tau$, strongly suggesting that
$\alpha$ and $\eta$ are likewise exchanged. Indeed, the same
arguments as in \cite{GW} (where the following formula appears as
eqn. (2.25)) strongly suggest that the transformation of
$(\alpha,\eta)$ under $S$ is
\begin{equation}\label{combo} (\alpha,\eta)\to (\eta,-\alpha).\end{equation}
The formulas (\ref{dinky}) and (\ref{combo}) can be extended, as
in \cite{GW}, to the full duality group.  However, since we will
restrict ourselves here to the most basic form of the geometric
Langlands duality, we will not need this generalization.

As in \cite{GW}, the main assumption of the present paper is that
the class of surface operators that we have introduced is mapped
to itself by $S$-duality, with the claimed transformation of the
parameters.  Once this is assumed, an elaboration of relatively
standard arguments leads to the geometric Langlands duality.  This
will be the focus of section \ref{wildram}.  But first, we will
reconsider isomonodromy from the point of view of supersymmetric
gauge theory.

\section{Supersymmetric Perspective On Isomonodromy}
\label{physper}

As explained in the introduction and in section \ref{iso}, one of
the key facts of this subject is invariance under isomonodromic
deformation.  As long as we constrain the leading coefficient
$u_n$ (or $T_n$) to be regular, which for $G_\C=SL(N,\C)$ means
that it is diagonalizable with distinct eigenvalues, the
parameters $u_2,\dots,u_n$ that characterize the irregular
singularity are irrelevant, both in the $B$-model of complex
structure $J$ and in the $A$-model of symplectic structure
$\omega_K$.  This is so because $\MH$, as a complex symplectic
manifold with complex structure $J$ and holomorphic symplectic
structure $\Omega_J$, is independent of the parameters noted. Our
next goal will be to understand this from the point of view of
supersymmetric gauge theory.  The argument will also show that the
process of changing $u_2,\dots,u_n$ commutes with duality, that
is, with the mirror symmetry between $\MH(G,C)$ in complex
structure $J$ and $\MH({}^L\neg G,C)$ in symplectic structure
$\omega_K$.

\subsection{Order And Disorder Operators}\label{ordisord}
In general, in quantization of a classical field theory, there are
two ways to define an operator (whether a local operator or an
operator supported on a line or a surface).  One may begin with a
classical expression and then quantize it.   Or one can define an
operator by prescribing the singularity that the fields should
have near a given point or line or surface.  The two cases
correspond to order and disorder operators, respectively. It is
sometimes  possible to mix the two constructions, and we will find
this useful.

For example, in twisted $\EUN=4$ super Yang-Mills theory, a
supersymmetric Wilson operator is constucted using the holonomy of
the complex connection $\CA=A+i\phi$:
$W_R(S)=\Tr_R\,P\exp\left(-\int_S\CA\right)$. Here $S$ is a loop
in spacetime, around which we take the holonomy, and $R$ is some
chosen representation of the gauge group.  When interpreted
quantum mechanically, $W_R(S)$ is a typical case of an order
operator.

An 't Hooft operator, instead, cannot be conveniently defined by
quantizing a classical expression.  Rather we modify the space in
which quantization is carried out by asking for the gauge field to
have a certain kind of singularity.  For simplicity, take the
gauge group to be $U(1)$ and the four-manifold $M$ to be
$\Bbb{R}^4=\Bbb{R}^3\times \Bbb{R}$, where the 't Hooft operator
is to be localized at the origin in the first factor, and the
second factor is parametrized by a ``time'' coordinate $s$. In
this particular case (as discussed for instance in \cite{KW}, eqn.
(6.9)), the 't Hooft operator is defined by considering fields
with the following sort of singularity:
\begin{align}\label{offus}F=&\frac{i}{2}\star_3d\frac{1}{|x|}\\
\nonumber
                           \phi=&
                           \frac{i}{2|x|}ds.\end{align}

There is no reasonable way\footnote{Incorporating a Dirac string
is generally unilluminating.} to add a source term to Maxwell's
equations that would generate the sort of singularity given by the
first line of eqn. (\ref{offus}). That is why 't Hooft operators
are understood as disorder operators; the appropriate singularity
is simply postulated, rather than being derived by quantization in
the presence of an appropriate source. However, the singularity in
$\phi$ actually can be usefully derived in that way.  This is not
a new result, but we will explain it in detail since it will serve
as a prototype for our study of surface operators.

In the conventions of \cite{KW}, the classical action for
$\phi=\sum_\nu \phi_\nu\,dx^\nu$ is
\begin{equation}\label{ofus} I_\phi=-\frac{1}{e^2}\int d^3x\,ds
\,\sum_{\mu,\nu}(\partial_\mu\phi_\nu)^2.\end{equation} We add a
source term
\begin{equation}\label{zobus} I'=\frac{4\pi
i}{e^2}\int_S \,\phi,\end{equation} where $S=\{0\}\times \R$ is
the locus of the 't Hooft operator in $\R^3\times \R$.  The
constant has been chosen so that the Euler-Lagrange equations for
the combined action $I_\phi+I'$, which read
\begin{equation}\label{oyo}(1/e^2)\partial_\mu\partial^\mu\phi(x,s)+(2\pi
i/e^2)\delta^3(\vec x)ds=0,\end{equation} are solved by
$\phi=(i/2|x|)ds$, the same singular behavior as in (\ref{offus}).
Adding a term $I'$ to the action is equivalent to including in the
path integral a factor
\begin{equation}\label{obus}\exp(-I')=\exp\left(-\frac{4\pi
i}{e^2}\int_S \,\phi(0,s)\right).\end{equation} The conclusion
then is that instead of simply postulating that $\phi$ has the
singular behavior in (\ref{offus}), we can generate this singular
behavior by including in the definition of the 't Hooft operator
the $\phi$-dependent term $\exp(-I')$.

We have carried out this discussion for gauge group $U(1)$, but
the general case is similar.  The 't Hooft operator is defined
using a homomorphism $\rho:U(1)\to G$, by means of which the
singular abelian solution (\ref{offus}) is embedded in $G$.  The
$\phi$-dependence of the 't Hooft operator is incorporated again
with a factor $\exp(-I')$.  $I'$ is defined as in (\ref{zobus}),
with $\phi$ replaced by $\Tr(\rho(1)\,\phi)$, where  $\rho(1)$ is
the image of $1\in \mathfrak{u}(1)$ under the homomorphism
$\rho:\mathfrak{u}(1)\to \mathfrak{g}$.  The Euler-Lagrange
equations now give
\begin{equation}\phi=\frac{i}{2|x|}\rho(1)ds.\end{equation}
This is the standard singular behavior of $\phi$ in the presence
of the 't  Hooft operator.

\subsection{Analog For Surface Operators}\label{surfan}

Now we want to work out an analog of this discussion for surface
operators.  We begin by reconsidering the surface operators relevant
to the tamely ramified case:
\begin{align}\label{lulgid}A&=\alpha\,d\theta\\ \nonumber \phi&=
\beta\,\frac{dr}{r}-\gamma
\,d\theta=(\beta+i\gamma)\frac{dz}{2z}+(\beta-i\gamma)\frac{d\bar
z}{2\bar z}
\end{align}

There is no reasonable way, in general, to add to the Lagrangian a
source such that the singularity in $A$ will appear upon solving
classical equations.  So in the usual spirit of disorder
operators, we will simply postulate this singularity, as was done
in \cite{KW}.  However, as in the case of the 't Hooft operator,
it is possible to write a classical source term that accounts for
the singularity in $\phi$.

It suffices to explain how to do this in a local model near the
singularity.  So we take $M=D\times C$, where $D\cong \C$ is the
complex $w$-plane, and $C\cong \C$ is the complex $z$-plane. The
locus of the singularity is the origin in the $C$, that is, it is
the locus $D\times \{0\}\subset M$, characterized by $z=0$.  So
the source term in the action will be supported on this locus. As
in the discussion of 't Hooft operators, we begin with the abelian
case and take the source term to be
\begin{equation}\tilde I=\frac{\pi}{e^2}\int_{D\times\{0\}}|d^2w|
\left((\beta-i\gamma)
\partial_{\bar z}\phi_z+(\beta+i\gamma)\partial_z\phi_{\bar
z}\right).\end{equation} Since  $\tilde I$ and the bulk action
$I_\phi$ from eqn. (\ref{ofus}) are both invariant under
translations of $w$, the resulting singularity is a function of
$z$ only.  From the combined action $I_\phi+\tilde I$, the
Euler-Lagrange equation for  a classical solution that depends
only on $z$ is
\begin{equation}\label{hobbo}\frac{\partial^2\phi_z}{ \partial
z\partial \bar z}=\pi (\beta+i\gamma)
\partial_z\delta^2(z).\end{equation} The solution is
\begin{equation}\label{obbo} \phi_z=\frac{\beta+i\gamma}{2z}.\end{equation}
Of course, $\phi_{\bar z}$ is minus the complex conjugate,
\begin{equation}\phi_{\bar z}=-\frac{\beta-i\gamma}{2\bar
z}\end{equation} and $\phi=\phi_z\,dz+\phi_{\bar z}\,d\bar z$ has
precisely the desired singular form given in eqn. (\ref{lulgid}).

As in the discussion of 't Hooft operators, it is straightforward
to extend this to the nonabelian case. We simply include a trace
in the definition of $\tilde I$:
\begin{equation}\label{fungo}\tilde
I=\frac{\pi}{e^2}\int_{D\times\{0\}}|d^2w|\,\Tr
\left((\beta-i\gamma)
\partial_{\bar z}\phi_z+(\beta+i\gamma)\partial_z\phi_{\bar
z}\right).\end{equation}

\subsubsection{A Detail}\label{adetail}
There is a detail to explain about this formula.  The surface
operators of \cite{GW} depend on the choice of a Levi subgroup
$\Bbb{L}$ of $G$.  The most basic case is that $\Bbb{L}$ is simply
the maximal torus $\Bbb{T}$.  Along the locus $D$ of a surface
operator, the structure group of the $G$-bundle $E$ is reduced
from $G$ to $\Bbb{L}$.  The connection $A$ and the field $\phi$,
along $D$, are valued in the Lie algebra $\mathfrak{l}$ of
$\Bbb{L}$; moreover, $\beta$ and $\gamma$ take values in the
center of $\mathfrak{l}$.  Given these facts, the component of
$\phi$ that contributes in the trace in (\ref{fungo}) similarly
takes values in the center of $\mathfrak{l}$; for
$\Bbb{L}=\Bbb{T}$, this simply means that it is
$\mathfrak{t}$-valued.

As one consequence, there is no need to replace
 the derivatives $\partial_z$ and $\partial_{\bar z}$ in
(\ref{fungo}) by covariant derivatives $D_z$ and $D_{\bar z}$.
With $A$ and $\phi$ being $\mathfrak l$-valued along $D$, and
$\beta,\gamma$ taking values in the center of $\Bbb{L}$,
$\Tr\,\beta[A,\phi]=\Tr\,\gamma[A,\phi]=0$ along $D$.  For the
same reason, if $\sigma$ and $\bar\sigma$ are the other scalars of
$\EUN=4$ super Yang-Mills theory, then
\begin{equation}\label{donkey}\Tr\,\beta[\sigma,\bar\sigma]=0\end{equation}
 along $D$.  This fact, which will be useful later,
  follows from the fact that $\sigma$ and
 $\bar\sigma$ are $\mathfrak l$-valued along $D$, while $\beta$ takes
 values in the center of $\mathfrak l$.

\subsection{Dependence On $\beta$}\label{varpar}

We are now going to use this perspective on the surface operators
to show that the topological field theories relevant to geometric
Langlands are independent of $\beta$, and moreover that varying
$\beta$ commutes with the geometric Langlands duality.  This was
already argued in \cite{GW}, but we will give a different
explanation that is a good starting point for the irregular case.

Let us consider only the part of (\ref{fungo}) that involves
$\beta$:
\begin{equation}\label{bungo}
I_\beta=\frac{\pi}{e^2}\int_{D\times\{0\}}|d^2w|\,\Tr \,\beta \left(
\partial_{\bar z}\phi_z+\partial_z\phi_{\bar
z}\right).\end{equation} We expand the one-form $\phi$ on $M=D\times
C$ as $\phi=dw\,\phi_w+d\bar w\,\phi_{\bar w}+dz\,\phi_z+d{\bar
z}\,\phi_{\bar z}$.  We have $d^*\phi=2(\partial_{\bar
w}\phi_w+\partial_w\phi_{\bar w}+\partial_z\phi_{\bar
z}+\partial_{\bar z}\phi_z)$.  So, after adding to $I_\beta$ the
total derivative $(\pi/e^2)\int |d^2w|\,\Tr\,(\partial_w\phi_{\bar
w}+\partial_{\bar w}\phi_w)$, we can write
\begin{equation}\label{hungox}
I_\beta=\frac{\pi}{2e^2}\int_{D\times\{0\}}|d^2w|\,\Tr\,\beta
\,d^*\phi.\end{equation}

In $\EUN=4$ super Yang-Mills, it is possible, at a generic value of
the twisting parameter $t$, to find a fermionic field $\hat \eta$
such that
\begin{equation}\label{juvani}\{Q,\hat\eta\}=d^*\phi,\end{equation}
 with $Q$
the topological supercharge of the theory. This follows from eqn.
(3.27) of \cite{KW}. ($\hat \eta$ is a $t$-dependent linear
combination of the fields $\eta$ and $\tilde\eta$ that appear in
that equation.) Hence, we get
\begin{equation}\label{ungox}I_\beta=\{Q,V\},\end{equation}
with
\begin{equation}V=\frac{\pi}{2e^2}\int_{D\times\{0\}}|d^2w|\,\Tr\,\beta
\,\hat\eta.\end{equation} At the important values $t=\pm i$, we
cannot achieve (\ref{juvani}), but we can pick $\hat\eta$ so that
$\{Q,\hat\eta\}=d^*\phi\mp i[\bar\sigma,\sigma].$   This leads
again to (\ref{ungox}), since according to (\ref{donkey}) the
extra term in $\{Q,\hat\eta\}$ does not contribute to $\{Q,V\}$.

Since terms in the action of the form $\{Q,V\}$ are irrelevant in
topological field theory, and are mapped by electric-magnetic
duality to terms of the same form, it follows that the topological
field theories related to geometric Langlands, and the dualities
between them, are independent of $\beta$. This was shown more
directly in \cite{GW}. In the $B$-model of complex structure $J$,
$\beta$ is irrelevant because it is a Kahler parameter; in the
$A$-model of symplectic structure $\omega_K$, $\beta$ is
irrelevant because it is a complex structure parameter.

\subsection{The Wild Case}\label{wildcase}

We can make a very similar argument in the case of wild
ramification.   Let us first explain the basic structure of the
argument for $G=U(1)$.

We consider a surface operator appropriate to wild ramification.
The local behavior near the singularity was described in section
\ref{locmod}:
\begin{align}\label{jougid}A&=\alpha \,d\theta\\
\nonumber \phi&=\frac{dz}{2}\left(
\frac{u_n}{z^n}+\frac{u_{n-1}}{z^{n-1}}+\dots+\frac{u_1}{z}\right)
 +\frac{d\bar z}{2}\left(\frac{\bar u_n}{\bar z^n}+
 \frac{\bar u_{n-1}}{\bar z^{n-1}}+\dots+\frac{\bar u_1}
 {\bar z}\right).
\end{align}
We want to show that the terms proportional to $u_k$ and $\bar
u_k$, $k>1$, can be generated by adding to the action for the
surface operator a term of the general form $\{Q,\dots\}$. It
turns out that, for $k>1$, this is true separately for the terms
linear in $u_k$ and in $\bar u_k$; we will just consider the
former. By contrast, for $k=1$, the above argument gave a weaker
result: only the part of the action linear in $\beta={\rm
Re}\,u_1$, not the part linear in $\gamma={\rm Im}\,u_1$, is of
the form $\{Q,\dots\}$.

To induce in $\phi_z$ a term $u_k/2z^k$, we add to the action of
the surface operator a term
\begin{equation}\label{turyo}I_k=  \frac{\pi}{e^2(k-1)!}\int_{D\times\{0\}}\,u_k
\partial_z^k\phi_{\bar z}\end{equation}
 In the presence of this term, the equation of motion
(\ref{hobbo}) receives a contribution proportional to $u_k$:
\begin{equation}\label{ohobbo}\frac{\partial^2\phi_z}{ \partial
z\partial \bar z}=\frac{(-1)^{k-1}}{(k-1)!}\pi u_k
\partial_z^{k}\delta^2(z)+\dots.\end{equation}
The solution gives the required contribution
$\phi_z=u_k/2z^k+\dots$.

In (\ref{turyo}), we can replace $\partial_z^k\phi_{\bar z}$ with
$\partial_z^{k-1}(\partial_z\phi_{\bar z}+\partial_{\bar
z}\phi_z)$ for the following reason.  The term we have added is
$\partial_z^{k-2}(\partial_z\partial_{\bar z}\phi_z)$.  The
equations of motion for $\phi_z$, in the abelian theory, read
\begin{equation}\label{dormo}\left(\partial_w\partial_{\bar
w}+\partial_z\partial_{\bar z}\right)\phi_z=0.  \end{equation} So
we can replace $\partial_z^{k-2}(\partial_z\partial_{\bar
z}\phi_z)$ by $-\partial_z^{k-2}\partial_w\partial_{\bar
w}\phi_z$. But this last term is a total derivative on the complex
$w$-plane, and so vanishes when inserted in (\ref{turyo}).  This
manipulation clearly only makes sense for $k\geq 2$, and that is
why we will get a stronger result in that case.

Just as in the case $k=1$, by adding another total derivative, we
can further replace $\partial_z\phi_{\bar z}+\partial_{\bar
z}\phi_z$ with $\partial_z\phi_{\bar z}+\partial_{\bar
z}\phi_z+\partial_w\phi_{\bar w}+\partial_{\bar
w}\phi_w=d^*\phi/2$.
 So we replace (\ref{turyo}) with
\begin{equation}\label{sturyo}I_k'= \frac{\pi}{2e^2(k-1)!}\int_{D\times\{0\}}\,u_k
\partial_z^{k-1}\left(d^*\phi\right)+c.c.\end{equation}
But now, using the existence of the field $\hat\eta$ with
$\{Q,\hat\eta\}=d^*\phi$ (in the abelian theory, such a field
exists even if $t=\pm i$, since $[\sigma,\bar\sigma]=0$), we see
that
\begin{equation}\label{guryo}I_k'=\{Q,V_k\}\end{equation}
with
\begin{equation}\label{curyo}
V_k=\frac{\pi}{2e^2(k-1)!}\int_{D\times\{0\}}\,u_k\partial_z^{k-1}\hat\eta.\end{equation}
So the parameters $u_2,\dots,u_k$ are entirely irrelevant in the
topological field theory.

\subsubsection{The  Nonabelian Case} We want to extend this to nonabelian
$G$. As in most of this paper (except section \ref{gcase}) we
assume that $u_n$ is regular and semisimple. We generalize the
abelian case by simply including a trace and replacing derivatives
with covariant derivatives:
\begin{equation}\label{curyoc}
V_k=\frac{\pi}{2e^2(k-1)!}\int_{D\times\{0\}}\Tr\,u_kD_z^{k-1}\hat\eta.\end{equation}
At first sight, it looks like there will be many problems in
repeating the above computation to show that $\{Q,V_k\}$ generates
precisely the desired singularity in $\phi_z$. In making the
argument, we freely integrated by parts and assumed that
derivatives commute with each other.  In nonabelian gauge theory,
the Yang-Mills curvature will generally appear in such
manipulations.  Moreover, we used the linear form (\ref{dormo}) of
the equations of motion, which of course does not hold in general
if $G$ is nonabelian. Finally, in the nonabelian case, if $t=\pm
i$, instead of $\{Q,\hat\eta\}=d^*\phi$, we have
$\{Q,\hat\eta\}=d^*\phi\mp [\sigma,\bar\sigma]$.  For the last
term not to affect the evaluation of $\{Q,V_k\}$, we need to have
$\partial_z^{k-1}[\sigma,\bar\sigma]=0$ at $z=0$.

All of these problems are resolved because of a key point that was
explained in section \ref{nearab}: as long as $u_n$ is regular and
semisimple,\footnote{More generally, this holds \cite{BB} as long
as $u_n,\dots,u_2$ can be simultaneously conjugated to $\mathfrak
t_\C$, the Lie algebra of a maximal torus $\Bbb{T}_\C$, and the
subgroup of $G_\C$ that commutes with all of them is precisely
$\Bbb{T}_\C$.  If the commuting subgroup is a more general Levi
subgroup $\Bbb{L}_\C$, the corresponding statement \cite{BB} is
that all fields are $\mathfrak l_\C$-valued modulo terms that
vanish near $z=0$ faster than any power of $z$. This will be good
enough for the argument since the $u_k$ take values in the center
of $\mathfrak l_\C$.} $A$ and $\phi$ are $\mathfrak t$-valued near
$z=0$ modulo ``off-diagonal'' terms that vanish faster than any
power of $z$. In the full $\EUN=4$ quantum theory, the same holds,
for essentially the same reasons, for $\sigma$ and all of the
other fields.  This is more than we need to resolve the problems
mentioned in the previous paragraph; there we only needed to know
that the fields are abelian up to order $z^n$. That ensures the
vanishing of all commutator and gauge curvature terms that would
appear in showing that $\{Q,V_k\}$ generates a singularity of the
desired form.

\subsubsection{Curved Spacetime}\label{helfo}
We have so far assumed that $D,C$, and $M=D\times C$ are flat, and
we have freely commuted derivatives with each other. However, once
we have arrived at the formula (\ref{curyoc}) for $V_k$, we can
immediately generalize to the case of a surface operator supported
on a general codimension two surface $D$ in a general
four-manifold $M$.
 We just extend the definition of
$V_k$ to make sense in the curved case, by including the
Riemannian connection in the definition of the covariant
derivatives and interpreting $u_k$ as a section of an appropriate
power of the normal bundle.\footnote{There is a topological
condition on the $u_k$. It comes from the fact that, if $u_n$ is
everywhere regular semi-simple, its discriminant trivializes a
certain power of the normal bundle to $D$ in $M$. Whenever the
classical geometry exists, the construction described in the text
gives the appropriate definition of the quantum surface operator
that preserves the topological symmetry.} By adding $\{Q,V_k\}$ to
the action, we induce the desired singularity in the $\phi$-field
along $D$, without modifying the topological field theory.  Of
course, $\{Q,V_k\}$ also contains less singular terms, involving
the curvatures of $D$ and $M$, that are absent in the original
local model for the singularity and ensure the topological
symmetry. In fact, we will study such terms later to understand
isomonodromy.

\subsubsection{Interpretation}\label{interpretation}
So, roughly speaking, the topological field theories obtained by
twisting $\EUN=4$ super Yang-Mills theory are independent of the
parameters $u_2,\dots,u_n$, as well as $\beta={\rm Re}\,u_1$.
There is one crucial caveat here. The space of fields in which the
theory is defined changes discontinuously if $u_n$ ceases to be
regular semisimple -- even the dimension of $\MH$ changes.  We can
change the $u_k$'s by adding terms like $\{Q,V_k\}$ to the action
for the surface operator, but in doing so we should avoid the
singularities that will occur if $u_n$ ceases to be regular
semisimple.

The space of regular semisimple $u_n$ is not simply-connected. For
example, for $G_\C=SL(N,\C)$, a semisimple element $x$ of the Lie
algebra can be diagonalized with eigenvalues
$\lambda_1,\dots,\lambda_N$. The condition for $x$  to be regular
is that the $\lambda_i$ are all distinct. When this condition is
imposed, it is possible to have monodromies in which the
$\lambda_i$ loop around each other.

\def\EUX{\eusm X}
So the statement that the topological field theory is independent
of $u_2,\dots,u_n$ and $\beta$ is incomplete.  A more precise
statement is that, as long as we keep $u_n$ regular semisimple,
the theory is locally independent of these parameters, since we
can vary them by adding a term $\{Q,\dots\}$ to the action for the
surface operator.  The proper formulation, as we will make
explicit in section \ref{connef}, is in terms of a flat
connection. Let $\EUX$ be the space of all $u_2,\dots,u_n\in
\mathfrak t_\C$, $\beta\in \mathfrak t$, with $u_n$ regular. Since
$\EUN=4$ super Yang-Mills theory can be defined and topologically
twisted for any point in $\EUX$, we get a family of topological
field theories parametrized by $\EUX$. (These theories depend
non-trivially on additional parameters $\alpha,\gamma,\eta$, and
$\Psi$, which we hold fixed in the present discussion.)  The
ability to change $u_2,\dots,u_n$ and $\beta$  by adding to the
action $\{Q,\dots\}$ means that we have a flat connection on this
family of topological field theories.

Such a flat connection, of course, may have global monodromies.
The fundamental group of $\EUX$ is called the braid group of $G$;
we denote it as $B(G)$.  The fundamental group is non-trivial only
because of the constraint that $u_n$ must be regular, and would be
unchanged if we hold fixed $u_{n-1},\dots,u_2$ and $\beta$. The
monodromies of the flat connection will give an action of $B(G)$
on the topological field theory defined in the presence of a
surface operator with wild ramification.

Our next task will be to make more explicit the flat connection
that governs the variation of the irrelevant parameters.
Interpreted in the $B$-model in complex structure $J$,\footnote{In
this context, $\beta$ does not play an essential role and is
usually not considered.} this flat connection is equivalent to the
classical isomonodromy connection \cite{JMU}. It will appear in a
differential-geometric formulation similar to that of \cite{B}.

\subsection{Flat Connection}\label{connef}

\begin{figure}[tb]
{\epsfxsize=3in\epsfbox{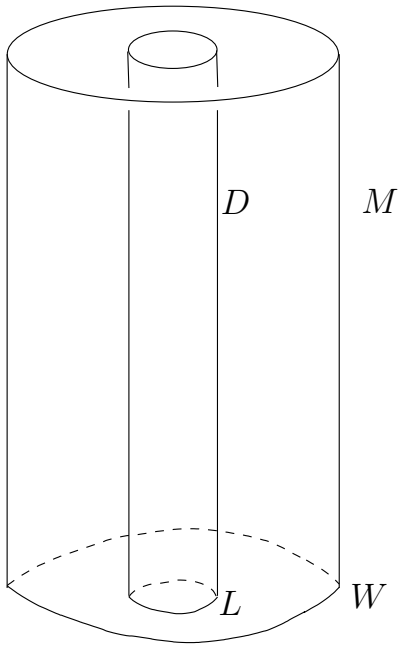}}
\begin{center}
\end{center}
 \caption{A surface operator whose support is a two-manifold $D=\Bbb{R}\times L$
 in a four-manifold $M=\Bbb{R}\times W$. Here $\Bbb{R}$ parametrizes the time direction,
 which runs vertically.  By endowing the surface operator with
 time-dependent couplings, we define a flat connection on the bundle $\hat{\mathcal
 H}\to \EUX$ of physical Hilbert spaces.}
 \label{zolgo}
\end{figure}

We take our four-manifold $M$ to be $\Bbb{R}\times W$, where $W$
is a three-manifold, possibly with boundary, and $\Bbb{R}$
parametrizes the ``time.'' We consider a surface operator (fig.
\ref{zolgo}) whose support is a product $D=\Bbb{R}\times L$ where
$L$ is a one-manifold in $W$. We suppose that $L$ is either closed
or else  terminates on the boundaries of $W$.

\def\A{\CA}
\def\F{\CF}
Quantization of twisted $\EUN=4$ super Yang-Mills theory on $W$ in
the presence of the surface operator (and possible additional
operators that preserve the topological symmetry) gives a
``physical Hilbert space'' ${\mathcal H}$ that {\it a priori}
depends on all the parameters of the theory.  Keeping fixed
$\alpha,\gamma,\eta,$ and $\Psi$, we allow only $u_2,\dots,u_n$
and $\beta$ to vary, and we think of ${\mathcal H}$ as the fiber
of a vector bundle  $\hat{\mathcal H}\to \EUX$.  We want to endow
this bundle with a flat connection.

To do this, we let $s$ denote a ``time'' coordinate on $\R$.  We
want to define parallel transport along a given path in $ \EUX$
that starts at a point $x_i$ and ends at $x_f$.  To do so, we
think of $u_2,\dots,u_n$ and $\beta$ as time-dependent functions
of $s$, constant in the far past and the far future, with initial
and final values determined by $x_i$ and $x_f$, and describing, as
$s$ varies from the past to the future, the chosen path in $\EUX$.
Topological symmetry can be preserved in the presence of this
time-dependence. We simply incorporate $ u_k$, for example, by
including in the surface operator action a term $\{Q,V_k\}$, where
$V_k$ is defined precisely as before
\begin{equation}\label{curyoco}
V_k=\frac{\pi}{2e^2(k-1)!}\int_{D\times\{0\}}\Tr\,u_kD_z^{k-1}\hat\eta,\end{equation}
but now with time-dependent $u_k$. (The ability to do this was
essentially already exploited in section \ref{helfo}.)

The path integral with this time-dependent action defines a map
from ${\mathcal H}_i$, the fiber of $\hat{\mathcal H}$ at $x_i$,
to ${\mathcal H}_f$, the fiber at $x_f$.  In fact, this map is
invariant under continuous displacements of the path $w$.  To
define the map from ${\mathcal H}_i$ to ${\mathcal H}_f$, we need
to pick some path.  But if we then change the path continuously,
remaining in $\EUX$ and without changing the values in the far
past and future, we merely add terms $\{Q,\dots\}$ to the action,
which will not change the transition amplitudes of the topological
field theory.  So the map from ${\mathcal H}_i$ to ${\mathcal
H}_j$ that comes from the path integral is invariant under
continuous changes in the path from $x_i$ to $x_f$.  This shows
that, in fact, the path integral defines a flat connection on the
bundle $\hat{\mathcal H}\to \EUX$.

We phrased this argument in terms of the physical Hilbert space, but
it applies more broadly to the full structure of the topological
field theory defined on the three-manifold $W$.  All operators,
branes, etc., that can be defined in the topological field theory
can be transported in the same way.

\subsection{Relation To Isomonodromy}

Our next goal is to make this more precise.  We start in the far
past with a supersymmetric field configuration that has a
singularity of the appropriate sort, reflecting the presence of a
surface operator.  Then we evolve the configuration into the
future solving the supersymmetric equations that were described in
\cite{KW}, but now with a singularity determined by the surface
operator. In the case of a time-dependent surface operator, the
resulting  solution will, of course, also be time-dependent.

Though we could carry out the discussion in four dimensions, for
brevity we will specialize to the case most relevant to geometric
Langlands -- compactification on a Riemann surface.  So we take
$W=S^1 \times C$, where $C$ is the Riemann surface on which we carry
out geometric Langlands.  The four-manifold is then $M=\Sigma\times
C$ where $\Sigma=\R\times S^1$. As explained in \cite{KW}, section
3.3, an
 irreducible solution of the supersymmetric equations is a pullback from the three-manifold
 $M'=\R\times \Sigma$.  This results in a slight simplification: the
 supersymmetric equations become independent of the twisting
 parameter $t$ introduced in \cite{KW}.  They assert that the
 complex-valued connection $\A=A+i\phi$ is flat
 \begin{equation}\label{hyto}0=\F=d\A+\A\wedge \A\end{equation}
 and obeys the ``gauge condition''
 \begin{equation}\label{pyto}0=d_A^*\phi.\end{equation}
 In fact \cite{Cos}, solutions of this pair of equations, modulo
 $G$-valued gauge transformations, are in natural correspondence
 with irreducible solutions of the first equation, modulo
 $G_\C$-valued gauge transformations.

\begin{figure}[tb]
{\epsfxsize=3in\epsfbox{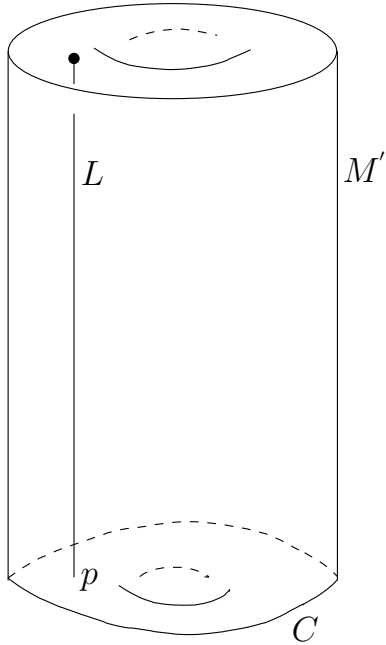}}
\begin{center}
\end{center}
 \caption{We consider supersymmetric fields with a singularity along
  $\R\times p\subset M'=\R\times C$.}
 \label{uzolgo}
\end{figure}
We are interested in solutions in the presence of a surface
operator whose support is of the form $S^1\times L$, where $L$ is
a line in the three-manifold $M'=\R\times C$.  We take initial
conditions that are pulled back from $M'$, in which case the full
time-dependent solution has this property.  Thus, instead of a
surface operator in $M=S^1\times M'$, we can simplify the
discussion slightly and think of a line operator in $M'$.
Moreover, we take $L$ to be of the form $L=\R\times p$, with $p$ a
point in $C$ (fig. \ref{uzolgo}).  As usual, $\R$ parametrizes the
time.

In this situation, the singular behavior of the fields near $L$
will be time dependent, reflecting the time-dependence of the
surface operator. Before trying to calculate, let us first guess
what might happen. We start with the tamely ramified case. At each
value of the ``time'' (which we call $s$), we expect to have a
flat connection with the usual tame singularity:
\begin{align}\label{pigid}A&=\alpha\,d\theta+\dots\\ \nonumber \phi&=
\beta\,\frac{dr}{r}-\gamma \,d\theta+\dots
\end{align}
In generalizing to the time-dependent case, it cannot be right to
simply give $\beta$ an $s$-dependence, because then the connection
would not be flat; indeed, the curvature would be
$\CF=(d\beta/ds)ds\wedge dr/r$.  But there is an obvious way to
modify (\ref{pigid}) to describe a flat connection even when $\beta$
is time-dependent.  We simply add another term:
\begin{align}\label{pigidi}A&=\alpha\,d\theta+\dots\\ \nonumber \phi&=
\beta\,\frac{dr}{r}-\gamma \,d\theta+(\ln
r)\frac{d\beta}{ds}ds+\dots.
\end{align}
The term that we have added is less singular than the terms that
were present already, which behave as $1/r$ rather than $\ln r$.
The addition ensures that the connection remains flat when $\beta$
varies.

This has a natural analog for the wildly ramified case.  The
time-dependent generalization of (\ref{jugid}) is
\begin{align}\label{tugid}A&=\alpha \,d\theta+\dots \\
\nonumber
\phi&=\frac{dz}{2}\left(\frac{u_n}{z^n}+\frac{u_{n-1}}{z^{n-1}}+\dots+\frac{u_2}{z^2}\right)
-
\frac{ds}{2}\left(\frac{du_n}{ds}\frac{1}{(n-1)z^{n-1}}+\frac{du_{n-1}}{ds}\frac{1}{(n-2)z^{n-2}}+\dots
+\frac{du_2}{ds}\frac{1}{z}\right)\\ \nonumber & +\frac{d\bar
z}{2}\left(\frac{\bar u_n}{\bar z^n}+\frac{\bar u_{n-1}}{\bar
z^{n-1}}+\dots+\frac{\bar u_2}
 {\bar z^2}\right)-
\frac{ds}{2}\left(\frac{d\bar u_n}{ds}\frac{1}{(n-1)\bar
z^{n-1}}+\frac{d\bar u_{n-1}}{ds}\frac{1}{(n-2)\bar z^{n-2}}+
\dots+\frac{d\bar u_2}{ds}\frac{1}{\bar z}\right)
\\ \nonumber
& +\beta\,\frac{dr}{r}-\gamma \,d\theta+(\ln
r)\frac{d\beta}{ds}ds+\dots.
\end{align}
Again, we have added less singular terms so that the behavior near
$z=0$ is consistent with the equations $\CF=0$ and $d_A^*\phi=0$.

These equations are first order in $s$ and have just the right
structure to uniquely determine a solution for all $s$ (modulo a
$G$-valued gauge transformation),  given initial data at a fixed
value of $s$ (which we take to lie in the far past). If we solve
those equations with the singular behavior specified in
(\ref{tugid}), the ordinary monodromies will certainly be
$s$-independent, since this is a consequence of flatness.  It is
also true that the generalized monodromies -- the Stokes matrices
-- are $s$-independent.  In fact, this is part of what is proved
in \cite{B}. What is shown there is that  the singular behavior
(\ref{tugid}), when combined with the equation $\F=0$, gives a way
to reformulate in terms of differential geometry the classic
isomonodromy connection \cite{JMU}. (In \cite{B}, the partial
gauge-fixing condition $d_A^*\phi=0$ was not imposed, so more
general complex-valued gauge transformations were allowed; also,
variation of $\beta$ was not considered, since this parameter is
not relevant in the complex symplectic geometry in complex
structure $J$.)

Now let us see how the above equations arise from quantum field
theory.  We begin with the parameter $\beta$.  According to the
above discussion, whether $\beta$ is time-independent or not, we
incorporate it by including in the action of a surface operator with
support $D$ a term $\{Q,V\}$ where
\begin{equation}V=\frac{\pi}{2e^2}\int_{D}|d^2w|\,\Tr\,\beta
\,\hat\eta.\end{equation} We have
\begin{equation}\label{trogo}\{Q,V\}=\frac{\pi}{e^2}\int_D
|d^2w|\,\Tr\,\beta\left(\partial_{\bar
w}\phi_w+\partial_w\phi_{\bar w}+\partial_{\bar
z}\phi_z+\partial_z\phi_{\bar z}\right).\end{equation} In our
present application, $D=\Bbb{R}\times S^1$, and we can parametrize
$D$ by, say, $w=s+i\theta$, where $\theta$ is an angular
coordinate on $S^1$ and as before $s$ is the time. When $\beta$ is
constant, we integrate by parts and discard the terms in
(\ref{trogo}) involving derivatives with respect to $w$ or $\bar
w$.  If, however, $\beta$ has an explicit $s$-dependence, then we
cannot discard these terms. Rather, writing $\partial_{\bar
w}\phi_w+\partial_w\phi_{\bar
w}=(\partial_s\phi_s+\partial_\theta\phi_\theta)/2$, and
integrating by parts, we see that the terms that were previously
dropped contribute
\begin{equation}\label{progo}-\frac{\pi}{2e^2}\int_D
|d^2w|\Tr\,\frac {d\beta}{ds}\phi_s.\end{equation} As a result,
rather as in (\ref{hobbo}), the equation of motion for $\phi_s$
becomes
\begin{equation}\frac{\partial^2\phi_s}{\partial z\partial\bar
z}=2\pi\frac{d\beta}{ds}\delta^2(z).\end{equation} The solution is
$\phi_s\sim \ln r (d\beta/ds)$, and this accounts for the
logarithmic term in (\ref{pigidi}).

We can similarly analyze the variation of the parameters
$u_2,\dots,u_n$ and their complex conjugates.  Whether $u_k$ is
time-dependent or not, it is incorporated by adding the term
$\{Q,V_k\}$ to the surface operator action, with $V_k$ as defined
in (\ref{curyoco}).  In the time-dependent case, the evaluation of
$\{Q,V_k\}$ gives rise to an extra term proportional to $du_k/ds$,
and this leads to the corresponding term in (\ref{tugid}).

\section{Geometric Langlands With Wild
Ramification}\label{wildram}

Our goal now is to describe the geometric Langlands program with
wild ramification, for the case that the coefficient $T_n$ of the
leading singularity of the connection is regular and semi-simple.
Given what we have explained so far, the arguments are rather
similar to those that have been given in \cite{KW}, \cite{GW} for
the unramified and tamely ramified cases. We will therefore give
only a brief overview of the construction, followed by more detail
on the points that are new.

The basic idea is to compare four-dimensional ${\cal N}=4$
supersymmetric gauge theory with gauge group $^L\neg G$ to gauge
theory with gauge group $G$. We work on a four-manifold
$M=\Sigma\times C$, with $\Sigma$ and $C$ being Riemann surfaces,
and with a surface operator supported on $D=\Sigma\times p$, for
$p$ a point in $C$. The surface operator is chosen to describe
wild ramification.  $C$ is the Riemann surface on which we study
geometric Langlands and $\Sigma $ will play an auxiliary role. The
parameters on which the surface operator depends are $\alpha$,
which appears in the ansatz $A=\alpha\,d\theta+\dots$ for the
gauge field, together with $\eta$, which is the coefficient of a
topological term in the effective action on $\Sigma$, and
$u_1,\dots,u_n\in \frak t_\C$, which  are the coefficients of the
polar part of the Higgs field, up to conjugacy. In this section,
we always assume that $u_n$ is regular. The relation between the
parameters $(\alpha,\eta,u_1,\dots,u_n)$ in the description with
gauge group $G$ and the corresponding parameters
$({}^L\neg\alpha,{}^L\neg\eta,{}^L\neg u_1,\dots,{}^L\neg u_n)$ in
the description with gauge group $^L\neg G$ is (as in section
\ref{quantpar})
\begin{align}
(\alpha,\eta) & = ({}^L\neg \eta,-{}^L\neg\alpha) \\ \nonumber u_i
& = |{}^L\neg\tau|\,{}^L\neg u_i^*.
\end{align}
Of these relations, the non-trivial one is the first one.  The
second relation simply reflects the usual gauge theory conventions
about coupling constants and metrics on the Lie algebras  $\frak g$
and $^L\neg\frak g$ and could be simplified if  different choices
were made. (The map $u_i\to u_i^*$ is a map between dual Lie
algebras $\frak t$ and $^L\neg\frak t$ that is determined by the
choice of metrics. And $\tau=\theta/2\pi+4\pi i/e^2$,
$^L\neg\tau={}^L\neg\theta/2\pi+4\pi i/{}^L\neg e^2$ are the gauge
coupling parameters.)

In this situation, the underlying four-dimensional theory reduces
for many purposes to a two-dimensional sigma model on $\Sigma$
with target space $\MH(C)$, the moduli space of Higgs bundles on
$C$ of the appropriate sort. Geometric Langlands duality comes
from\footnote{From a gauge theory point of view, we take the
twisting parameter $t$ of \cite{KW} to be $i$ and $1$,
respectively, in the descriptions via $^L\neg G$ and $G$. We also
take $ \tau$ and $^L\neg \tau$ to be imaginary. So the values of
the canonical parameter $\Psi$ in the $^L\neg G$ and $G$
descriptions are respectively $\infty$ and 0.} electric-magnetic
duality in four dimensions, which in two dimensions is the mirror
symmetry or $S$-duality between the $B$-model in complex structure
$J$ for gauge group $^L\neg G$ and the $A$-model in symplectic
structure $\omega_K$ for gauge group $G$.

The $B$-model is completely determined by the complex structure $J$
on $\MH$, that is, the complex structure in which $\MH$ parametrizes
flat bundles with wild ramification.  In this complex structure,
$\MH$ varies holomorphically with the exponent of formal monodromy
$^L\neg T_1=-i({}^L\neg\alpha-i^L\neg\gamma)$ (where
$^L\neg\gamma={\rm Im}\,^L\neg u_1$). It is independent of the
Kahler parameters $^L\neg\beta={\rm Re}\,{}^L\neg u_1$ and
$^L\neg\eta$. The dependence of $\MH$  on the coefficients $^L\neg
T_i={}^L\neg u_i$, $i=2,\dots,n$, of the higher order poles in the
connection is a little more subtle, as explained most fully at the
end of section \ref{strategy}. $\MH$ appears at first sight to vary
holomorphically with these parameters (in complex structure $J$).
But via the theory of isomonodromic deformation, $\MH$ is actually
independent of these parameters from a complex-analytic point of
view, though only infinitesimally independent of them from an
algebraic point of view.

The dual of this $B$-model is the $A$-model in symplectic
structure $\omega_K$. For the same reasons as in \cite{GW} (see
also the discussion of (\ref{yfosto})), $\MH$ as a symplectic
variety with symplectic structure $\omega_K$ is independent of
$\alpha$ and $\beta$.  The complexified Kahler class of $\MH$
varies holomorphically with $({\rm Im}\,\tau)\gamma^*-i\eta$, into
which the exponent of formal monodromy transforms under duality.
Since $\omega_K$ is the imaginary part of the holomorphic
symplectic form $\Omega_J$,  the symplectic nature of
isomonodromic deformation \cite{B}, described in sections
\ref{backgr} and \ref{physper}, implies that $\MH$ as a symplectic
variety with symplectic form $\omega_K$ is independent of
$u_2,\dots,u_n$. (There should also be an ``algebraic'' point of
view in which this independence only holds infinitesimally.)

The basic geometric Langlands duality, as usual, maps a $B$-brane
of $\MH({}^L\neg G)$, in complex structure $J$, to an $A$-brane of
$\MH(G)$, with symplectic structure $\omega_K$.  In particular, a
flat bundle with wild ramification described by coefficients
$T_2,\dots,T_n$ determines a point in $\MH({}^L\neg G)$ and hence
a zero-brane supported at that point.  $S$-duality, which as in
\cite{SYZ} amounts to $T$-duality on the fibers of the Hitchin
fibration, maps this zero-brane to an $A$-brane supported on the
appropriate fiber of the Hitchin fibration for $\MH(G)$.

\def\M{\EUM}
Because of an analogy with number theory, however, the geometric
Langlands duality is usually formulated in a different way. The
right hand is usually described not in terms of $A$-branes on
$\MH(G)$, but in terms of ${\cal D}$-modules (that is, modules for
the sheaf of differential operators) on $\M(G)$, the moduli space
of $G$-bundles on $C$, perhaps with some additional structure. As
explained in section 11 of \cite{KW}, and in section 4.4 of
\cite{GW} for the tamely ramified case, one can in fact identify
$A$-branes on $\MH(G)$ with ${\cal D}$-modules on $\M(G)$. This
depends upon two facts: (i) the existence of a canonical
coisotropic $A$-brane whose support is all of $\MH(G)$; (ii) the
fact that, from the point of view of complex structure $I$, a
dense open set in $\MH(G)$ is the cotangent bundle of $\M(G)$.

To implement this reasoning in the wildly ramified case, we will
first examine step (ii) and then, after a discussion of the
topology of $\MH$ in the wildly ramified case, return to step (i).

\subsection{Relation To A Cotangent Bundle}\label{cotbun}

We want to approximate the moduli space of Higgs bundles in the
wildly ramified case as the cotangent bundle of something.  This
identification, which will make it possible to relate $A$-branes
to ${\cal D}$-modules, should be holomorphic in complex structure
$I$. We will adopt a special notation in this section. We write
$\M_H^{\{0\}}$ for the moduli space of Higgs bundles in the
absence of ramification, and similarly we write $\M^{\{0\}}$ for
the moduli space of $G_\C$-bundles $E\to C$ with no additional
structure. We write $\MH(\vec u)$ for the moduli space of ramified
Higgs bundles with poles determined up to conjugacy by
$u_1,\dots,u_n$. ($\alpha$ will play no role for the moment, so we
omit it in the notation.) In complex structure $I$, a point in
$\M_H^{\{0\}}$ represents a Higgs bundle, that is a pair
$(E,\varphi)$, where $E\to C$ is a holomorphic $G_\C$-bundle and
$\varphi\in H^0(C,K_C\otimes {\rm ad}(E))$. Away from a set of
high codimension, $\M_H^{\{0\}}$ is \cite{H} the cotangent bundle
of $\M^{\{0\}}$. The map from $\M_H^{\{0\}}$ to $\M^{\{0\}}$ is
simply the forgetful map $(E,\varphi)\to E$. The complex
dimensions of $\M_H^{\{0\}}$ and of $\M^{\{0\}}$ are
\begin{align}{\rm dim}\,\M_H^{\{0\}}& = 2(g-1){\rm dim}\,G\\
\nonumber
              {\rm dim}\,\M^{\{0\}}& = (g-1){\rm dim}\,G \end{align}
The fact that the dimension of $\M_H^{\{0\}}$ is twice that of
$\M^{\{0\}}$ is consistent with the relation between
$\M_H^{\{0\}}$ and the cotangent bundle of $\M^{\{0\}}$.

Our goal is to compare $\MH(\vec u)$ to the cotangent bundle of
some space that we will provisionally call $\tilde \M$, until we
understand what it is.  $\tilde \M$ will be a moduli space of
$G_\C$-bundles over $C$ with some additional structure near the
ramification point $p$. The complex dimension of $\MH(\vec u)$ is,
according to eqn. (\ref{covec}),
\begin{equation}{\rm dim}\,\MH(\vec u)= 2(g-1){\rm dim}\,G+n({\rm
dim}\,G-r),\end{equation} with $r$ the rank of $G$. The dimension
of $\tilde\M$ will therefore have to be
\begin{equation}\label{oxley}{\rm dim}\,\tilde\M=(g-1){\rm dim}\,G+\frac{n}{2}({\rm
dim}\,G-r).\end{equation}

As in the absence of ramification, a point in $\MH(\vec u)$ still
represents in complex structure $I$  a pair $(E,\varphi)$.  But
now $\varphi$ has a pole at the point $p$:
\begin{equation}\label{upoles}\varphi=\frac{dz}{2}\left(
\frac{u_n}{z^n}+\frac{u_{n-1}}{z^{n-1}}+\dots+\frac{u_1}{z}+\dots\right)
\end{equation}
where regular terms are omitted. Here $u_1,\dots,u_n$ take values in
$\frak t_\C$, the Lie algebra of the maximal torus $\Bbb{T}_\C$ of
$G_\C$. A gauge transformation
\begin{equation}\varphi\to g\varphi g^{-1}\end{equation}
preserves this form of $\varphi$ if and only if $g$ takes values in
$\frak t_\C$ modulo terms of order $z^n$.  (This assertion depends
on our assumption that $u_n$ is regular and so commutes precisely
with $\Bbb{T}_\C$.)  Let $\M^{[n]}$ be the moduli space of
$G_\C$-bundles with a reduction of structure group to $\T_\C$ near
$p$ up to order $z^n$. The pair $(E,\varphi)$ determines a point in
$\M^{[n]}$.

Could $\M^{[n]}$ be the space $\tilde \M$  whose cotangent bundle
is related to $\MH(\vec u)$? The answer to this question is
``no,'' since the dimension is wrong. In fact, the dimension of
$\M^{[n]}$ is
\begin{equation}\label{likely}{\rm dim}\,\M^{[n]}=(g-1){\rm dim}\,G+n({\rm
dim}\,G-r).\end{equation} This formula arises as follows. The
complex codimension of $\Bbb{T}_\C$ in $G_\C$ is ${\rm dim}\,G-r$,
and reducing the structure group of $E$ from $G_\C$ to
$\Bbb{T}_\C$ up to order $z^n$ increases the dimension of the
moduli space by $n$ times this. Comparing (\ref{likely}) to
(\ref{oxley}), we see that reducing the structure group of $E$
from $G_\C$ to $\Bbb{T}_\C$ up to order $z^n$ is precisely twice
as much structure as we want. In other words, we want a reduction
of the structure group that will increase the dimension of the
moduli space by $\frac{n}{2}({\rm dim}\,G-r)$, not by $n({\rm
dim}\,G-r)$.

\def\MM{\M^{\{n\}}}
This suggests that we consider a reduction of the structure group
to a Borel subgroup $\EUB$, since the codimension of $\EUB$ is
half that of $\T_\C$.  For $G_\C=SL(N,\C)$ or $GL(N,\C)$, we
pick\footnote{The geometric Langlands duality, when expressed in
terms of ${\cal D}$-modules, will depend on this choice, while as
a relation between $B$-branes and $A$-branes, it does not.  So in
this sense, the formulation of the duality as a mirror symmetry
may be more natural.} an ordering of the eigenvalues of $u_n$, and
relative to this ordering we let $\EUB$ denote the group of upper
triangular matrices.  For any $G$, we pick a Borel subgroup $\EUB$
that contains $u_n$. (Since we assume $u_n$ to be regular
semi-simple, the number of possible choices of $\EUB$ is $\# {\cal
W}$, the order of the Weyl group $\cal W$ of $G$.) The codimension
of $\EUB$ in $G_C$ is $\frac{1}{2}({\rm dim}\,G-r)$. So if we
define $\MM$ to be the moduli space of $G_\C$-bundles with a
reduction of the structure group to $\EUB$ near the point $p$ up
to order $z^n$, then the dimension of $\MM$ is precisely what is
written on the right hand side of (\ref{oxley}).  $\MM$ will turn
out to be the desired space $\tilde \M$.

Since a Higgs field $\varphi$ of the form described in
(\ref{upoles}) determines a reduction of the structure group of
$E$ to $\Bbb{T}_\C$, modulo terms of order $z^n$, it certainly
determines a reduction of the structure group to the larger group
$\EUB$. However, different $\varphi$'s will determine the same
reduction to $\EUB$ if their polar parts differ by a strictly
upper triangular matrix.  We consider two Higgs fields $\varphi$
and $\varphi_0$ both with poles of order $n$. We suppose that
$\varphi_0$ obeys (\ref{upoles}) and that (for $G_\C=SL(N,\C)$)
\begin{equation}\label{showform}\varphi=\varphi_0+\begin{pmatrix} 0& * & \dots & * \\
                                   0 & 0 & \dots & * \\
                                    0 & 0& \ddots & * \\
                                   0 & 0 & \dots &
                                   0\end{pmatrix}+{\rm
                                   regular}.\end{equation}
The matrix is a strictly upper triangular matrix whose entries
have poles at most of order $n$.  The regular terms are not
required to be upper triangular.  Since the diagonal terms of
$\varphi$ and $\varphi_0$ are equal, and moreover the leading
coefficient $u_n$ has distinct eigenvalues, it is possible, by
changing the trivialization of $E$ (that is by a gauge
transformation $\varphi\to g\varphi g^{-1}$), to transform away
the strictly upper triangular polar part of $\varphi$ and  put
$\varphi$ in the form of (\ref{upoles}). Moreover, the required
change of trivialization is $\EUB$-valued, that is, $g$ is upper
triangular. So $\varphi$ and $\varphi_0$ determine the same
reduction of the structure group to $\EUB$.

It is convenient to write $\frak b$ for the Lie algebra of $\EUB$
(consisting of upper triangular matrices) and $\frak n$ for the
subalgebra of strictly upper triangular matrices.  Two Higgs
fields $\varphi$ and $\varphi_0$ determine the same reduction to
$\EUB$ if, relative to some choice of trivialization of $E$, their
polar parts are both $\frak b$-valued.  Once this is done, the
diagonal parts of $\varphi$ and $\tilde \varphi$ are the
``eigenvalues,'' and the condition that up to a gauge
transformation they each are of the form (\ref{upoles}) (with the
same $\vec u$) is that the difference $\varphi-\varphi_0$ should
be $\frak n$-valued.

\def\ande{{\rm ad}^{\{n\}}(E)}
Now let us describe the cotangent bundle of $\MM$.  First of all,
the tangent space to the moduli space $\M^{\{0\}}$ of bundles is
$H^1(C,{\rm ad}(E))$, where ${\rm ad}(E)$ is the sheaf of sections
of the adjoint bundle ${\rm ad}(E)$ derived from $E$.   To get the
tangent space to $\MM$, we must replace ${\rm ad}(E)$ by its
subsheaf consisting of sections that are $\frak b$-valued near $p$
up to order $z^n$. So let $\ande$ be this subsheaf. For example,
for $G_\C=SL(N,\C)$, a section $f$ of $\ande$ is a section of
${\rm ad}(E)$ of the form
\begin{equation}\label{noone} f=\begin{pmatrix} * & * & \dots & * \\
                                   0 & * & \dots & * \\
                                   0  &0 & \ddots & * \\
                                   0 & 0 & \dots & *\end{pmatrix}+{\cal O}(z^n). \end{equation}
The matrix entries are regular at $z=0$ and the terms not written
are of order $z^n$.  The cotangent space to $\MM$ will be, by
Serre duality, $H^0(C,K_C\otimes (\ande)^*)$, where $(\ande)^*$ is
a sheaf that is dual to $\ande$. $\ande$ was defined as in
(\ref{noone}), by requiring zeroes of order $n$ strictly below the
main diagonal. The dual of this is to allow poles of order $n$
strictly above the main diagonal. Hence $(\ande)^*$ is the sheaf
of sections of ${\rm ad}(E)$ of the form
\begin{equation}\label{ngoone} h=\begin{pmatrix} 0 & * & \dots & * \\
                                   0 & 0 & \dots & * \\
                                   0  &0 & \ddots & * \\
                                   0 & 0 & \dots & 0\end{pmatrix}+{\rm regular} \end{equation}
where the matrix is strictly upper triangular and its entries may
have poles of order $n$.  The regular terms are  not required to
be upper triangular.  Differently put, the polar part of $h$ is
$\frak n$-valued.

A point in the cotangent bundle $T^*\MM$ is therefore a pair
$(E,\varphi)$, where $E\to C$ is a $G_\C$-bundle with a reduction
of structure group to $\EUB$ up to $n^{th}$ order, and $\varphi$
is a Higgs field with $\frak n$-valued poles of order at most $n$.

\subsubsection{Affine Deformation}
Now we are going to define an ``affine deformation'' of this
cotangent bundle.  The affine deformation will depend on a choice
of $u_1,\dots,u_n\in \frak t_\C\subset \frak b$, and we will
denote it as $T^*\MM(\vec u)$.  A point in $T^*\MM(\vec u)$ is a
pair $(E,\varphi)$ where $E$ is as before and $\varphi$ is a Higgs
field with poles of order $n$. Instead of being $\frak n$-valued,
we now take the polar part of $\varphi$ to be $\frak b$-valued,
but with diagonal (or $\frak t_\C$-valued) part that is required
to agree with $\vec u$.  This means that (relative to the
reduction of the structure group of $E$ to $\EUB$ near the point
$p$), $\varphi$ is of the form (\ref{showform}).

The space $T^*\MM(\vec u)$ has a natural holomorphic map to $\MM$,
in which we forget $\varphi$ (remembering only the reduction in
structure group to $\EUB$ that it determines) and map the pair
$(E,\varphi)$ to $E$.  The fiber is a copy of $\C^n$, where $n$ is
the dimension of $\MM$.  If $\varphi$ and $\varphi_0$ are two
points in the fiber, their difference $\varphi-\varphi_0$ has a
strictly upper triangular polar part that thus determines a point
in the fiber of the cotangent bundle $T^*\MM$.  Thus, if
$\varphi_0$ were given, we could identify $T^*\MM(\vec u)$ with
$T^*\MM$ by mapping $\varphi$ to the cotangent vector
$\varphi-\varphi_0$.  Since there is no natural way to do this,
$T^*\MM(\vec u)$ cannot naturally be identified with $T^*\MM$;
rather we call it an affine deformation of $T^*\MM$.

In complex structure $I$,  once we pick a Borel subgroup that
contains $u_n$,  $\MH(\vec u)$ is the same as $T^*\MM(\vec u)$
away from a set of high codimension (where stability conditions
come into play).  A point in $\MH(\vec u)$ is a pair $(E,\varphi)$
with certain conditions on the poles of $\varphi$.  $\varphi$
determines the reduction of structure group of $E$ to $\EUB$
modulo $z^n$, and then the pair $(E,\varphi)$ determines a point
in $T^*\MM(\vec u)$.  This gives a holomorphic map
\begin{equation}\label{projform}\pi:\MH(\vec u)\to T^*\MM(\vec u)\end{equation}
that is an isomorphism away from singularities (whose codimension is
large if the genus of $C$ is large).  Thus, we have succeeded in
relating $\MH(\vec u)$ to an affine deformation of a cotangent
bundle.

This particular affine deformation has the important property of
being symplectic.  Thus, $\MH(\vec u)$ is hyper-Kahler and in
particular is a complex symplectic manifold in complex structure
$I$.  Its complex symplectic structure agrees, for any choice of
local Lagrangian section, with the natural complex symplectic
structure of $T^*\MM$.

\subsection{Topology}\label{topology} Next we will describe some
useful facts about the topology of $\MH(\vec u)$ and $\MM$.  The
discussion will roughly parallel section 3 of \cite{GW}, in an
abbreviated form.  As in \cite{GW}, some statements only hold away
from singularities of the moduli spaces.  The singular set is of
high codimension if the genus of $C$ is large.

\def\mone{{\eusm M}^{\{1\}}}
The first basic fact is that if we set $n=1$, we are in the tamely
ramified case, and in particular $\mone$, which is commonly called
the moduli space of parabolic bundles, was considered in some
detail in \cite{GW}.  Topologically, it is a fiber bundle over
$\M=\M^{\{0\}}$ with fiber the flag manifold $G_\C/\EUB$, which
parameterizes possible reductions of structure group from $G_\C$
to $\EUB$ at a given point $p\in C$:
\begin{equation}\label{yefto}\begin{matrix}G_\C/\EUB&\to& \mone\\
                                &   & \downarrow \\
                                & & ~\M. \end{matrix}\end{equation}
However, when we increase $n$, no further topology is involved;
$\MM$ is contractible onto $\mone$.  In fact, to give a reduction
of the structure group of $E$ from $G_\C$ to $\EUB$, up to order
$z^n$, means that (relative to a trivialization of $E$ near $p$)
we are given a function $\Phi(z)$ valued in $G_\C/\EUB$, defined
up to $z^n$.   $\Phi(0)$ takes values in $G_\C/\EUB$, which is
topologically non-trivial, but the derivatives of $\Phi$ take
values in contractible spaces (for example, $d\Phi/dz|_{z=0}$
takes values in the tangent space to the flag manifold at the
point defined by $\Phi(0)$).

One useful consequence of this is that, just as in the case $n=1$
which is described in \cite{GW}, the second cohomology group of
$\MM$ can be identified with the affine weight lattice of $G$:
\begin{equation}\label{sturkey}H^2(\MM;\Z)=\Z\oplus
\Lambda_{wt}.\end{equation}

Since\footnote{We restore the $\alpha$-dependence in the notation as
it will be relevant in this discussion.} $\MH(\alpha,\vec u)$ is a
deformation of the cotangent bundle of $\MM$ (away from an
exceptional set of high codimension), it is contractible onto $\MM$
and in particular has the same second cohomology group.  As a
hyper-Kahler manifold, $\MH(\alpha,\vec u)$ has symplectic forms
$\omega_I$, $\omega_J$, and $\omega_K$ that are Kahler,
respectively, with respect to the complex structures $I,J$, and $K$.
As in \cite{GW}, it is useful to determine the cohomology classes of
the symplectic forms.

First of all, according to \cite{B} and as we explained in section
\ref{physper}, if $\MH$ is understood as a moduli space of flat
bundles in complex structure $J$, then the holomorphic symplectic
form $\Omega_J=\omega_K+i\omega_I$ is independent of
$u_2,\dots,u_n$. Hence in particular the cohomology classes
$[\omega_K]$ and $[\omega_I]$ are independent of these variables.

On the other hand, if we think of $\MH(\alpha,\vec u)$ as a moduli
space of Higgs bundles in complex structure $I$, then the
cohomology class of the holomorphic symplectic form
$\Omega_I=\omega_J+i\omega_K$ is manifestly holomorphic in
$u_2,\dots,u_n$.  Since we already know that the cohomology class
$[\omega_K]$ is independent of $u_2,\dots,u_n$, it follows that
the same is true for $[\omega_J]$.

So all three cohomology classes are independent of the variables
$u_2,\dots,u_n$
 that are absent in the tamely ramified case.  Given this, it is perhaps not
surprising that the arguments of \cite{GW} can be carried over to
show that the cohomology classes $[\omega_I]$, $[\omega_J]$, and
$[\omega_K]$, as functions of $\alpha$, $\beta={\rm Re}\,u_1$, and
$\gamma={\rm Im}\,u_1$ are given by the formulas obtained in that
paper (in eqns. (3.76) and (3.77)):
\begin{align}\label{yfosto}\left[{\omega_I\over 2\pi}\right]&=e\oplus
(-\alpha^*)\\ \nonumber \left[{\omega_J\over
2\pi}\right]&=0\oplus(-\beta^*),~~\left[{\omega_K\over 2\pi}\right]
=0\oplus(-\gamma^*).\end{align} To get the first result, we use as
in the discussion of eqn. (3.4) of \cite{GW} the explicit formula
\begin{equation}\label{nostro}\omega_I=-{1\over 4\pi}\int_C\,\Tr\left(\delta A\wedge \delta A
-\delta\phi\wedge\delta \phi\right)
\end{equation}
The term involving $\delta\phi$ is exact, so we can replace
$\omega_I$ with $\omega_I'=-{1\over 4\pi}\int_C\,\Tr \,\delta
A\wedge \delta A$, which is a pullback from $\MM$. Then we can
borrow, for example, the reasoning in eqn. (3.53) of \cite{GW}, or
one of the references cited there, to arrive at the claimed result
for $[\omega_I]$.  The derivation of eqns. (3.8)-(3.10) of
\cite{GW} also carries over directly and leads to the formulas for
$[\omega_J]$ and $[\omega_K]$ that are given in (\ref{yfosto}).

\subsection{From $A$-Branes To $\cal D$-Modules}\label{intdm}   The fact
that $\MH(\vec u)$ is an affine deformation of the cotangent
bundle of $\MM$ makes it possible to identify $A$-branes on
$\MH(\vec u)$ with twisted ${\cal D}$-modules on $\MM$.

\def\bcc{{\cal B}_{c.c.}}
 The main steps are as in section 11 of \cite{KW}.
We consider the $A$-model  of the hyper-Kahler manifold $\MH(\vec
u)$ in symplectic structure $\omega_K$.  Provided $[\omega_J]=0$,
this model admits a special $A$-brane, the canonical coisotropic
brane ${\cal B}_{c.c.},$ whose Chan-Paton bundle is a complex line
bundle with curvature a multiple of $\omega_J$.  According to
(\ref{yfosto}), the condition to have $[\omega_J]=0$ is that
$0=\beta={\rm Re}\,u_1$. So we make that restriction in this
section.  Since the variable $\beta$ is irrelevant in the
$A$-model, this is not a serious restriction.

Because $\MH(\vec u)$ is an affine deformation of a cotangent
bundle, the space of $(\bcc,\bcc)$ strings can be partially
sheafified.  To be more precise, for every open set ${\cal
U}\subset \MM$, one can define the space of open string states
that are regular in $\pi^{-1}({\cal U})$, where $\pi:\MH(\vec
u)\to \MM$ is the projection.  And in addition, these spaces fit
together to make a sheaf over $\MM$. As is usual in the theory of
branes, the space of $(\bcc,\bcc)$ strings in any such open set
actually forms a ring (the multiplication law comes from joining
of open strings). The ring structure is compatible with
restrictions to open subsets. So we actually get a sheaf of rings
over $\MM$.

As is further shown in \cite{KW}, the sheaf of rings that one
obtains in this situation is a sheaf of differential operators
acting on some ``line bundle'' over $\MM$. (We put the phrase
``line bundle'' in quotes, because the relevant structure is a
little less than a line bundle; it is permissible to take complex
powers and to ignore torsion.  The reason for this will be
recalled in section \ref{parameters}.) For example, in the case of
the canonical coisotropic brane of a hyper-Kahler manifold $\MH$
that actually is a cotangent bundle of some space $\M$ (rather
than an affine deformation of one), time reversal symmetry was
used in \cite{KW} to show that the sheaf of $(\bcc,\bcc)$ strings
is the sheaf of differential operators acting on $K_\M^{1/2}$, the
square root of the canonical bundle $K_\M$ of $\M$.  (This sheaf
of rings is well-defined, regardless of whether a square root of
$K_\M$ exists or is unique globally.) More generally, if $\MH$ is
an affine deformation of the cotangent bundle of $\M$, one gets
the sheaf of differential operators on $K_\M^{1/2}\otimes {\cal
L}$, where ${\cal L}\to \M$ is some line bundle.  This
generalization was used in section 11.3 of \cite{KW} to describe
the role in the geometric Langlands program of the canonical
parameter $\Psi$.

The canonical coisotropic brane may seem rather special, but its
existence has a very general implication for the $A$-model.  Let
$\cal B$ be any $A$-brane.  Then the  $(\bcc,\cal B)$ strings
(which can always be sheafified along $\M$) give a sheaf of
modules for the sheaf of $(\bcc,\bcc)$ strings.  So in other
words, there is a natural way to associate to any $A$-brane ${\cal
B}$ a corresponding module for the sheaf of differential operators
acting on $K_\M^{1/2}\otimes {\cal L}$.  Hence, the geometric
Langlands duality can be restated in terms of twisted ${\cal
D}$-modules, that is, modules for the sheaf of differential
operators acting on some line bundle.

This general picture applies to our problem, since $\MH(\vec u)$ is
an affine deformation of a cotangent bundle of a space $\MM$. So the
$(\bcc,\bcc)$ strings are the differential operators acting on
sections of $K_\M^{1/2} \otimes {\cal L}$ for some ${\cal L}$. And
to an $A$-brane on $\MH(\vec u)$, there is a naturally associated
module for the algebra of such operators.

\subsubsection{Parameters}\label{parameters}

To go farther, let us understand the relation between the
parameters involved in an affine deformation of the cotangent
bundle of $\MM$ and the parameters involved in choosing a ``line
bundle'' ${\cal L}$.

\def\EUV{\eusm V}
\def\CO{{\cal U}}
\def\CU{{\cal U}}
Let us first explicitly describe in general how to construct an
affine deformation $\EUV$ of the cotangent bundle of $\MM$. More
specifically, since $\MH(\vec u)$ is hyper-Kahler, we will analyze
affine deformations that are also symplectic (in complex structure
$I$). Thus, we consider a complex symplectic manifold $\EUV$ with
a holomorphic symplectic form $\Omega_I$ and a holomorphic map
$\pi:\EUV\to \MM$.  We also suppose that locally, after picking a
local section $s$ that is Lagrangian (that is, $\Omega_I$ vanishes
when restricted to $s$), $\EUV$ can be identified with $T^*\MM$
with its usual symplectic struture.

Now cover $\MM$ with small open sets $\CO^\alpha$ on each of which
$\pi$ admits a Lagrangian section $s^\alpha$. On an intersection
$\CO^\alpha\cap\CO^\beta$, we have a pair of Lagrangian sections
$s^\alpha$ and $s^\beta$.
  Pick local coordinates
$q^i$ on $\CO^\alpha\cap\CO^\beta$. We would like to pick conjugate
coordinates $p_i$ that vary linearly on the fibers of the cotangent
bundle.  Such coordinates, which we will call $p_i^\alpha$, are
uniquely determined if we ask that they should vanish at $s^\alpha$
and that
\begin{equation}\label{holog} \Omega_I=\sum_i dp_i^\alpha\wedge
dq^ i.\end{equation} Likewise we can define  coordinates $p_i^\beta$
that vanish on $s^\beta$ and such that
\begin{equation}\label{olog} \Omega_I=\sum_i dp_i^\beta\wedge
dq^ i.\end{equation} Compatibility of these relations imply that
we must have
\begin{equation}\label{yurty}p_i^\beta- p_i^\alpha=\frac{\partial
\phi^{\alpha\beta}}{\partial q^i}\end{equation} with
$\phi^{\alpha\beta}$ a holomorphic function on
$\CO^\alpha\cap\CO^\beta$.

The one-form
$\lambda^{\alpha\beta}=\partial\phi^{\alpha\beta}=\sum_i (\partial
\phi^{\alpha\beta}/\partial q^i)\,dq^i$ is closed.  It is
therefore a section over $\CO^\alpha\cap \CO^\beta$ of what we may
call $\Omega^1_{\rm cl}(\MM)$, the sheaf of closed (and in
particular holomorphic) $(1,0)$-forms. Moreover, consistency of
the gluing operation of (\ref{yurty}) implies that in triple
overlaps, we have
$\lambda^{\alpha\beta}+\lambda^{\beta\gamma}+\lambda^{\gamma\alpha}=0$.
Finally a transformation $\lambda^{\alpha\beta}\to
\lambda^{\alpha\beta}+\mu^\alpha-\mu^\beta$ could be eliminated by
redefining the sections $s^\alpha$ with which we began by
$s^\alpha\to s^\alpha-\mu^\alpha$.

\def\hone{H^1(\MM,\Omega^{1}_{\rm cl}(\MM))}
All this means that the $(1,0)$-forms $\lambda^{\alpha\beta}$
determine an element of the sheaf cohomology group
$H^1(\MM,\Omega^{1}_{\rm cl}(\MM))$. So every affine symplectic
deformation $\EUV$ of the cotangent bundle $T^*\MM$ determines a
class $\lambda\in H^1(\MM,\Omega^{1}_{\rm cl}(\MM))$. Conversely,
given such a $\lambda$, one can construct $\EUV$ by reversing the
construction. So affine symplectic deformations of $T^*\MM$ are
classified by $\hone$.

 In triple
overlaps $\CO^\alpha\cap\CO^\beta\cap \CO^\gamma$, we have
$\lambda^{\alpha\beta}+\lambda^{\beta\gamma}+\lambda^{\gamma\alpha}=0$
(since $\lambda^{\alpha\beta}=\lambda^\alpha-\lambda^\beta$), so
$\partial(\phi^{\alpha\beta}+\phi^{\beta\gamma}+\phi^{\gamma\alpha})=0$
Hence the quantities
$c^{\alpha\beta\gamma}=\phi^{\alpha\beta}+\phi^{\beta\gamma}+\phi^{\gamma\alpha}$
are complex constants. If the $c^{\alpha\beta\gamma}$ are integer
multiples of $2\pi i$, we can use the objects
$\exp(\phi^{\alpha\beta})$ as transition functions defining a
complex line bundle ${\cal L}$. In general, this is not the case,
but one can define differential operators acting on ${\cal L}$ or
(more pertinently) on ${K_{\M}^{1/2}}\otimes {\cal L}$. The point is
that the sheaf of such differential operators can be defined
globally, even though the transition functions defining ${\cal L}$
only close up to complex constants $\exp(c^{\alpha\beta\gamma})$.
The reason for this is simply that such constants commute with
differential operators.  This is also why torsion in ${\cal L}$ does
not affect the sheaf of differential operators acting on ${\cal L}$.

So to every affine symplectic deformation $\EUV$ of $T^*\MM$ there
corresponds a ``line bundle'' $\cal L$ and a sheaf of differential
operators acting on sections of $K_{\MM}^{1/2}\otimes {\cal L}$.
It is then a natural conjecture that if $\bcc$ is the canonical
coisotropic brane over $\EUV$, then the sheaf of $(\bcc,\bcc)$
strings is the sheaf of differential operators acting on
$K_{\MM}^{1/2}\otimes {\cal L}$. In section 11.3 of \cite{KW},
this was shown for a particular case  by a special argument. But
actually it is a general fact that can be demonstrated by studying
the sheaf of $(\bcc,\bcc)$ strings in $\sigma$-model perturbation
theory. (The lowest non-trivial order of perturbation theory
determines the full result, as explained in \cite{KW}.  In that
order, the only possible answer is a linear map from $\lambda$ to
$c_1({\cal L})$, so a special case really determines the general
answer.)

The quantities $c^{\alpha\beta\gamma}$ can be interpreted as  a
\v{C}ech cocycle defining a class in $H^2(\MM,\C)$.  This class
has an alternative interpretation: it is the cohomology class of
$\Omega_I$.  To see this, one traces through the usual relation
between de Rham cohomology and \v{C}ech cohomology, to identify
the de Rham cohomology class of the closed two-form $\Omega_I$
with the \v{C}ech cohomology class represented by the collection
$c^{\alpha\beta\gamma}$.  According to \ref{yfosto}, given that we
have set $\beta=0$, the cohomology class of
$\Omega_I=\omega_J+i\omega_K$ is $0\oplus (-i\gamma^*)$.

On the other hand, by another standard construction, the
two-cocycle $c^{\alpha\beta\gamma}/2\pi i$ represent the first
Chern class $c_1({\cal L})$.  The only subtlety is that since we
have not required the $c^{\alpha\beta\gamma}$ to take values in
$2\pi i\Bbb{Z}$, we must interpret $c_1({\cal L})$ as an element
of the complex cohomology of $\MM$, not the integral cohomology.
Combining these statements, the first Chern class of ${\cal L}$ in
complex cohomology is
\begin{equation}\label{goofo}c_1({\cal L})=0\oplus
(-i\gamma^*).\end{equation} This formula is notably independent of
the parameters $u_2,\dots,u_n$, showing that when these parameters
are varied, ${\cal L}$ changes by tensoring by a line bundle that
is topologically trivial, though in general holomorphically
non-trivial.

We conclude with a comment on the non-integrality of the formula
(\ref{goofo}). If ${\cal L}$ is a line bundle in the generalized
sense (for example, the complex power of a line bundle), then it
does not make sense to define the sheaf of sections of ${\cal L}$,
but one can define the sheaf of differential operators acting on
${\cal L}$. Suppose, however, that ${\cal N}$ is an honest line
bundle. Then\footnote{For brevity, in this explanation we absorb the
usual $K_{\M}^{1/2}$ in the definition of ${\cal L}$.} ${\cal
D}_{\cal L}$ and ${\cal D}_{{\cal L}\otimes {\cal N}}$, the sheaves
of differential operators acting respectively on ${\cal L}$ and on
${\cal L}\otimes{\cal N}$, are Morita equivalent, meaning that they
have equivalent categories of modules.  The equivalence is
established using a bimodule consisting of the differential
operators mapping from ${\cal L}\otimes {\cal N}$ to ${\cal L}$. So
not only can one define ${\cal D}_{\cal L}$ where ${\cal L}$ is not
an honest line bundle, but in a sense this is the essential case.

\subsection{Restatement Of The Duality}\label{jungo}

At last we can restate the geometric Langlands duality in terms of
${\cal D}$-modules. The $n$-plet $\vec u=(u_1,\dots,u_n)$ (with
${\rm Re}\,u_1=0$) determines an affine symplectic deformation
$T^*\MM(\vec u)$ of the cotangent bundle of $\MM$.  This
deformation determines a class $\lambda\in H^1(\MM,\Omega^1_{\rm
cl})$, and an associated ``line bundle'' ${\cal L}\to \MM$.  An
$A$-brane ${\cal B}$ over $\MH$ determines a module for this sheaf
of rings, namely the sheaf of $(\bcc,{\cal B})$ strings. Composing
this with the mirror symmetry between the $B$-model and the
$A$-model, we associate to a $B$-brane over $\MH$ in complex
structure $J$ a twisted ${\cal D}$-module over $\MM$.

This statement needs to be generalized slightly to include the
$\theta$-like parameter $\eta$, upon which the $A$-model of $\MH$
depends. $\eta$ was not part of the above discussion because it is
not part of the classical geometry.  Just as in \cite{GW}, the
quickest way to restore the $\eta$-dependence is to simply use the
fact that the $A$-model varies holomorphically in $\eta+i({\rm
Im}\,\tau) \,\gamma^*$.

In the $B$-model of $\MH$, the exponent of formal monodromy,
according to (\ref{ery}), is
$T_1=-i({}^L\neg\alpha-i{}^L\neg\gamma)$.  Under $S$-duality,
$-iT_1$ transforms into $\eta+i({\rm Im}\,\tau)\,\gamma^*$. This
quantity is the first Chern class of ${\cal L}$.
\def\wt{{\rm wt}}

The last statement fixes the normalization of the map from $\vec u$
to the ``line bundle'' ${\cal L}$, about which we have been
imprecise so far. The statement can be justified by the same
arguments as in \cite{KW}. In general, $\eta$ takes values in $\frak
t^*/\Lambda_\wt$, where $\Lambda_\wt$ is the weight lattice of $G$.
Hence, a shift $\eta\to \eta+x$ should be a symmetry for $x\in
\Lambda_\wt$. Moreover, as the cohomology of $\MM$ is the affine
weight lattice, a choice of $x$ determines a line bundle\footnote{In
the underlying quantum field theory, $\eta$ is a $\theta$-like
angle.  A shift of $\eta$ by a lattice vector induces in the sigma
model of target $\MH$ a corresponding shift of the $B$-field by an
integral cohomology class. This acts on branes by the tensor product
with a line bundle.}
 ${\cal L}^x\to \MM$. The shift
$\eta\to\eta+x$ acts on ${\cal L}$ by ${\cal L}\to {\cal L}\otimes
{\cal L}^x$, and as ${\cal L}^x$ is an honest line bundle, this maps
the sheaf of differential operators acting on $K^{1/2}_{\MM}\otimes
{\cal L}$ to an equivalent one.

\subsubsection{Isomonodromy Of $\cal D$-Modules}\label{isomd}

A  final question here is to understand the counterpart in terms of
$\cal D$-modules of the variation of the parameters $u_2,\dots,u_n$
via isomonodromic deformation.

Such deformation gives a symmetry of the $B$-model in complex
structure $J$, and of the mirror $A$-model with symplectic
structure $\omega_K$.  But what does it mean in terms of twisted
$\cal D$-modules on $\MM$?

I do not know the answer, but will offer a speculation.  First of
all, some things are clear from section (\ref{strategy}). {}From a
complex analytic point of view, the category of modules for the
sheaf of algebras ${\cal D}_{{K^{1/2}_{\MM}}\otimes {\cal L}}$
should be independent of $u_2,\dots,u_n$.  From an algebraic point
of view, this category should be independent of $u_2,\dots,u_n$ only
infinitesimally.

Such infinitesimal independence sounds peculiar, but is naturally
produced by isomonodromic deformation.  Isomonodromic deformation
gives a way of varying ${\cal D}$-modules.  One of our main concerns
in this paper has been isomonodromic deformation of ${\cal
D}$-modules -- that is, flat bundles with singularities -- over the
Riemann surface $C$.  It would be very natural if variation of the
parameters $u_2,\dots,u_n$ is accomplished by an analogous process
of isomonodromic deformation on $\MM$.  In that case, one would
expect the independence of $u_2,\dots,u_n$ to hold as an actual
statement complex analytically and as an infinitesimal statement
from an algebraic point of view.

Part of the story is surely  that (as we see in eqn. (\ref{goofo}))
the first Chern class of ${\cal L}$ is independent of
$u_2,\dots,u_n$.

In the study of a flat bundle  with irregular singularities on a
Riemann surface $C$, one often makes an analogy between varying
the complex structure of $C$ and varying the irregular parameters
$u_2,\dots,u_n$.  It is plausible that this analogy extends to the
present discussion.  The interpretation of the geometric Langlands
program via four-dimensional topological field theory implies that
complex analytically the $B$-model, the $A$-model, and the
associated ${\cal D}$-modules must all be invariant under local
deformations of the complex structure of $C$. (There are global
monodromies.)  From an algebraic point of view, these statements
may  hold only  infinitesimally.

\subsection{Symmetries}\label{symmetries}

An important feature of geometric Langlands duality is that it
commutes with the action on branes of certain ``line operators.'' In
the unramified case, Wilson operators classified by a representation
of $^L\neg G$ act on the $ B$-branes of $^L\neg G$ gauge theory.
Electric-magnetic duality maps them to 't Hooft operators, also
classified by representations of $^L\neg G$, that act on the
$A$-branes of $G$ gauge theory -- and on the associated ${\cal
D}$-modules.

In the presence of tame or wild ramification, the same algebra of
Wilson or 't Hooft operators acts at an unramified point. In the
case of tame ramification, however, the algebra of line operators
that can act at a ramification point is more complicated. One way to
describe it, explained in section 4.5 of \cite{GW} (for a very
different point of view about a closely related problem, see
\cite{Bez}), involves monodromies in the space of parameters.  A
surface operator governing tame ramification at a point $p\in C$ is
labeled by parameters $(\alpha,\beta,\gamma,\eta)\in \T\times \frak
t^2\times {}^L\neg \T$.  A local singularity (that is, a singularity
that only depends on the behavior near $p$, and not on global
properties) develops when this quartet of parameters ceases to be
regular, that is when some element $w\not= 1$ of the Weyl group
${\cal W}$ leaves fixed $(\alpha,\beta,\gamma,\eta)$.

The $B$-model, for example,\footnote{The $A$-model of course is
similar with $\alpha$ and $\eta$ exchanged.} depends on $\alpha$ and
$\gamma$, which determine the monodromy
$U=\exp(-2\pi(\alpha-i\gamma))$ around the ramification point $p$.
Let ${\cal W}_U$ be the subgroup of the Weyl group of $^L\neg G$
that fixes $U$.  We say that a pair $(\beta,\eta)\in \frak t\times
{}^L\neg \T$ is ${\cal W}_U$-regular if it is not left fixed by any
element of ${\cal W}_U$ other than the identity.  Let ${Z}_U$ be the
space of ${\cal W}_U$-regular pairs. The $B$-model can be regarded
as a locally constant family of models parametrized by ${ Z}_U/{\cal
W}_U$.  The fundamental group of this quotient acts as a group of
symmetries of the model.

What this group is depends very much on $U$.  For $U=1$, which
means that $\alpha=\gamma=0$, it is the affine braid group of
$^L\neg G$. If, however, $U$ is regular semisimple, then ${\cal
W}_U$ is trivial and $ Z_U/W_U$ is topologically the same as
$\T_\C$. Its fundamental group is abelian, generated by lattice
shifts in $\eta$, and can be identified as the character group
of\footnote{The corresponding symmetries of the $B$-model can be
interpreted in terms of Wilson operators associated with
representations of the torus $^L\neg \T_\C$.  At a generic point
in $C$, the structure group is $^L\neg G_\C$ and the natural
Wilson operators are associated with a representation of this
group; but at the point $p$, the structure group is reduced from
$^L\neg G$ to $^L\neg \T_\C$ by the monodromy $U$, and one can
consider Wilson operators associated with a representation of the
torus. The representation ring of the torus is larger than that of
the full group, but is still abelian.} $^L\neg \T_\C$.

In short, a nonabelian symmetry of branes arises if $U$ is not
regular semisimple because, while keeping fixed $\alpha$ and
$\gamma$, one can vary $\beta$ and $\eta$ to get a local
singularity.  The nonabelian symmetry involves monodromies around
such singularities.

Now let us determine the analog for wild ramification. In the wild
case, the parameters are $u_2,\dots,u_n$ as well as
$(\alpha,\beta,\gamma,\eta)$.  In the $B$-model, for example, we
must hold fixed $u_2,\dots,u_n$ as well as
$U=\exp(-2\pi(\alpha-i\gamma))$. (In section (\ref{excase}), we
will refine the analysis to incorporate isomonodromy.)   Because
of our assumption that $u_n$ is regular semisimple, the whole
collection of parameters does not commute with any non-trivial
Weyl transformation.  A local singularity does not arise even if
we set $\alpha=\beta=\gamma=\eta=0$.   Indeed, nothing goes wrong
in the analysis in \cite{BB} if we set $\alpha=\beta=\gamma=0$
(and $\eta$ is anyway not part of that classical analysis) as long
as $u_n$ remains regular semisimple.  Since there are no local
singularities when $\beta$ and $\eta$ are varied, the only
monodromies come from lattice shifts of $\eta$ and the relevant
monodromy group is the character group of $^L\neg \T_\C$.
Effectively, having $u_n$ regular semisimple is similar to having
$U$ regular semisimple in the tame case.

Traditionally, in the Langlands program, one is most interested in
symmetries that can act on branes associated with a given flat
$^L{}G_\C$-bundle $E\to C$.  If we make this restriction (we will
relax it in section (\ref{excase})), we can explain more simply
why the algebra of line operators is commutative in the case of
wild ramification.  As long as $E$ is irreducible, it corresponds
in the wild case (with $u_n$ regular semisimple for the moment) to
a smooth point in $\MH$ and thus to a canonical zerobrane.  The
algebra of line operators that acts on branes associated with $E$
must be commutative, since there is only one object for this
algebra to act on.

By contrast, for tame ramification with $U$ not  semisimple, it is
possible to have a flat bundle $E\to C$ that corresponds to a
local singularity of $\MH$.  In this case, $E$ corresponds not to
a single brane, but to a whole ``category'' of branes supported at
the singularity.  A nonabelian algebra of line operators can act
on this category.

It may seem that this result simply reflects the fact that we have
constrained $u_n$ to be regular semisimple. In section
\ref{gcase}, it will hopefully become clear that this is not the
case. In a certain sense, wild ramification never presents
complications that do not occur for tame ramification, and
completely wild ramification (not reducing to tame ramification in
a subgroup; see section \ref{hint}) is always analogous to the
easy case of tame ramification with regular semi-simple monodromy.

\subsubsection{Action Of The Braid Group}\label{excase}
We have considered so far the symmetries that are analogous to those
that are present in the tame case.  However, we should also look for
new constructions of symmetries.

In fact, in the presence of wild ramification, the $B$-model and the
dual $A$-model do have a new kind of symmetry, already described in
section (\ref{interpretation}), which results from the fact that
these models are invariant under deformations of $u_2,\dots,u_n$ via
isomonodromy. The constraint that $u_n$ should be semisimple means
that $u_n$ takes values in a space ${\mathfrak t}_{\C}^{\rm reg}$
that is not simply-connected. The fundamental group of ${\mathfrak
t}_{\C}^{\rm reg}$ is called the braid group of $G$; we denote it as
$B(G)$.  (From this definition, it is clear that $B(G)=B({}^L\neg
G)$.)  Because of the complex symplectic nature of isomonodromy,
$B(G)$ acts as a group of symmetries of the $B$-model and the
$A$-model.  The argument of section (\ref{physper}) shows that it
commutes with the duality between them.

The action of $B(G)$ on $\MH$ has been described in an important
special case ($C=\Bbb{CP}^1$ with one point of wild ramification and
one point of tame ramification) in \cite{Btwo}.  An interesting
question is whether the action of $B(G)$ is algebraic.  The
isomonodromy equations themselves are algebraic, and while their
solutions are not algebraic, their monodromy may be.

A brane associated with a particular flat $^L{}G_\C$-bundle $E\to C$
is not an eigenbrane for the action of $B(G)$.  The action of $B(G)$
maps $E$ to other flat bundles (generally to infinitely many of
them) with the same values of $U$ and of $u_2,\dots,u_n$.

\section{Relaxing The Main Assumption}\label{gcase}
\label{ugo}

In studying flat connections with an irregular singularity
\begin{equation}\CA_z=\frac{T_n}{z^n}+\frac{T_{n-1}}{z^{n-1}}+\dots+\frac
{T_1}{z}+\dots,\end{equation} we have always, until this point,
assumed that the coefficient $T_n$ of the leading singularity is
regular and semi-simple.  It follows then that up to a gauge
transformation, one can assume  the singular part of the
connection to be $\frak t_\C$-valued.

Our goal in the present section is to relax this key assumption.  In
section \ref{semisimp}, we relax the assumption that $T_n$ is
regular.  In section \ref{sltwo}, we relax the assumption that $T_n$
is semi-simple.  We carry out that discussion in detail for
$G=SU(2)$ and indicate some of the ideas for the general case.

\def\LL{{\Bbb{L}}}
\subsection{Semi-Simple Singularity}\label{semisimp}

The starting point for much of our work has been the  analysis by
Biquard and Boalch \cite{BB} of Hitchin's differential equations
with irregular singularities.  The assumption made in that work is
weaker than assuming that $T_n$ is regular semi-simple.  They
assume that in some gauge the polar part of the connection, or
equivalently the objects\footnote{Moreover, they  indicate how the
assumption on $T_1$ can be removed by combining their analysis
with that in \cite{Biq}.  Thus, it is really only  necessary to
assume that the irregular coefficients $T_n,\dots,T_2$ are $\frak
t_\C$-valued in some gauge.} $T_n,\dots,T_1$, take values in
$\frak t_\C$. (It is equivalent to assume simply that
$T_n,\dots,T_1$ are semi-simple. In that case, one can conjugate
$T_n$ to $\frak t_\C$, after which, following logic explained in
section \ref{prelim}, one can make a gauge transformation to
ensure that $T_{n-1}$  commutes with $T_n$. Then one conjugates
$T_{n-1}$ to $\frak t_\C$, after which one repeats the procedure
until $T_n,\dots,T_1$ are all $\frak t_\C$-valued.)

The main theorem of \cite{BB} can be stated in essentially the
same way under this more general hypothesis. To a $G_\C$-valued
flat connection $\CA$ with a singularity of this kind, together
with a choice of $\beta\in \frak t$, corresponds a solution
$(A,\phi)$ of Hitchin's equations in which $\varphi$ has the sort
of singularity that we would by now expect
\begin{equation}\label{sortof}\varphi=\frac{u_n}{z^n}+\frac{u_{n-1}}{z^{n-1}}
+\dots+\frac{u_1}{z}+\dots.\end{equation} The relation between the
$u$'s and the $T$'s is the familiar one; thus, $u_k=T_k$ for
$k>1$, while $u_1$ and $T_1$ are expressed in the usual fashion in
terms of $\alpha,\beta,\gamma\in \frak t$, that is
$T_1=-i(\alpha-i\gamma)$, $u_1=\beta+i\gamma$.

Biquard and Boalch construct a hyper-Kahler moduli space
$\MH(\alpha;\vec u)$ that varies smoothly with these parameters as
long as the unbroken symmetry groups do not jump. To be precise
here, we introduce a sequence of subgroups of $G$ defined as
follows.  We let $\Bbb{L}_0$ be the subgroup that commutes with
all variables $u_1,\dots,u_n$ and $\alpha$ (or $T_1,\dots,T_n$ and
$\beta$), and for $k=1,\dots,n-1$, we let $\Bbb{L}_k$ be the
subgroup that commutes with $u_{k+1},\dots, u_n$.  Thus
$\Bbb{L}_0\subset\dots\subset\Bbb{L}_{n-1}$ are an ascending chain
of Levi subgroups of the gauge group. (This chain of Levi
subgroups is implicit in \cite{BB}, for example in equations (2.2)
and (2.3).) If the collection $T_1,\dots,T_n$ is regular, we have
$\Bbb{L}_0=\Bbb{T}$.

In defining the moduli space, we pick such an ascending chain of
Levi subgroups, and allow the parameters $\alpha$ and
$u_1,\dots,u_n$ to vary freely, subject only to the condition that
the $\Bbb{L}_i$ remain fixed. Under this restriction,
$\MH(\alpha;\vec u)$ varies smoothly. Moreover, it has all of the
properties that are familiar from the case that $u_n$ is regular
semi-simple. For example, as a complex symplectic manifold in
complex structure $J$, it is independent of $u_2,\dots,u_n$.  The
cohomology classes of the symplectic forms are still as presented in
(\ref{yfosto}).

Following the logic of sections \ref{supersurf} and \ref{quantpar},
we introduce surface operators characterized by the classical
parameters $\alpha$ and $\vec u$ plus the theta-like parameters
$\eta$. The familiar arguments motivate the hypothesis that they
transform under electric-magnetic duality as described in section
\ref{quantpar}.  This leads to  a quantum field theory picture of
wild ramification precisely like what was described in section
\ref{wildram}.

Most statements that we have made for the case that $u_n$ is regular
semi-simple have straightforward analogs in the present situation.
For example, the $B$-model in complex structure $J$, and $A$-model
with symplectic structure $\omega_K$, and the duality between them
are all independent of the irregular parameters $u_2,\dots,u_n$.  In
fact, the physical treatment of isomonodromy given in section
\ref{physper} carries over immediately to this situation.

The Hitchin fibration can be constructed as usual, and the usual
arguments show that electric-magnetic duality acts by $T$-duality on
the fibers of this fibration.

\subsubsection{Interpretation Via ${\cal D}$-Modules}\label{realint}
One important point  really does require some explanation. To
describe a duality between the $B$-model in complex structure $J$
and the $A$-model with symplectic structure $\omega_K$, everything
proceeds in the usual way. However, we do require some explanation
of how to relate $A$-branes to ${\cal D}$-modules. We can
introduce in the usual way the canonical coisotropic brane on
$\MH(\alpha;\vec u)$. Given this, the main step to relate
$A$-branes to ${\cal D}$-modules is to approximate
$\MH(\alpha;\vec u)$ by the cotangent bundle (or more precisely by
an affine deformation of a cotangent bundle) of some space
${\tilde \M}$.

Let $E\to C$ be a holomorphic $G_\C$-bundle.  Locally, $E$ can be
trivialized and its structure group is the group of holomorphic
maps $g:C\to G_\C$.  We say that $E$ is endowed with structure of
type ${\Bbb{L}}^*$ at a point $p\in C$ if we are given a reduction
of the structure group of $E$ to the subgroup consisting of maps
to $G_\C$ that take values in $\Bbb{L}_{i,\C}$ mod $z^{i+1}$, for
$i=0,\dots,n-1$. (As usual, $z$ is a local complex coordinate that
vanishes at $p$.)   Thus, if $E$ is endowed with ${\Bbb{L}}^*$
structure, a gauge transformation $g(z)$ preserves this structure
if and only if $g(0)\in \Bbb{L}_{0,\C}$, $(g^{-1}(dg/dz))|_{z=0}$
is valued in $\frak l_{1,\C}$, and so forth.

If $E$ is endowed with a Higgs field $\varphi$ with a polar part
as described in eqn. (\ref{sortof}), then this determines a
natural ${\Bbb{L}}^*$ structure, given by the subgroup of gauge
transformations that leave fixed the polar part of $\varphi$. Thus
a Hitchin pair $(E,\varphi)$ determines a point in what we might
call $\M({\Bbb{L}}^*,p)$, the moduli space of $G_\C$-bundles with
${\Bbb{L}}^*$ structure at $p$.  Just as in section \ref{cotbun},
it is not true that $\MH$ can be approximated as the cotangent
bundle to $\M({\Bbb{L}}^*,p)$.  The dimensions are wrong; we want
to endow $E$ with half as much structure.

\def\EUB{{\eusm B}}
In section \ref{cotbun}, what we did at this stage was to pick a
Borel subgroup $\EUB$ that contains $u_n$.  This was in the
context of assuming $u_n$ to be regular semi-simple (which implies
that $\Bbb{L}_0=\dots =\Bbb{L}_{n-1}=\Bbb{T}$).  In general,
 we pick an ascending chain of Borel subgroups
$\EUB_0\subset\EUB_1\subset\dots\subset \EUB_{n-1}$ such that
$\Bbb{L}_k$ is a maximal compact subgroup of $ \EUB_k$ for all
$k$. This can be done by picking a maximal torus contained in
$\Bbb{L}_0$, choosing a Weyl chamber, and saying that the Lie
algebra of $\EUB_k$ is spanned by that of $\Bbb{L}_k$ plus the
positive roots.  There are some choices to be made here, so the
geometric Langlands duality, when expressed in terms of ${\cal
D}$-modules, is possibly not quite as natural as it is when
expressed in terms of $A$-branes.

We say that $E$ is endowed with structure of type $\EUB^*$ at the
point $p\in C$ if we are given a reduction of the structure group
 of $E$ to the subgroup consisting of maps to $G_\C$ that take
values in $\Bbb{L}_{i,\C}$ mod $z^{i+1}$, for $i=0,\dots,n-1$. Let
$\M(\EUB^*,p)$ be the moduli space of $G_\C$-bundles endowed with
structure of type $\EUB^*$ at $p$. The same arguments as in
section (\ref{cotbun}) show that $\MH(\alpha,\vec u)$ can be
approximated as an affine deformation of the cotangent bundle
$T^*\M(\EUB^*,p)$.

As in sections (\ref{intdm}) and (\ref{jungo}), this leads to a
restatement of the geometric Langlands duality, since it enables
us to naturally map branes in the $A$-model of $\MH$ with
symplectic structure $\omega_K$ to twisted ${\cal D}$-modules on
$\M(\EUB^*,p)$.

\subsubsection{Symmetries}

We should likewise re-examine the discussion in section
(\ref{symmetries}) of symmetries of the category of branes. The
operators that act at unramified points are the usual Wilson and
't Hooft operators, but what happens at points of ramification?

Nothing really changes in our discussion in section
(\ref{symmetries}) if the coefficients $u_2,\dots, u_n$, taken
together, commute only with $\T_\C$.  This condition is equivalent
to $\Bbb{L}_1=\T$.  When this is the case, nothing goes wrong with
the analysis in \cite{BB} even if we set $\alpha=\beta=\gamma=0$;
$\MH$ remains as a complete hyper-Kahler manifold.  As there are
no singularities to be avoided, the only monodromies (in the
$B$-model) come from lattice shifts of $\eta$, and the natural
operations at the point $p$ are Wilson lines associated with
representations of $\T_\C$.

The result is different if $\Bbb{L}_1\not=\T$.  In that case, the
local analysis  near the ramification point leads to singularities
if the triple $(\alpha,\beta,\gamma)$ ceases to be
$\Bbb{L}_1$-regular (that is, if the subgroup of $\Bbb{L}_1$ that
commutes with this triple is larger than a maximal torus).  The
justification of this claim is precisely as in  the tamely
ramified case \cite{GW}, but now with $\Bbb{L}_1$ taking the place
of the gauge group $G$.  Interesting nonabelian groups of
monodromies, such as the affine braid group of $\Bbb{L}_1$, can
definitely act at the point $p$.

\subsection{Non-Semi-Simple Singularity}\label{sltwo}

So far we have relaxed the assumption that the coefficient $T_n$ or
$u_n$ of the leading singularity is regular, while retaining an
assumption of semi-simplicity.  Our next goal is to relax the
assumption of semi-simplicity.  For example, we will allow $T_n$ to
be nilpotent.

One might think that this would involve completely new
complications, analogous to what happens in the tame case when the
monodromy is unipotent. However, classical facts about irregular
singularities ensure that this is not the case. In a sense, wild
ramification is always analogous to the easy case of tame
ramification, namely the case of regular semi-simple monodromy.

To make things easy, we will discuss this primarily for the case
$G=SU(2)$, $G_\C=SL(2,\C)$, deferring some simple remarks on the
general case to section (\ref{hint}).   As in section
(\ref{backgr}), we begin with a flat $SL(2,\C)$-bundle $E\to
C\backslash p$, where $p$ is a point at which we will allow
ramification.  After picking an extension of $E$ as a holomorphic
bundle over $p$, we trivialize its holomorphic structure  near $p$
and describe it by a connection $\CA=dz\,\CA_z$. Here $dz\, \CA_z$
is a holomorphic section of $K_C\otimes {\rm ad}(E)$, with a pole
at $p$.

Let us suppose that $\CA_z$ has a pole of order $n$:
$\CA_z=T_n/z^n+\dots$.  If $T_n$ is regular semi-simple, we are back
in the case that we have analyzed in most of this paper.  The only
alternative, for $G_\C=SL(2,\C)$, is to suppose that $T_n$ is
nilpotent. After conjugating $T_n$ to an upper triangular form,
$\CA_z$ looks like
\begin{equation}\CA_z=\begin{pmatrix} a& z^{-n}b\\ c&
-a\end{pmatrix},\end{equation} where by hypothesis, $a$ and $c$
have poles at most of order $n-1$ at $z=0$ and $b$ is regular
there. Now a gauge transformation of the form
\begin{equation}\label{ggg}g=\begin{pmatrix} 1 & 0 \\ f(z) &
1\end{pmatrix},\end{equation} with $f$ regular at $p$, can set $a$
to zero.  After doing this, suppose that $c$ has a pole of order
$k<n$, so
\begin{equation}\label{uhazy}\CA_z=\begin{pmatrix} 0 & z^{-n}b\\ z^{-k}\tilde c & 0
\end{pmatrix},\end{equation}
where $\tilde c$  is regular at $z=0$.  If $n-k\geq 2$, then we
can reduce $n$ and also reduce $n-k$ by making a further gauge
transformation with\footnote{Though $g$ is not single-valued near
$z=0$, its action on connections by $d_A\to g d_A g^{-1}$ is
well-defined.  The square root in $g$ means that a gauge
transformation by $g$ changes the topology of the bundle $E\to C$
that is obtained by extending over $p$ the holomorphic structure
of the flat bundle $E\to C\backslash p$. Needing to make this
gauge transformation means that the initial choice of extension of
$E$ was not optimal.}
\begin{equation}g=\begin{pmatrix} z^{1/2} & 0 \\ 0 &
z^{-1/2}\end{pmatrix}\end{equation} (followed by a triangular
gauge transformation as in (\ref{ggg}) to set the diagonal terms
to zero). Since $g$ is not invertible at $z=0$, this gauge
transformation has the effect of changing the extension over the
point $p$ of the holomorphic structure of the flat bundle $E$.

After repeating this process, we reduce to a connection of the form
(\ref{uhazy}), possibly with a smaller value of $n$, and with either
$k=n$ or $k=n-1$.  For $k=n$, we are back in the familiar situation
that $T_n$ is semisimple.  So the only really new case to consider
is a connection of the form
\begin{equation}\label{hazy}\CA_z=\begin{pmatrix} 0 & z^{-n}b\\ z^{-n+1}c & 0
\end{pmatrix}.\end{equation}

Even this case is not really new if we are allowed to extract a
square root of $z$.  If we take a double cover of a neighborhood of
the point $p$, by introducing a new variable $t$ with $t^2=z$, then
we can reduce to the previous case via a gauge transformation
\begin{equation}\label{elme}g=\begin{pmatrix} t^{1/2} & 0 \\ 0 &
t^{-1/2}\end{pmatrix}.\end{equation}   In terms of $t$, we write
$\CA=dt\,\CA_t$, and since $dz=2t\,dt$, we have $\CA_t=\CA_z/2t$.
After making the gauge transformation (\ref{elme}), followed by a
gauge transformation of the form (\ref{ggg}) to set the diagonal
part of the connection to zero, we get
\begin{equation}\label{azy}\CA_t=\begin{pmatrix} 0 & t^{-2n}b\\ t^{-2n}c & 0
\end{pmatrix},\end{equation}
with new functions $b$ and $c$.  Again, the leading coefficient of
$\CA_t$ is regular semi-simple.  $\CA_t$ is a function only of
$z=t^2$, and so is even under $t\to -t$.  So $\CA=dt\,\CA_t$ is odd
under $t\to -t$.

We are not really allowed to take such a double cover of the
$z$-plane in quantum field theory, but we can do so in the
classical analysis of  conformally invariant partial differential
equations, such as Hitchin's equations.  So the above observation,
which is a standard one in the study of irregular singularities,
tells us how to adapt the analysis in \cite{BB} so as to apply to
a connection of the form (\ref{hazy}).

The local model of a solution of Hitchin's equations that one should
start with is essentially the model described in section
(\ref{locmod}).  Since $\CA_t$ as written in eqn (\ref{azy}) is
regular semi-simple, we can conjugate it to a diagonal form and
essentially borrow the ansatz of eqn. (\ref{jugid}).  The only
modification we need is to set $\alpha$ and some of the other
coefficients to zero in order to ensure that $\CA=A+i\phi$ is odd
under $t\to -t$.  So on the $t$-plane, the local model solution of
Hitchin's equations is
\begin{align}\label{ulgid}A&=0  \\
\nonumber
\phi&=\frac{dt}{2}\left(\frac{v_{n-1}}{t^{2(n-1)}}+\frac{v_{n-2}}{t^{2(n-2)}}+\dots+\frac{v_1}{t^2}\right)
 +\frac{d\bar t}{2}\left(\frac{\bar v_{n-1}}{\bar t^{2(n-1)}}+\frac{\bar v_{n-2}}{\bar t^{2(n-2)}}
 +\dots+\frac{\bar v_1}
 {\bar t^2}\right).
\end{align}

We would like to express this model solution on the $z$-plane.  We
cannot just divide by $t\to -t$ because $\CA$ is odd, rather than
even, under this transformation.  Instead,  we must accompany the
operation $t\to -t$ with a gauge transformation that anticommutes
with $\CA$.  Such a gauge transformation is
\begin{equation}\label{hobob}M=\begin{pmatrix}0&1\\
-1&0\end{pmatrix}.\end{equation}  On the $z$-plane, the local model
solution can be written
\begin{align}\label{pugid}A&=0   \\
\nonumber \phi&=
\frac{dz}{4}\left(\frac{v_{n-1}}{z^{n-1/2}}+\frac{v_{n-2}}{z^{2(n-3/2)}}+\dots+\frac{v_1}{z^{3/2}}\right)
 +\frac{d\bar z}{4}\left(\frac{\bar v_{n-1}}{\bar z^{n-1/2}}+\frac{\bar v_{n-2}}{\bar z^{n-3/2}}
 +\dots+\frac{\bar v_1}
 {\bar z^{3/2}}\right).
\end{align}

The meaning of the half-integral powers  of $z$ in (\ref{pugid})
is of course that the formulas are only valid away from a ``cut''
in the $z$-plane. In crossing this cut, we must make the gauge
transformation (\ref{hobob}).  At the cost of making the formulas
less transparent (and making it less obvious that they represent a
solution of Hitchin's equations), we can eliminate the cut by
making a unitary or $SU(2)$-valued gauge transformation on the
$z$-plane that removes the discontinuity, for example a gauge
transformation by $g=(z/\bar z)^{i M/8}$. This gauge
transformation puts the local model solution (\ref{pugid}) of
Hitchin's equations into a single-valued form. Being unitary, it
is a symmetry of Hitchin's equations.

\subsubsection{Surface Operators}

As in section \ref{supersurf}, the next step is to define surface
operators in ${\cal N}=4$ super Yang-Mills theory by requiring
that the local behavior near a codimension two surface $D$
coincides with the model solution that we have just described.
What parameters do these surface operators depend on?  The usual
parameters $\alpha$ and $\eta$ are absent because they are not
compatible with the gauge transformation (\ref{hobob}).  We
already noted that it is not possible to add a term
$A=\alpha\,d\theta$ to (\ref{ulgid}). Similarly, there is no
parameter analogous to $\eta$. The fields in (\ref{pugid}) and the
gauge transformation (\ref{hobob}) do not commute with any $SU(2)$
gauge transformations except the central elements $1$ and $-1$.
So along the support $D$ of a surface operator of this type, the
structure group of the bundle $E$ is reduced to the group $\{\pm
1\}$.  This leaves no possibility to introduce a $\theta$-like
parameter $\eta$ along the surface.  This contrasts, of course,
with the cases studied in \cite{GW} and in section
\ref{supersurf}; in those examples, the structure group along $D$
is $U(1)$, so it is possible to add a $\theta$-like angle.

So the usual parameters $\alpha$ and $\eta$ have no analogs for
$SU(2)$ surface operators of the type considered here. These two
statements are related to each other by duality, since $\alpha$
and $\eta$  transform into each other under duality.

These surface operators therefore depend only on the parameters
$\vec v=(v_1,\dots,v_{n-1})$.  These parameters control the
singularity of the Higgs field, so they transform under duality as
that field does. They therefore transform ``trivially,'' that is,
they transform precisely like the parameters $u_k$ in eqn.
(\ref{dinky}).

Another difference between the present case and the regular
semi-simple case is that there is no analog of the exponent of
formal monodromy.  There is no term in (\ref{pugid}) that is
precisely of order $1/z$.  Terms in the fields that are less
singular than $1/z$ are free to fluctuate and terms that are more
singular can be varied by isomonodromy.  The fact that $\alpha$,
$\eta$, and the exponent of formal monodromy are all absent means
that all parameters $v_1,\dots,v_{n}$ of a surface operator of this
type can be varied by isomonodromy.  Presumably, the usual
mathematical constructions \cite{JMU}, \cite{B} of isomonodromy can
be adapted to this situation.  From a physical point of view, the
isomonodromy operation can be justified as in section \ref{physper}.

A last comment along these lines is that symmetries that can act on
branes at a ramification point $p$ of this type are very limited.
Since the structure group along $D$ reduces to the center of
$SU(2)$, which is isomorphic to $\Bbb{Z}_2$, the Wilson operators
that can act at $p$ are associated with representations of
$\Bbb{Z}_2$.  Dually, the same is true for 't Hooft operators on the
other side.

As the only parameters are the irregular parameters $\vec v$, the
only monodromy that we can consider is the monodromy of the
isomonodromy connection.  Here we observe that $\MH(\vec v)$
varies smoothly only as long as the leading coefficient $v_n$ is
nonzero -- a statement precisely analogous to the usual
requirement in the semi-simple case that the leading coefficient
$u_n$ should be regular. So we can define the monodromy around the
origin in the complex $v_n$ plane.  This gives an automorphism of
the theory, but (just as in the semi-simple case) not one that
would usually be considered in the geometric Langlands program,
since a $B$-brane associated with a wildly ramified flat
$SL(2,\C)$-bundle is not an eigenbrane.

\subsubsection{Description By ${\cal D}$-Modules}\label{descripto}
The Hitchin fibration and the duality between the $B$-model and the
$A$-model go through in the usual way.  But as in section
\ref{realint}, a point that requires some discussion is the relation
of $A$-branes to ${\cal D}$-modules.  This depends upon relating
$\MH(\vec v)$, the moduli space of Higgs bundles in the presence of
the singularity, to the cotangent bundle of something.

To describe how to do this, we consider Hitchin pairs $(E,\varphi)$,
where $E$ is a holomorphic $G_\C$-bundle, and $\varphi$ is a Higgs
field with a singularity of the type that we have described.  In
some local trivialization of $E$, and with some choice of the local
coordinate $z$, $\varphi$ looks like
\begin{equation}\label{gurky}\varphi=dz\,\begin{pmatrix} 0 & z^{-n}\\ z^{-n+1} & 0
\end{pmatrix}+\dots\end{equation}
where the ellipses refer to regular terms.  (The choice of
trivialization and of the local parameter $z$ can be used to
eliminate from $\varphi$ additional polar terms of order less than
$n$.)  The choice of $\varphi$ endows $E$ with some additional
structure near the point $p$. In particular, the structure group of
$E$ is naturally reduced from the group of all holomorphic bundle
automorphisms to the subgroup consisting of those that commute with
the polar part of $\varphi$.  The condition for a holomorphic
section of ${\rm ad}(E)$ given by a matrix of functions
\begin{equation}\label{highly}\epsilon=\begin{pmatrix} a & b\\ c &
-a\end{pmatrix}\end{equation} to commute with the polar part of
$\varphi$, modulo regular terms, is easily seen to be that $a$ and
$zb-c$ are both divisible by $z^n$.

So $\MH(\vec v)$ has a natural map to a space $\M_0$ that
parametrizes holomorphic $G_\C$-bundles with a reduction of the
structure group to the subgroup just described.  As usual, this
endows $E$ with too much structure; to approximate $\MH(\vec v)$
as the cotangent space to something, we must define a space
${\M}^*$ that parametrizes bundles with just half as much
structure. Suppose that $n$ is even, say $n=2k$.  We can impose
half as much structure by requiring $a$ and $zb-c$ to be of order
$z^k$.  Thus we allow only infinitesimal gauge transformations
with generators of the form
\begin{equation}\label{hgy}
\epsilon=\begin{pmatrix} z^k\epsilon_1 & \epsilon_2\\
z\epsilon_2+z^k\epsilon_3 &
-z^k\epsilon_1\end{pmatrix}.\end{equation} Here $\epsilon_i$,
$i=1,2,3$ are regular at $z=0$.  Let ${\rm ad}_0(E)$ be the sheaf
of sections of ${\rm ad}(E)$ that are of this form. Such gauge
transformations  generate a Lie algebra. The cotangent space to
${\M}^*$ is $H^0(C,K_C\otimes {\rm ad}_0(E)^*)$, where ${\rm
ad}_0(E)^*$ is the dual to ${\rm ad}_0(E)$. Thus, the cotangent
space is spanned by differentials of the form
\begin{equation}\varphi=dz\,\begin{pmatrix}z^{-k}\varphi_1& z^{-k}\varphi_3\\
-z^{-k+1}\varphi_3+\varphi_2&-z^{-k}\varphi_1\end{pmatrix},\end{equation}
where $\varphi_i$, $i=1,2,3$ are regular at $z=0$.

We define an affine deformation  of the cotangent space to ${
\M^*}$ by shifting the cotangent space by the desired singularity
(\ref{gurky}). So a point in the fiber over $E$ of the affine
deformation is a pair $(E,\varphi)$, where the local form of
$\varphi$ is
\begin{equation}\varphi=dz\,\left\{\begin{pmatrix} 0 & z^{-n}\\ z^{-n+1} & 0
\end{pmatrix}  +\begin{pmatrix}z^{-k}\varphi_1& z^{-k}\varphi_3\\
-z^{-k+1}\varphi_3+\varphi_2&-z^{-k}\varphi_1\end{pmatrix}\right\}.\end{equation}
By a gauge transformation of the allowed form (\ref{hgy}), we can
in a unique fashion eliminate the polar part of the second term in
this formula and reduce $\varphi$ to the original form
(\ref{gurky}). This gives an embedding of this affine deformation
of the cotangent bundle of ${\M^*}$ as a dense open subset of
$\MH(\vec v)$. So, by the usual reasoning, $A$-branes on $\MH(\vec
v)$ map naturally to twisted ${\cal D}$-modules on ${\M^*}$.

In this discussion, to keep the formulas simple, we picked the
local coordinate $z$ to eliminate the parameters $\vec v$ from the
starting point (\ref{gurky}).  We can restore these parameters by
making the same argument in a more general coordinate system, or
equivalently by replacing the starting point (\ref{gurky}) with a
more general ansatz involving the parameters $\vec v$.

If $n$ is odd, say $n=2k+1$, we make a similar construction, taking
$a$ to be of order $z^k$ and $zb-c$ of order $z^{k+1}$ in
(\ref{highly}).

\subsubsection{A Hint About The General Case}\label{hint}

We can approach the general case of an irregular singularity for
$G$ of rank greater than 1 in much the same way.  The key
classical fact \cite{Wa} is that, given an irregular singularity
at $z=0$, after possibly passing to a finite cover of the
$z$-plane, one can reduce to the case that the irregular part of
the singularity can be diagonalized.

For simplicity, we consider $G_\C=SL(N,\C)$ or $GL(N,\C)$. For
other groups, one proceeds in much the same way, with conjugation
to the maximal torus playing the role of diagonalizing a matrix.

The connection $\CA_z$ is an $N\times N$ matrix-valued function
with a possible pole at $z=0$.  It has $N$ possibly multiple
eigenvalues $\lambda_i$, $i=1,\dots,N$.  The eigenvalues behave
for small $z$ as $\lambda_i\sim z^{-r_i}$, with rational numbers
$r_i$.  Tame ramification is the case that all $r_i$ are equal to
or less than 1.  What one might call completely wild ramification
is the case that $r_i>1$ for all $i$.  The general case is a
mixture of these two possibilities.

Let $k$ be the least common multiple of the denominators of all
those $r_i$ that are greater than 1.  Then if we pass to a
$k$-fold cover of the $z$-plane by $t^k=z$, we reduce to the case
that the irregular eigenvalues are all integers.  A gauge can then
be chosen in which the irregular part of the connection is
diagonal.  The procedure here generalizes what we did explicitly
for $G_\C= SL(2,\C)$.  (Recall that this procedure applies even
though to begin with the leading singularity may be nilpotent.) A
local model solution of Hitchin's equations can then be written
along the lines of eqns. (\ref{ulgid}) or (\ref{pugid}), and the
analysis of \cite{BB} can be applied.

Surface operators appropriate to this situation can be defined. As
in \cite{GW}, where the discussion was limited to the tame case
$r_i\leq 1$, in defining them, one specifies the coefficients of
those singular terms in the connection with $r_i\geq 1$. (Modes
with $r_i<1$ are square-integrable and can fluctuate quantum
mechanically; they are not specified as part of the definition of
a surface operator.) Mirror symmetry or electric-magnetic duality
can be invoked in the usual way to get a duality between the
$B$-model of $\MH({}^L\neg G,C)$ and the $A$-model of $\MH(G,C)$.

However, one key point is unclear.  In section \ref{descripto}, we
saw for $G_\C=SL(2,\C)$ how to approximate $\MH(G,C)$ by a
cotangent bundle, and therefore how to interpret $A$-branes of
$\MH(G,C)$ as twisted ${\cal D}$-modules on an appropriate
variety.  For general $G_\C$, this step is unclear.

\appendix
\section{Examples Of Stokes Matrices}

The purpose of this appendix is to briefly describe a few examples
in which Stokes matrices can be computed easily.

One very simple example is a differential equation in triangular
form, for example
\begin{equation}\left(\frac{d}{dz}+\frac{1}{z^n}\begin{pmatrix}1 & 0 \\ 0
& -1\end{pmatrix} +\begin{pmatrix} 0 & {h(z)}/{z^{n-1}} \\ 0 &0
\end{pmatrix}\right)\Psi=0,\end{equation}
where we assume that $h(z)$ is a polynomial.  (For a systematic
study of this type of example, see \cite{BJL}.)  One solution is
\begin{equation}\Psi_1=\exp(1/nz^{n-1})\begin{pmatrix} 1\\ 0
\end{pmatrix}.\end{equation}  For all values of ${\rm Arg}\,z$, this
solution obeys the standard asymptotic behavior of eqn. (\ref{asis})
as $z\to 0$.  This will ensure that (for a connection of the
triangular form considered here) the lower-triangular Stokes
matrices are all trivial. A second solution with the standard
asymptotic behavior  exists in a suitable angular sector.
 For the second solution, we can take
\begin{equation}\Psi_2=\begin{pmatrix}\exp(1/nz^{n-1})f(z)\\
                                     \exp(-1/nz^{n-1})
                                     \end{pmatrix}\end{equation}
where
\begin{equation}f(z)=-\int_0^z
dt\frac{h(t)}{t^{n-1}}\exp(-2/nt^{n-1}).\end{equation}  The
integration must be taken over a contour that approaches the
origin (at the lower limit) in a direction such that the integral
converges. We would like to pick this contour so that the
asymptotic behavior of $\Psi_2$ near $z=0$ will be
\begin{equation}\label{asbe}\Psi_2\sim \exp(-1/nz^{n-1})\begin{pmatrix}0 \\
1\end{pmatrix}.\end{equation} Then $\Psi_1,\Psi_2$ are a standard
pair of asymptotic solutions as specified in eqn. (\ref{asis}). In
any sufficiently small angular region in the complex $z$-plane, it
is possible to pick the contour to ensure the desired asymptotic
behavior of $\Psi_2$, but after analytic continuation to a larger
sector, the asymptotic behavior will be different.

\begin{figure}[tb]
{\epsfxsize=3in\epsfbox{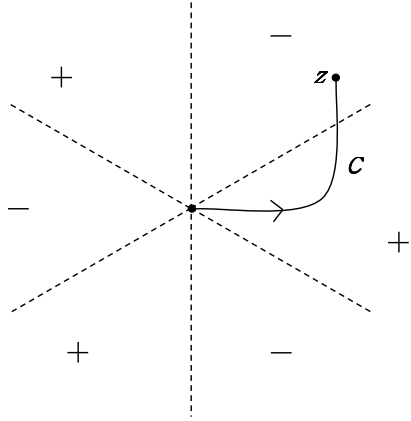}}
\begin{center}
\end{center}
 \caption{The dotted lines are the Stokes lines that divide the plane into $n-1$
 positive sectors and $n-1$ negative sectors, as sketched here for
 $n=4$.  The contour $\cal C$ runs from the origin to the point
 $z$; near the origin, it
  is a straight line
 in one of the positive sectors.
 }
 \label{uzlog}
\end{figure}
\begin{figure}[tb]
{\epsfxsize=3in\epsfbox{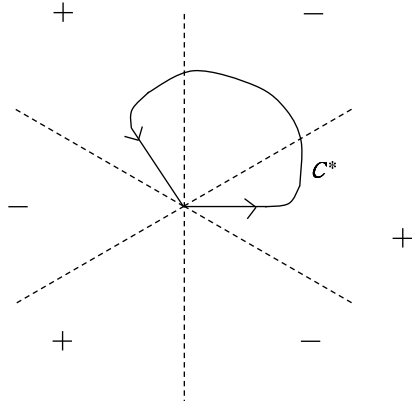}}
\begin{center}
\end{center}
 \caption{The contour ${\cal C}^*$ connects one positive sector to the next one.}
 \label{blog}
\end{figure}

 In this problem,
the Stokes rays are the rays with ${\rm Re}(1/z^{n-1})=0$. They
divide the $z$-plane into $n-1$ angular sectors with ${\rm
Re}(1/z^{n-1})>0$ and $n-1$ sectors with ${\rm Re}(1/z^{n-1})<0$
(fig. \ref{uzlog}). Let us call them positive and negative
sectors. If $\Psi_2$ is defined in a given positive sector using a
contour $\cal C$ that is a straight line from 0 to $z$, then it
has the asymptotic behavior of (\ref{asbe}).  It continues to have
this asymptotic behavior  when continued into an adjacent negative
sector. (In making this continuation, we keep the contour
unchanged near the lower limit, as in fig. \ref{uzlog}.) This
means that half of the Stokes matrices (the ones that would be
lower-triangular) are trivial for a differential equation of this
triangular type. However, when $\Psi_2$ is further continued to
the next positive sector, its asymptotic behavior no longer agrees
with (\ref{asbe}). Instead, we have
\begin{equation}\label{nasbe}\Psi_2\sim \exp(-1/nz^{n-1})\begin{pmatrix}0 \\
1\end{pmatrix}+w\exp(1/nz^{n-1})\begin{pmatrix}
1\\0\end{pmatrix},\end{equation} where \begin{equation}w=\int_{
{\cal C}^*}dt\frac{h(t)}{t^{n-1}}\exp(-2/nt^{n-1}).\end{equation}
Here $ {\cal C}^*$ is an integration contour that emerges from the
origin in one positive sector and ends by approaching the origin
in the next positive sector (fig. \ref{blog}). For solutions in
the second positive sector that do have the standard asymptotic
behavior of (\ref{asis}),  we can take
\begin{equation}\begin{pmatrix}\Psi_1' & \Psi_2'\end{pmatrix}=
\begin{pmatrix}\Psi_1 & \Psi_2\end{pmatrix}\begin{pmatrix}1& -w\\
0&1\end{pmatrix},\end{equation} showing the form of the Stokes
matrix.

One noteworthy fact is that the Stokes matrices depend on all
terms in the connection, including regular terms.  Thus, if
$h(z)=\sum_{k=0}^Na_k z^k$, then $w$ in general depends on all
coefficients $a_k$, including those with $k\geq n-1$.  The Stokes
matrices thus cannot be computed knowing only the singular part of
the connection.

For a second example, we consider the Airy equation
\begin{equation}\left(\frac{d^2}{dx^2}+x\right)\Psi=0.\end{equation}
This equation is equivalent to
\begin{equation}\left(\frac{d}{dx}+\begin{pmatrix} 0 & -1\\ x &
0\end{pmatrix}\right)\begin{pmatrix}u\\
v\end{pmatrix}=0,\end{equation} with $u=\Psi$, $v=d\Psi/dx$. This
equation has an irregular singularity at $x=\infty$; if we make the
change of variables $z=1/x$, we get a differential equation with a
singularity at $z=0$ that is of the type considered in eqn.
(\ref{hazy}), with $n=2$.

\begin{figure}[tb]
{\epsfxsize=3in\epsfbox{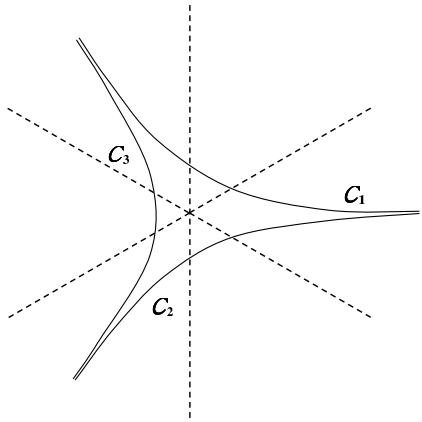}}
\begin{center}
\end{center}
 \caption{The contours ${\mathcal C}_i$ begin at infinity in the $i^{th}$ positive sector and end
 at infinity in the $i+1^{th}$. Their sum is homologous to zero.}
 \label{uzolog}
\end{figure}
This particular example of a differential equation with irregular
singularity can be analyzed rather explicitly.  Indeed, Airy's
equation can be solved by
\begin{equation}\Psi=\int_{\mathcal C}dp \,\exp(-p^3/3
-px),\end{equation} where ${\mathcal C}$ is a suitable contour. As
the function being integrated is an entire function on the complex
$p$-plane, there are no suitable closed contours.  We must select
a contour that begins and ends at infinity.  There are three
angular sectors in the $p$-plane in which ${\rm Re}\,p^3>0$.
${\mathcal C}$ must begin and end at infinity in one of these
contours.  If we denote as ${\mathcal C}_i$ a contour beginning in
the $i^{th}$ positive sector and ending in the $i+1^{th}$ (fig.
\ref{uzolog}), then we get three solutions of Airy's equations,
namely
\begin{equation}\Psi_i=\int_{\mathcal C_i}dp \,\exp(-p^3/3
-px),\end{equation} But they obey $\Psi_1+\Psi_2+\Psi_3=0$, since
${\mathcal C}_1+{\mathcal C}_2+{\mathcal C}_3$ is a closed contour
with no singularities inside.

To determine the asymptotic behavior of the solutions near the
irregular singularity, that is for $x\to\infty$, we first note that
the function $f(p)=p^3/3+px$ that appears in the exponent has two
critical points, at
\begin{equation}p_\pm=\pm \sqrt{-x}.  \end{equation}
Naively, the contribution of the critical point $p_\pm$ to the
integral will be of order $\exp(\pm (2/3)(-x)^{3/2})$ (times an
asymptotic series in negative powers of $x$).  We expect each of
the solutions $\Psi_i$ to have an asymptotic behavior for
$x\to\infty$ that will be a linear combination of such
exponentials.  The Stokes phenomena mean, in this context, that
the coefficients in these linear combinations will change when we
cross certain lines.

A simple way to proceed is actually (as in the appendix to
\cite{MMS})  to start with a given critical point, say $p_+$, and
ask what integral it dominates.
 We look for
a steepest descent contour ${\mathcal C}_+$ through $p_+$ with the
following properties: along ${\mathcal C}_+$, ${\rm Im}\,f$ is
constant, and ${\rm Re}\,f$ is minimized at $p_+$.  Because
$f'(p_+)=0$ and $f''(p_+)\not=0$, these conditions uniquely
determine what ${\mathcal C}_+$ should look like near $p_+$.
 Requiring the right behavior near $p_+$ and imposing
the condition that ${\rm Im}\, f$ is constant, we get a unique
contour $\mathcal C_+$ which passes through $p_+$. Generically, it
cannot be a closed contour; if it is closed, then ${\rm Re}\,f$
has a minimum on $\mathcal C_+$, which would be at a critical
point of $f$. The only critical point other than $p_+$ is $p_-$;
but generically ${\rm Im}\,f(p_+)\not={\rm Im}\,f(p_-)$ and a
contour $\mathcal C_+$ passing through $p_+$ and with ${\rm
Im}\,f$ constant cannot pass through $p_-$. So the contour
$\mathcal C_+$ generically extends to infinity and connects two of
the regions at infinity with ${\rm Re}\,p^3>0$. (It cannot start
and end in the same region, since then the integral would vanish,
while instead we will see momentarily that it is dominated by a
single critical point.)  Hence $\mathcal C_+$ is equivalent to one
of $\mathcal C_1,$ $\mathcal C_2$, or $\mathcal C_3$ (with one
orientation or the other). The function that must be integrated
over $\mathcal C_+$ has a constant phase and a maximum at $p_+$,
so it is dominated for large $x$ by the contribution of $p_+$. So,
as long as ${\rm Im}\,f(p_+)\not={\rm Im}\,f(p_-)$, one of the
three solutions $\Psi_1$, $\Psi_2$, and $\Psi_3$ is asymptotic to
$\exp((2/3)(-x)^{3/2})$. Similarly, one of the three solutions is
dominated by the critical point $p_-$ and is asymptotic to
$\exp(-(2/3)(-x)^{3/2})$.

Under $x\to \omega x$ with $\omega^3=1$, the three contours
$\mathcal C_i$ and three solutions $\Psi_i$ are permuted.  So
there is no natural way to pair up two of them with the two
critical points while omitting the third.  The association of
critical points $p_\pm$ with  contours $\mathcal C_i$ changes in
crossing the Stokes lines with ${\rm Im}\,f(p_+)={\rm
Im}\,f(p_-)$, where our analysis of the behavior of $\mathcal C_+$
breaks down.  By analyzing this process, one can compute the
Stokes matrices explicitly.

For more on this, see the appendix to \cite{MMS}.  For much more
on Airy's equation, along with other examples, see \cite{Wa}.

\bibliographystyle{amsplain}

\end{document}